\documentclass[twocolumn]{aastex61}
\pdfoutput=1 %for arXiv submission
\usepackage{amsmath,amstext}
\usepackage[T1]{fontenc}
\usepackage{apjfonts} 
\usepackage[figure,figure*]{hypcap}

\shorttitle{The Origin of Faint Tidal Features Around Galaxies in the RESOLVE Survey}
\shortauthors{Hood et al.}

\begin{document}

\title{The Origin of Faint Tidal Features Around Galaxies in the RESOLVE Survey}

\author{Callie E. Hood}
 \affiliation{Department of Physics and Astronomy, University of North Carolina, Chapel Hill, NC 27516, USA}
 \author{Sheila J. Kannappan}
 \affiliation{Department of Physics and Astronomy, University of North Carolina, Chapel Hill, NC 27516, USA}
 \author{David V. Stark}
 \affiliation{Department of Physics and Astronomy, University of North Carolina, Chapel Hill, NC 27516, USA}
 \affiliation{Kavli Institute for the Physics and Mathematics of the Universe (WPI),The University of Tokyo Institutes for Advanced Study, The University of Tokyo, Kashiwa, Chiba 277-8583, Japan}
 \author{Ian P. Dell'Antonio}
 \affiliation{Department of Physics, Brown University, Box 1843, Providence, RI 02912, USA}
 \author{Amanda J. Moffett}
 \affiliation{Department of Physics and Astronomy, University of North Carolina, Chapel Hill, NC 27516, USA}
 \affiliation{Department of Physics and Astronomy, Vanderbilt University, Nashville, TN 37240 USA}
 \author{Kathleen D. Eckert}
 \affiliation{Department of Physics and Astronomy, University of North Carolina, Chapel Hill, NC 27516, USA}
 \affiliation{Department of Physics and Astronomy, University of Pennsylvania, Philadelphia, PA 19104, USA}
 \author{Mark A. Norris}
 \affiliation{Jeremiah Horrocks Institute, University of Central Lancashire, Preston, Lancashire, PR1 2HE, UK}
 \author{David Hendel}
 \affiliation{Department of Astronomy, Columbia University, 550 W 120th St, New York, NY 10027, USA}

\begin{abstract}
We study tidal features (TFs) around galaxies in the REsolved Spectroscopy of a Local VolumE (RESOLVE) survey. Our sample consists of 1048 RESOLVE galaxies that overlap with the DECam Legacy Survey, which reaches an \textit{r}-band 3$\sigma$ depth of $\sim$27.9 mag arcsec$^{-2}$ for a 100 arcsec$^{2}$ feature. Images were masked, smoothed, and inspected for TFs like streams, shells, or tails/arms. We find TFs in 17$^{\pm 2} \%$ of our galaxies, setting a lower limit on the true frequency. The frequency of TFs in the gas-poor (gas-to-stellar mass ratio $<$ 0.1) subsample is lower than in the gas-rich subsample (13$^{\pm 3} \%$ vs. 19$^{\pm 2} \%$). Within the gas-poor subsample, galaxies with TFs have higher stellar and halo masses, $\sim 3\times$ closer distances to nearest neighbors (in the same group), and possibly fewer group members at fixed halo mass than galaxies without TFs, but similar specific star formation rates. These results suggest TFs in gas-poor galaxies are typically streams/shells from dry mergers or satellite disruption. In contrast, the presence of TFs around gas-rich galaxies does not correlate with stellar or halo mass, suggesting these TFs are often tails/arms from resonant interactions. Similar to TFs in gas-poor galaxies, TFs in gas-rich galaxies imply 1.7x closer nearest neighbors in the same group; however, TFs in gas-rich galaxies are associated with diskier morphologies, higher star formation rates, and higher gas content. In addition to interactions with known neighbors, we suggest that TFs in gas-rich galaxies may arise from accretion of cosmic gas and/or gas-rich satellites below the survey limit.
\end{abstract}


\section{Introduction}
In the paradigm of $\Lambda CDM$ cosmology, galaxy growth is driven by cosmological gas accretion as well as major and minor mergers. For decades, numerical simulations have shown how baryonic matter can trace merging events through the formation of discernible tidal features \citep[e.g.][]{toomre1972,dubinski1996,cooper2010}. A robust search for tidal features can help construct a fossil record of the history of recent galaxy accretion events, as well as test theories of long-term galaxy evolution \citep[e.g.][]{johnston2008}. Although the first simulations that demonstrated the formation of tidal features focused on major mergers, tidal features can also be formed by minor merger events \citep[e.g.][]{ebrova2013}. In simulations of the accretion of satellite dwarf galaxies by a larger host galaxy, \citet{bullock2005} successfully reproduced substructure similar to that observed around the Milky Way. Deep studies of individual nearby galaxies have detected faint features consistent with satellite accretion \citep[e.g.][]{martinez2008,martinez2015}. 

\indent Certain types of tidal features have been connected to particular types of merging events; we will use the same terminology as \citet{duc2015}. Tidal \textit{tails} or \textit{arms}, in which material from the primary galaxy is pulled out due to an interaction with a companion, are dependent on the primary's rotation (prograde for maximum effect) and thus are expected to come from progenitors with a gas-rich, disc dominated component \citep[e.g.][]{byrd1992,oh2008}. In contrast, tidal \textit{streams} consist of stripped material from a low-mass companion orbiting or being consumed by the primary galaxy; these features can be found around a variety of primary morphologies. Furthermore, broader \textit{fans} and \textit{plumes} are expected to result from dry, major mergers \citep{vandokkum2005,feldmann2008}. Shell systems found around elliptical galaxies \citep[e.g.][]{malin1983, schweizer1992} are expected to form from intermediate mass mergers. Recognition of merging systems based on detection of tidal features has been used in multiple studies of the merger fraction and rate at various redshifts \citep{conselice2008, bridge2010}. However, flyby interactions (in which no merger occurs) have also been shown to produce similar morphological disturbances, complicating the use of tidal features to identify mergers \citep[e.g.][]{richer2003, donghia2010}. 
 
\indent The varied origins of tidal features may lead to diverse results
from attempts to link those features, vs. other evidence of interactions, to galaxy properties such as gas fraction, star formation rate, and morphology. While some observational studies have found evidence that galaxy interactions can spur the conversion of HI into H$_{2}$ and the subsequent depletion of gas through star formation \citep[e.g.][]{lisenfeld2011, stark2013}, \citet{ellison2015} find no signs of gas depletion, despite enhanced star formation, in post-merger galaxies identified by tidal features. Interestingly,  simulations show gas-rich galaxy mergers exhibit tidal features for longer periods of time than their gas-poor counterparts \citep{lotz2010}. \citet{darg2010} studied merging galaxies with tidal features identified through the Galaxy Zoo project and concluded that the effects of mergers on spiral galaxies were much more dramatic (eroding their gas and angular momentum supplies and strongly enhancing their SFRs), so disturbed spiral galaxies were more easily observable than disturbed ellipticals. This result may also reflect the fact that modest gas fractions are expected to lead to more spheroidal remnant morphologies \citep{naab2006, duc2013}, while gas-rich mergers should yield disk-dominated remnants \citep{lotz2008,hopkins2009,lagos2017}.   

\indent Asymmetries in disks of galaxies have also been explained by gas accretion \citep{bournaud2005,mapelli2008,jog2009}. If the accretion is asymmetric, the stellar disk may be lopsided when the accreted gas turns into stars, producing features similar to those from galaxy interactions \citep{mapelli2008}. \citet{bournaud2005} studied the origin of disk asymmetry in spiral galaxies with N-body simulations and found that a large fraction of lopsidedness must come from cosmological accretion of gas to explain observational results.  However, the difference between accretion of gas and of a gas-dominated satellite, and between the resulting asymmetries in the primary galaxy, is not yet clear.

\indent Previous studies have produced discrepant estimates of the frequency of tidal features in the nearby universe from 3$\%$ to 70$\%$ \citep{malin1983,schweizer1988,vandokkum2005,tal2009,bridge2010,nair2010,miskolczi2011,kim2012,sheen2012,adams2012,atkinson2013,duc2015}. These discrepancies most likely stem from differences in detection criteria, sample selection, and surface brightness limits. 

\indent The varied nature of the origin of tidal features calls for a systematic study of their frequency in a statistically complete galaxy survey within which we can characterize gas content, environment, mass, morphology, and star formation. In this work, we present a uniform search for faint substructure within the highly complete volume-limited REsolve Spectroscopy of a Local VolumE (RESOLVE) survey. A census of the low surface brightness components of RESOLVE galaxies serves as the basis for investigating key questions about faint tidal features in the local universe. Which galaxy properties correlate most strongly with tidal features? Are these global trends, or are there separate subpopulations? Can the environmental dependence of tidal features illuminate their origins? 

This work is structured as follows. Section 2 discusses the data sets and methods used to identify faint tidal features around RESOLVE galaxies. Section 3 details the results of our census, characterizing our surface brightness limitations while investigating the relationship of detected features to various parameters recorded by the RESOLVE survey. A discussion of our results and comparison with previous works is presented in Section 4, and our conclusions are outlined in Section 5.

\section{Data and Methods}
\subsection{The RESOLVE Survey}
\indent Our main goal is to detect faint tidal features around galaxies in the RESOLVE survey, a volume-limited census of stellar, gas, and dynamical mass encompassing more than 50,000 cubic Mpc of the nearby universe \citep{kannappan2008}. A more thorough description of the survey design will given in S.J. Kannappan et al.\ (in preparation), but we briefly outline the key aspects of the survey below. 

\subsubsection{Survey Definition}
\indent RESOLVE is split into two equatorial strips, RESOLVE-A and RESOLVE-B. RESOLVE A spans 8.75 hr $<$ R.A. $<$ 15.75 hr and 0$^{\circ}$ $<$ Dec. $<$ 5$^{\circ}$, while RESOLVE-B spans from 22 hr $<$ R.A. $<$ 3 hr and -1.25$^{\circ}$ $<$ Dec. $<$ 1.25$^{\circ}$ (overlapping the deep SDSS Stripe-82 field). Both areas are also bounded in Local Group-corrected heliocentric velocity between $V_{LG}$=4500-7000 km s$^{-1}$. RESOLVE is an approximately baryonic mass limited survey, where baryonic mass ($M_{bary}$) is defined as the stellar mass ($M_{*}$) plus the atomic gas mass corrected for helium contributions ($1.4M_{HI}$). Galaxies in both RESOLVE regions are initially selected on \textit{r}-band absolute magnitude since this property tightly correlates with the total baryonic mass \citep{kannappan2013}. 

\indent The RESOLVE survey makes use of the SDSS redshift survey \citep{abazajian2009} as well as additional redshifts from the Updated Zwicky Catalog \citep{falco1999}, HyperLEDA \citep{paturel2003}, 2dF \citep{colless2001}, 6dF \citep{jones2009}, GAMA \citep{driver2011}, Arecibo Legacy Fast ALFA \citep{haynes2011}, and new observations with the SOAR and SALT telescopes (S.J. Kannappan et al., in preparation).  RESOLVE reaches \textit{r}-band completeness limits of $M_{r}$ $<$ -17 for RESOLVE-B and $M_{r}$ $<$ -17.33 for RESOLVE-A, the former limit being fainter as a result of RESOLVE-B's overlap with the deep Stripe-82 SDSS region. \citet{eckert2016} estimate the baryonic mass completeness limit by considering the range of possible baryonic mass-to-light ratios at the \textit{r}-band absolute magnitude completeness limit, obtaining $\sim$ 98\% completeness limits in baryonic mass at $M_{bary}$= $10^{9.3}$ $M_{\sun}$ and $M_{bary}$= $10^{9.1}$ $M_{\sun}$ in RESOLVE-A and RESOLVE-B, respectively.

\indent One advantage of the RESOLVE survey is the variety of multiwavelength data available. Several overlapping photometric surveys span near-infrared to ultraviolet wavelengths, which are used to estimate stellar masses and colors \citep{eckert2015}. The optical spectroscopic survey underway enhances redshift completeness over the SDSS \citep{eckert2016} for superior environment metrics and will in the future enable analysis of kinematics and metallicities. Importantly for the present work, 21 cm observations presented in \citet{stark2016} provide an unusually complete, adaptive sensitivity HI mass inventory for RESOLVE galaxies. 

\subsubsection{Custom Photometry and Stellar Masses}\label{photom}
\indent The extra-deep $r$-band imaging we use to identify tidal 
features is described in section 2.2; here we describe the baseline
multi-band photometry for RESOLVE. A full description of the photometric analysis for RESOLVE can be found in \citet{eckert2015}. All available photometric data, including SDSS \textit{ugriz} \citep{aihara2011}, 2MASS JHK \citep{skrutskie2006}, UKIDSS YHK \citep{hambly2008}, and GALEX NUV \citep{morrissey2007} were reanalyzed with custom pipelines to produce uniform magnitude measurements and improve the recovery of low surface brightness emission. This new uniform photometry was used to calculate stellar masses and other stellar population parameters using the spectral energy distribution fitting code described in \citet{kannappan2007} and modified in \citet{kannappan2013}. \citet{eckert2015} uses the second grid from \citet{kannappan2013}, which combines old simple stellar populations with ages ranging from 2 to 12 Gyr and young stellar populations described either by continuous star formation from 1015 Myr ago until between 0 and 195 Myr ago, or by simple stellar populations with ages 360, 509, 641, 806, or 1015 Myr.   This model grid includes four possible metallicities (Z = 0.004, 0.008, 0.02, and 0.05) and adopts a Chabrier IMF.  Likelihoods and stellar masses are computed for all models in the grid, and the median of the likelihood-weighted stellar mass distribution provides the most robust final stellar mass estimate. These stellar masses are given in \citet{eckert2015}.\footnote{We have used the updated stellar masses from the erratum to \citet{eckert2015}.} In addition, the short-term star formation rate of each galaxy is calculated from the GALEX NUV band using the calibration from \citet{wilkins2012}.  The ratio of the short-term star formation rate to the stellar mass is calculated to provide the specific star formation rate (SSFR) for each galaxy.

\subsubsection{HI Masses}\label{HI}
\indent The HI mass inventory for RESOLVE is fully described in \citet{stark2016}. That 21 cm data release is $\sim$ 94\% complete overall (94\% in RESOLVE-A, 95\% in RESOLVE-B), counting all galaxies with detections with S/N $>$ 5 or upper limits stronger than $1.4M_{HI}$/$M_{*}$ $\sim 0.05-0.1$. The HI masses and upper limits for RESOLVE come from the blind 21 cm ALFALFA survey \citep{haynes2011} as well as new observations with the GBT and Arecibo telescopes.  To increase the yield from the basic ALFALFA data products, \citet{stark2016} extracted 140 lower S/N detections and upper limits for RESOLVE galaxies within the ALFALFA grids. In addition, they acquired pointed observations with the GBT and Arecibo telescopes obtaining HI data for 290 galaxies in RESOLVE-A and 337 galaxies in RESOLVE-B, targeting those with either no HI measurements or weak upper limits from ALFALFA. 

\indent The HI masses and upper limits were measured according to the algorithms in \citet{kannappan2013} and \citet{stark2016}. Confused sources were identified based on the telescope used, with a search radius of 4$'$ for the ALFALFA smoothed resolution element, 9$'$ for the GBT, and 3.5$'$ for Arecibo pointed observations.  \citet{stark2016} deconfused the HI profiles when possible, building on the methods used in \citet{kannappan2013}.  For galaxies with confused detections, without HI observations, or with weak upper limits,  $M_{HI}$ is estimated using the relationship between color, axial ratio, and gas-to-stellar mass ratio as calibrated in \citet{eckert2015}.

\subsubsection{Environment Metrics}\label{envmetrics}
One of the benefits of studying tidal features in RESOLVE is the environmental context provided by the survey. \citet{moffett2015} identified groups of RESOLVE galaxies using the friends-of-friends (FoF) technique described in \citet{berlind2006}, which were then updated by \citet{eckert2016}.  \citet{eckert2016} used tangential and line-of-sight linking lengths of 0.07 and 1.1 times the mean spacing between the galaxies to optimize the group-finding procedure as recommended by \citet{duarte2014} and confirmed for RESOLVE-A. For RESOLVE-B, the volume is too small and subject to cosmic variance to use linking lengths from the survey's own mean galaxy density. Instead, a density appropriate for the depth of RESOLVE-B was constructed from the larger Environmental COntext (ECO) catalog (which encompasses RESOLVE-A, see \citealt{moffett2015}) by enforcing enforcing an \textit{r}-band luminosity $L_{r}$ selection limit of $<$ $-$17.0 mag to match RESOLVE-B. \citet{eckert2016} then determined the physical linking lengths and the relationship between the group halo mass and L$_{r}$, which were applied to RESOLVE-B. They assigned each group a halo mass by halo abundance matching (HAM) with the theoretical halo mass function from \citet{warren2006} using the total group \textit{r}-band luminosity \citep[for more details see][]{eckert2016}.

\indent Furthermore, for each galaxy in RESOLVE, we find nearest neighbors in two ways. First, we perform a cylindrical search for nearest neighbor galaxies also in RESOLVE within a radius of 100 Mpc in the plane of the sky.  Candidate neighbors are limited to $|V_{gal}$-$V_{neighbor}| \leq 500 $ km/s, where the local group corrected velocities are used to define galaxy redshift. Second, we also compute the distance to each galaxy's nearest neighbor using the algorithm described by \citet{man1999} as implemented in cKDTree in the scikit-learn package in Python \citep{pedregosa2011}. We implement this algorithm by converting RA, DEC, and redshift to physical units both with and without peculiar motions (so allowing false $\Delta z$ or zero $\Delta z$ within groups). The differences in results obtained from these methods are discussed in Section \ref{nnd}.

\subsection{Deep Imaging Sources}
This project utilizes \textit{r}-band images from both Data Release 3 of the DECam Legacy Survey\footnote{\url{http://legacysurvey.org/decamls/}} \citep[DECaLS;][]{blum2016} and the IAC Stripe 82 Legacy Project \citep{fliri2016}. The DECaLS survey covers both RESOLVE subvolumes, with deep \textit{r}-band image stacks of variable depths with a median 5$\sigma$ point source depth of 23.6 mag. Assuming Poisson noise, this depth yields a median surface brightness limit of $\sim$27.9 mag arcsec$^{-2}$ for a 3$\sigma$ detection of a 100 arcsec$^{2}$ low surface brightness feature \citep{schlegel2015}, allowing us to probe moderately faint tidal features around RESOLVE galaxies. However, scattered light and other image artifacts may contribute to functionally shallower surface brightness limits. DECaLS also provides depth maps for each stacked image in units of the canonical galaxy flux inverse-variance for each pixel, where the canonical galaxy is defined as an exponential profile with effective radius 0.45'' (much smaller than our features of interest).  For the specific DECaLS images used in this study, the distribution of 5$\sigma$ depths per pixel has a standard deviation of 0.6 mag around a median of 23.5 mag (similar to the point-source limit due to the small size of the ``canonical" galaxy). 

\begin{figure*}
\centering
\includegraphics[width=.95\linewidth]{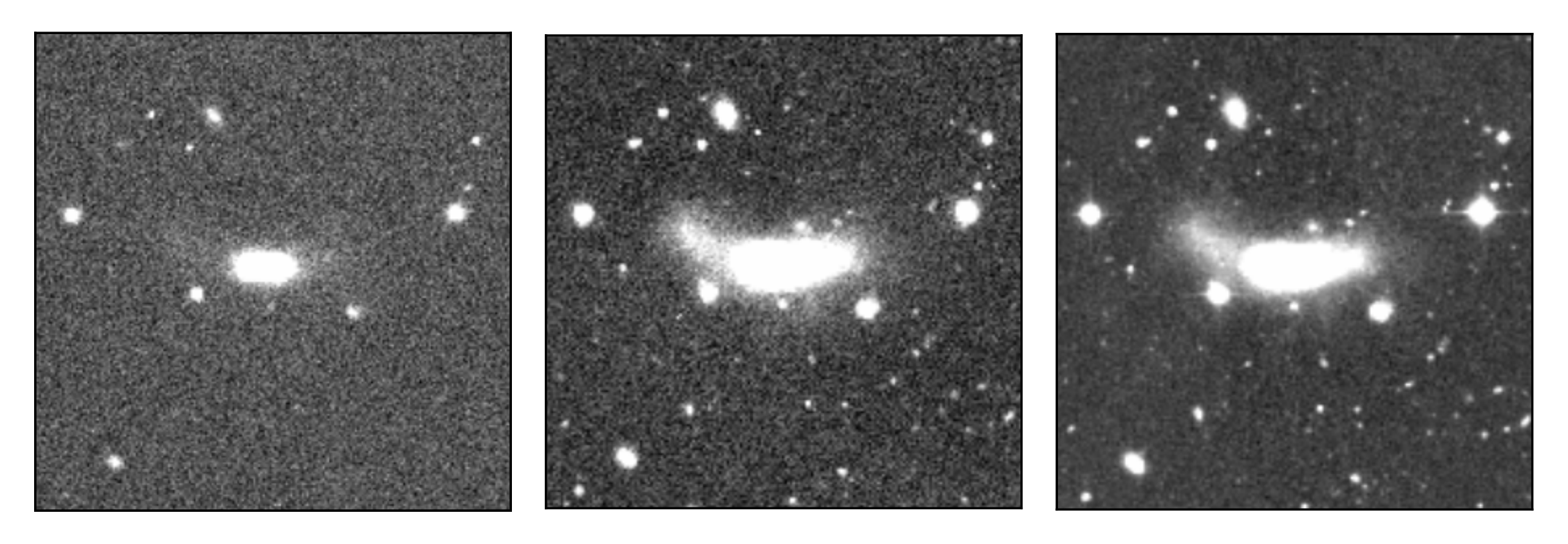}
  		\caption{A comparison of the SDSS (left), DECaLS (middle), and IAC (right) images of the galaxy rf0358.  The deeper DECaLS and IAC images reveal the faint extension on the left side of the galaxy.}
\label{fig:sdsscomp}
\end{figure*}

\indent In contrast, the IAC Stripe 82 Legacy Project provides deep \textit{r}-band co-adds for only RESOLVE-B, but reaches a surface brightness limit of 28.3 mag arcsec$^{-2}$ for a 3$\sigma$ detection of a 100 arcsec$^{2}$ feature \citep{fliri2016}. See Figure~\ref{fig:sdsscomp} for a side-by-side comparison of the galaxy rf0358 in SDSS, DECaLS, and IAC.  Though our analysis mainly focuses on results from the DECaLS images, the IAC co-adds allow for study of the dependence of our classifications on the surface brightness limits of our data. 

\subsection{Detection Process}\label{detection}
Although automating the detection of faint tidal features is an appealing goal, no  such current methods can surpass visual inspection for detecting faint and subtle substructures as well as avoiding false detections \citep[see][]{adams2012}. While human detection is the main driver of this project, various software packages were used in order to facilitate and improve the identification process. First, cutouts of 512 x 512 pixels (134 x 134 arcsec) in the \textit{r}-band were extracted of each galaxy from DECaLS DR3 image stacks. Due to an imperfect overlap of the RESOLVE-A sample and the DECaLS \textit{r}-band coverage, an initial 331 out of 1501 galaxies were not inspected with the DECaLS \textit{r}-band images. In addition, another 122 galaxies from both semesters were later removed from the detection sample during visual inspection due to various problems such as only part of the galaxy falling in the image, bright image defects, or fringing that would prevent any confident identification of tidal features. Thus, at the beginning of the detection process the sample of galaxies decreased from 1501 to 1048.

\begin{figure}[htb!]
\centering
\includegraphics[width=.95\linewidth]{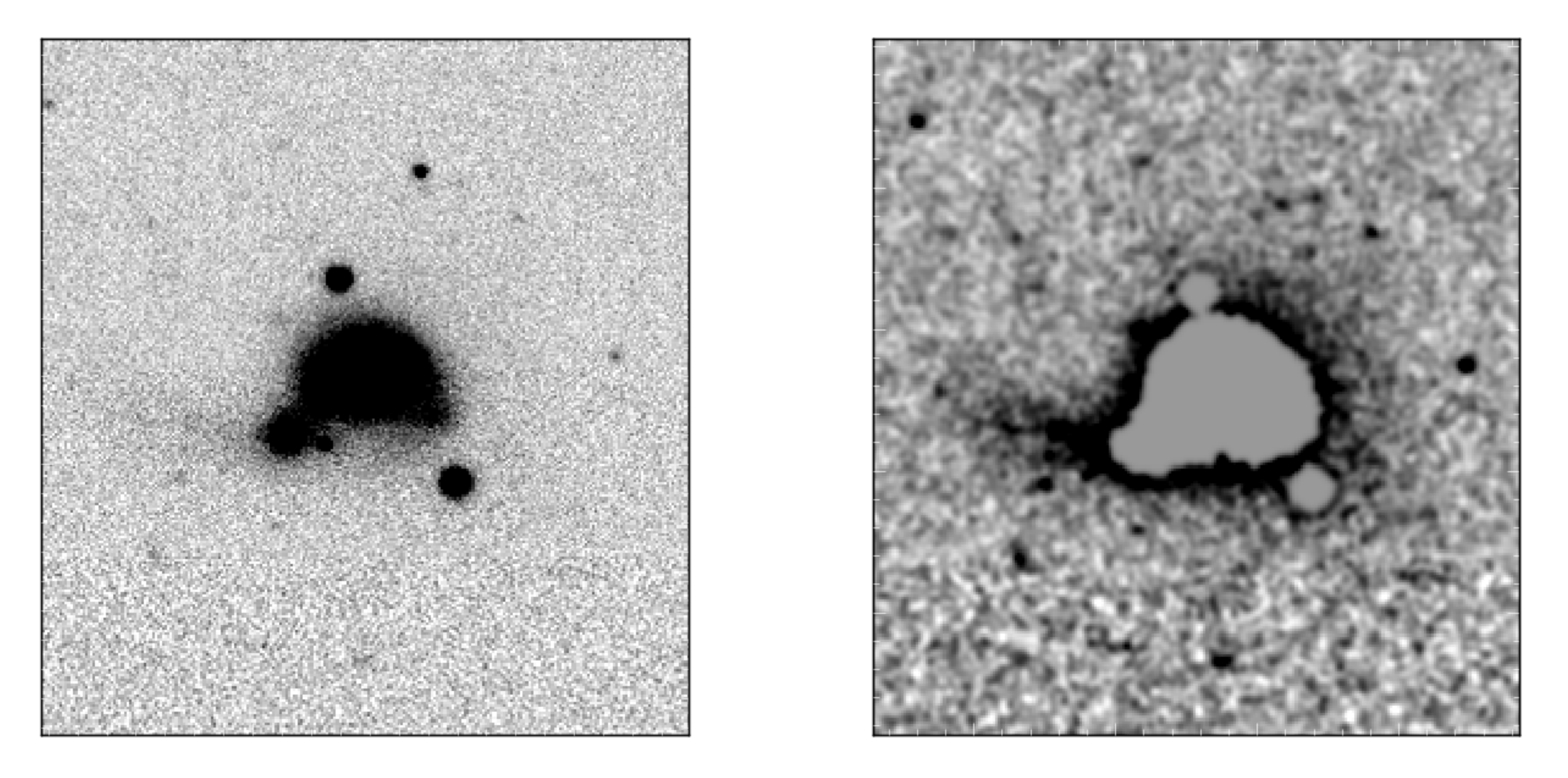}
  		\caption{Original DECaLS image of rf0464, as well as the smoothed and masked thumbnail. A tidal stream coming off of the satellite being accreted is enhanced in the smoothed image.}
\label{fig:rf0464}
\end{figure}

\begin{figure}[htb!]
\centering
\includegraphics[width=.95\linewidth]{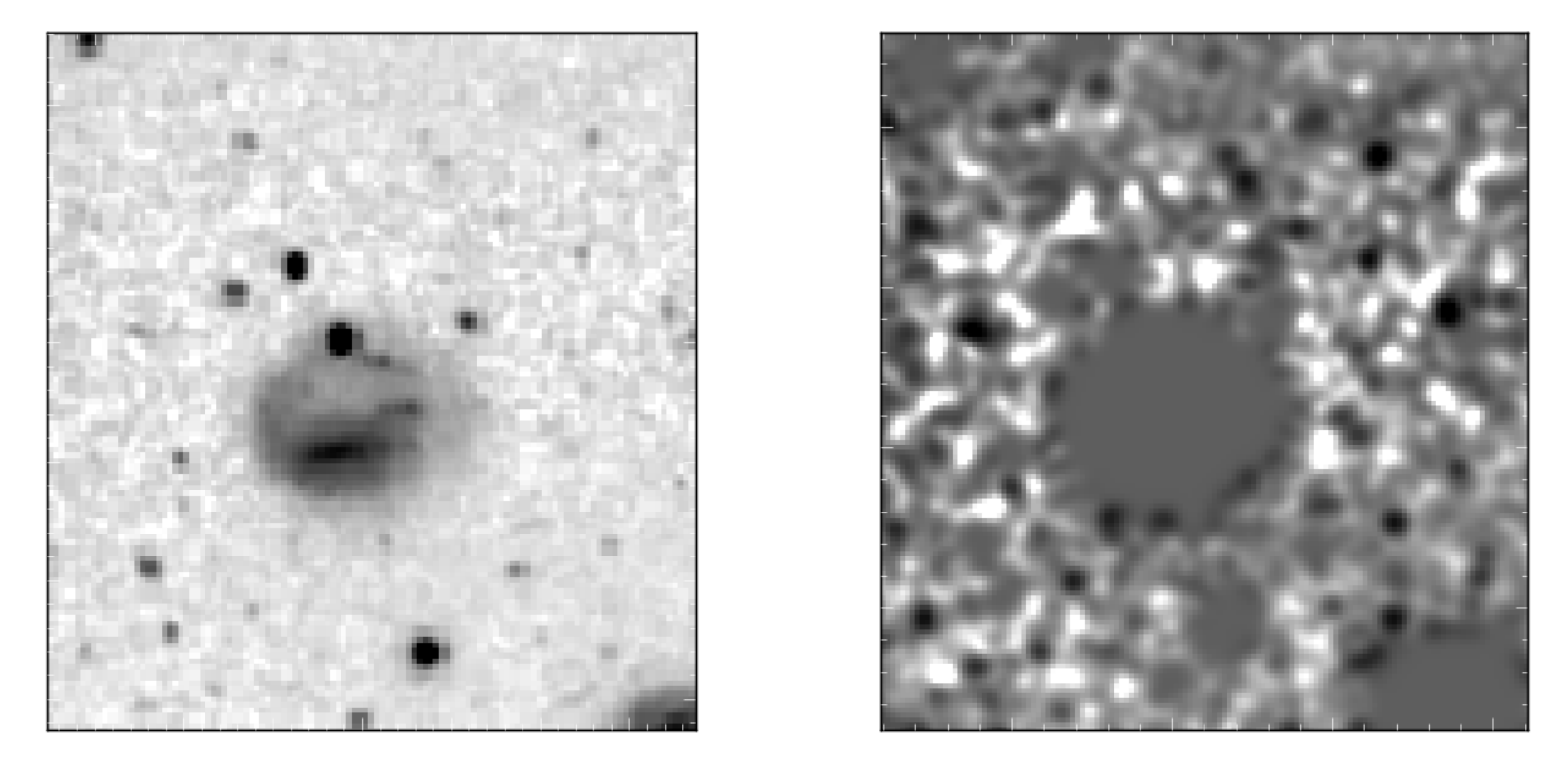}
  		\caption{Original DECaLS image of rf0305, which showcases a clear arm arcing above the galaxy. However, this feature is masked out of the smoothed image.}
        \label{fig:rf0305}
\end{figure}

\indent	To improve detection of particularly faint features, each galaxy thumbnail was run through a modification of a previously-developed masking and smoothing code (I. Dell$'$Antonio). For each image, SExtractor \citep{bertin1996} was used to detect sources within the image, and the resulting catalog was used to create binary masks. These masks were then applied to the original image using IRAF, nulling out these pixel locations and allowing for only faint and small-scale features to remain nonzero.  The masked thumbnail was then smoothed with a Gaussian filter with a sigma of 7 pixels (2 arcsec), chosen to enhance these features without smoothing them out entirely.  One such final image is shown in Figure~\ref{fig:rf0464}, with the faint tidal stream coming off of the smaller galaxy on the left much clearer in the smoothed image. However, while this masking and smoothing process greatly enhanced some of the features, some other substructure was possibly masked out in this process as confirmed in the test case shown in Figure~\ref{fig:rf0305}. As a result, both the original and the smoothed, masked images needed to be inspected for faint tidal features.

\indent    For the visual inspection, a code previously used for by-eye morphological classifications of RESOLVE-A galaxies \citep[see][]{moffett2015} was adapted with the APLpy Python package \citep{aplpy} to display images of galaxies as well as images with overlaid contours revealing larger-scale galactic structure. In addition, the masked and smoothed image of each galaxy, as well as the original, were displayed in a DS9 \citep{joye2003} window to allow for interactive adjustment of the contrast and scaling for each thumbnail. For each of the galaxies in the identification sample, a single classifier (Hood) recorded a flag after visual inspection indicating a positive or negative detection of faint tidal features, as well as any comments on particularly interesting substructures.  A similar process was used to classify tidal features around RESOLVE galaxies in the IAC images  for comparison with the DECaLS results (see Section \ref{comparison}).

\defcitealias{atkinson2013}{A13}

\indent All DECaLS images of our sample were then reclassified according to the detection classes defined by \citet{atkinson2013}, hereafter \citetalias{atkinson2013}, in order to aid comparison between the two studies. Each galaxy was put into a category from 0$-$4, with 4 denoting the ``certain" detection of a tidal feature, 3 denoting a ``probable" ($\sim$ 75\% certain) detection, 2 denoting a ``possible" ($\sim$ 50\% certain) detection, 1 denoting a ``hint" ($\sim$ 25\% certain) of a detection, and 0 denoting no evidence for tidal features. In addition, each galaxy in classes 1$-$4 was also assigned a flag to indicate the type of tidal feature seen.  We categorized our features as either ``narrow'' or ``broad.'' Our ``narrow'' category would encompass the ``streams,'' ``arms,'' and ``linear features'' categories from \citetalias{atkinson2013}, while ``broad'' includes their ``shells,'' ``fans,'' and ``miscellaneous diffuse structure'' labels.

\section{Results and Analysis}\label{results}
\subsection{Frequency of Tidal Features}\label{frequency}

\begin{table}[htbp]
\centering
\caption{Tidal Feature Detections} \label{tab:detect}
\begin{tabular}{ c c c}
\hline
\hline
Confidence & Number & Percentage \\
\hline
4 & 46 & $4.4\substack{+1.4 \\ -1.2}$ \\
3 & 134 & $12.8\substack{+2.2 \\ -2.0}$ \\
2 & 65 & $6.2\substack{+1.6 \\ -1.4}$ \\
1 & 59 & $5.6\substack{+1.6 \\ -1.3}$ \\
0 & 744 & $71.0\substack{+2.7 \\ -2.9}$ \\
\hline
\end{tabular}
\end{table}

% \begin{table*}[htbp]
% \centering
% \caption{Classes of Tidal Features} \label{tab:detect}
% \begin{tabular}{ c c c c c}
% \hline
% \hline
% Confidence & \multicolumn{2}{c}{\underline{Type 1}} & \multicolumn{2}{c}{\underline{Type 2}} \\ 
%  & Number & Percentage & Number & Percentage \\
% \hline
% 4 & 35 & $3.3\substack{+1.3 \\ -1.0}$ & 11 & $1.0\substack{+0.8 \\ -0.5}$\\
% 3 & 97 & $9.3\substack{+1.9 \\ -1.7}$ & 37 & $3.5\substack{+1.3 \\ -1.0}$\\
% 2 & 40 & $3.8\substack{+1.3 \\ -1.1}$ & 25 & $2.4\substack{+1.1 \\ -0.8}$\\
% 1 & 31 & $3.0\substack{+1.2 \\ -0.9}$ & 28 & $2.7\substack{+1.2 \\ -0.9}$\\
% \hline
% \end{tabular}
% \end{table*}

\begin{figure}[htbp]
\centering
\includegraphics[width=\linewidth]{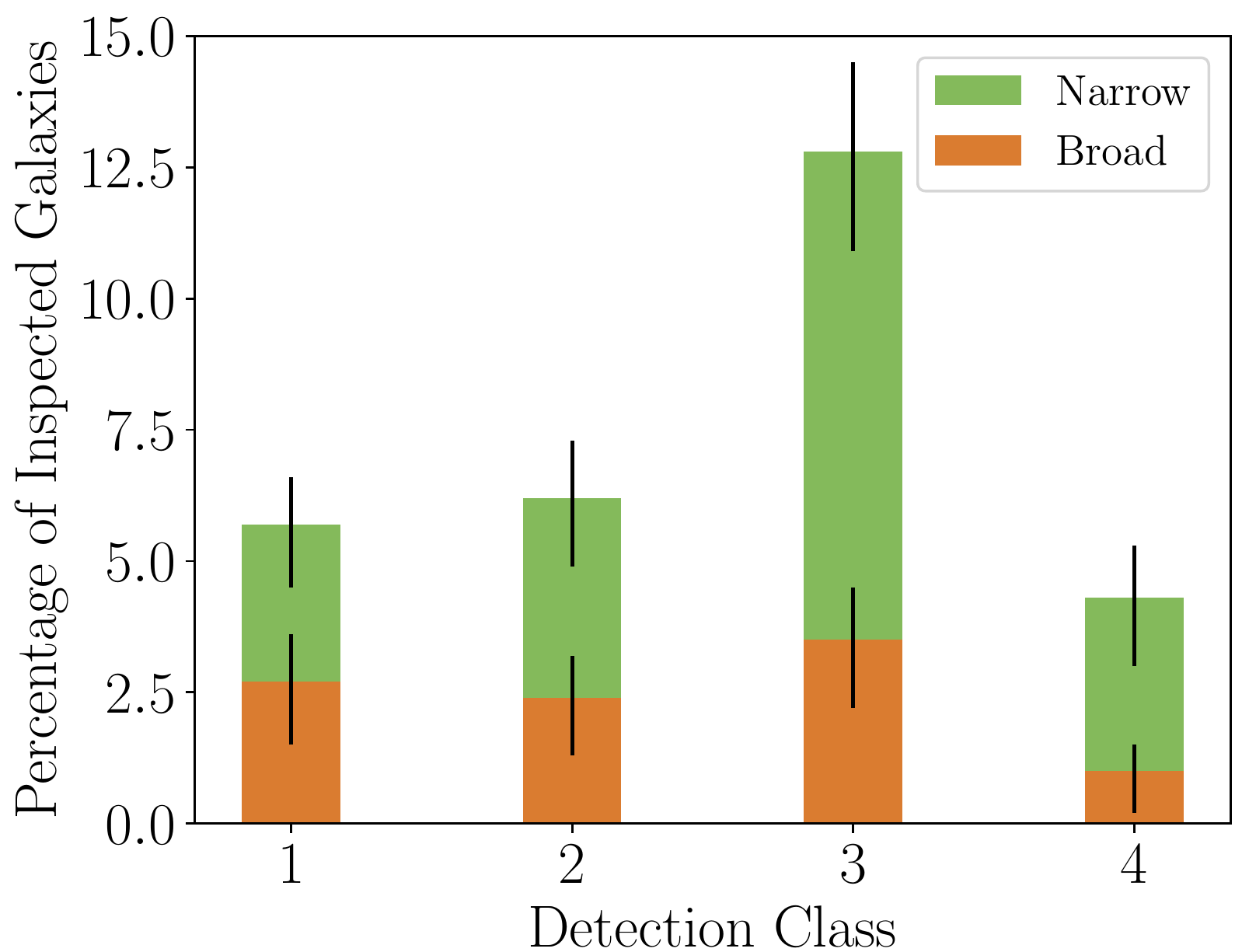}
  		\caption{Distribution of tidal feature detection classes, subdivided into ``narrow'' and ``broad'' features.}
\label{fig:typeperclass}
\end{figure}

The results of our tidal feature classifications are summarized in Table \ref{tab:detect}. We find that 17$^{\pm 2} \%$ of the 1048 RESOLVE galaxies inspected with DECaLS have tidal features detected with high confidence (classes 3 and 4). The uncertainty given is the range of possible percentages within a two-sided 1$\sigma$ confidence interval given small number binomial statistics.  As seen in Figure \ref{fig:typeperclass}, most of the features detected in all classes are narrow; however, some broader faint features are detected as well.  Though possibly visually indistinguishable, our sample presumably includes both tidal tails extending from the primary galaxy as well as tidal streams indicating a companion being stripped. We did not attempt to distinguish between different types of tidal features within our two categories, but future simulations of interactions causing these features would be useful to determine criteria for this level of sub-classification. A selection of features from each category found around RESOLVE galaxies is shown in the Appendix.  For simplicity, we will refer to galaxies with confident detections of tidal features as ``TF'' galaxies, while the rest (classes 0$-$2) will be ``NTF'' galaxies.  However, we will also discuss the effect on our results of treating class 2 galaxies as ``TF'' rather than ``NTF'' galaxies.

\indent Although streams and other substructures are predicted to be common in the current cosmological framework, they are not easy to detect. Depending on many factors, such as the relative masses of the interacting galaxies or the geometry of the interaction, a majority of substructure resulting from accretions can be expected to have surface brightnesses of 30 mag arcsec$^{-2}$ or fainter \citep{bullock2005}. This range is much lower than the limits of most imaging surveys, which in turn limits our ability to explore the statistical properties of these features. With the surface brightness limits of our DECaLS images being $\sim$27.9 mag arcsec$^{-2}$, we are able to probe the brightest of these features, but ultimately only put a lower limit on their frequency.  In addition, the \textit{lifetime} of a tidal feature resulting from a merger depends heavily on its surface brightness limit. \citet{ji2014} find that the major-merger feature lifetime goes up by a factor of two for a surface brightness limit of 28 mag arcsec$^{-2}$ when compared to a shallower limit of 25 mag arcsec$^{-2}$. Unfortunately, their models also provide evidence that tidal features from relatively minor mergers (mass ratios less than about 1/6) may be difficult to detect even in surveys that reach below 28 mag arcsec$^{-2}$.

\subsubsection{Comparison to Classifications with IAC Stripe 82 Co-adds}\label{comparison}
\indent For the 446 RESOLVE-B galaxies inspected with both DECaLS and IAC images, we find 20$^{\pm 2} \%$ and 24$^{\pm 2} \%$ have tidal features, respectively. The difference in frequency comes from 18 galaxies whose tidal features are detectable in the IAC co-adds but not the DECaLS images, most likely due to the $\sim$ 0.4 mag arcsec$^{-2}$ difference in surface brightness depth between the two image sources. We compared the stellar-mass, color, and gas-to-stellar mass ratio distributions of the two sets of galaxies with tidal features, finding no statistically significant differences between the two samples according to the Kolmogorov-Smirnov (K-S) test (p$_{KS} \sim 0.99$ for each comparison). Thus, the results presented in the following sections would likely not change significantly if we were using classifications from the deeper Stripe 82 co-adds. 

\subsubsection{Noise Degradation Experiment}\label{noise}
\indent In order to further test the dependence of our classifications on the surface brightness limit of our data, we have artificially degraded the IAC images to the same surface brightness limit as our DECaLS cutouts. When we re-classify these noisy IAC images, we find that $\sim$ 18$^{\pm 2} \%$ of the inspected galaxies host tidal features. We find that 21 of the TF galaxies in the DECaLS images are not similarly classified in the degraded IAC images; conversely, 9 of the TF galaxies in the degraded images are not flagged with the DECaLS images. The exact source of these discrepancies is unclear; some of these galaxies have barely discernible tidal features whose classification may be unreliable. We quantify the differences in these classifications with Cohen's kappa statistic, which is traditionally used to measure inter-classifier agreement while taking into account the probability of classifiers agreeing by chance. Cohen's kappa is calculated according to the following formula, where $p_{o}$ is the relative observed agreement between classifiers and $p_{e}$ is the hypothetical probability of chance agreement. 

\begin{equation}
\kappa \equiv {\frac{p_{o}-p_{e}}{1-p_{e}}}
\end{equation}

Our DECaLS and noisy IAC classifications have a Cohen's kappa value of 0.79 on a scale from $-$1 to 1, where 1 indicates perfect agreement. According to the hierarchy established by \citet{landis1977}, our value corresponds to a ``substantial" strength of agreement. As in 3.1.1, we do not find any statistically significant differences between the distributions of various parameters (stellar mass, color, etc.) for the TF galaxies from the two imaging sources. Therefore, any marginally detectable tidal features or false detections should have a minimal effect on the results below.

\subsection{Galaxy Properties and Tidal Features}\label{prop}

\begin{figure}
\centering
\includegraphics[width=\linewidth]{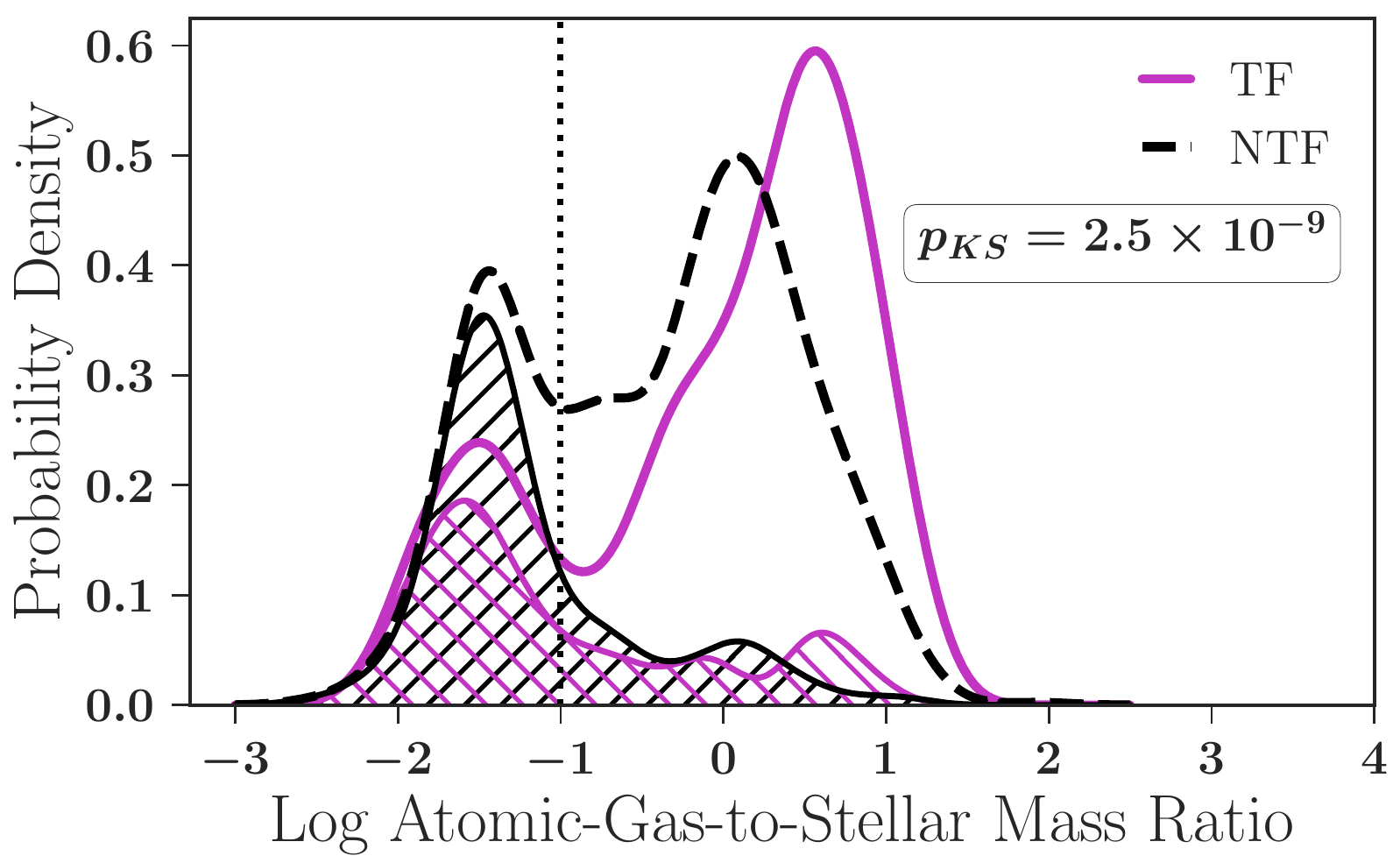}
  		\caption{Distributions of gas-to-stellar mass ratios (G/S) for galaxies identified with tidal features in magenta as well as without tidal features in black. The contributions of ``constrained" gas-to-stellar mass ratios (strong upper limits or ratios calculated with the photometric gas fractions technique) to each kernel density estimation are shown by the cross-hatched distributions. In Section \ref{GPGRsep}, we separate our galaxies into gas-poor and gas-rich populations.  We make this divide at G/S = 0.1 in order to include all strong upper limits in the gas-poor population, as marked by the dashed line. The G/S values above this line (those calculated with the photometric gas fractions technique) are not used in Figures~\ref{fig:ssfrgr} and~\ref{fig:SSFRresids} as explained in Section \ref{SF}. The probability densities shown are kernel density estimations in logarithmic space created with a Gaussian kernel and a cross-validated optimal bandwidth of h=0.16 dex, and thus have units of 1/ the x-axis unit.}
\label{fig:gasfractionkde}
\end{figure}
 
% \begin{table}[htbp]
% \caption{RESOLVE Tidal Feature Catalog} \label{tab:TFcat}
% \begin{tabular}{c l}
% \hline
% \hline
% Column & Description \\
% \hline
% 1 & RESOLVE ID \\
% 2 & Detection Confidence\tablenotemark{a}\\
% 3 & Feature Type \tablenotemark{b}\\
% 4 & Star Formation Rate \tablenotemark{c}\\
% 5 & $\mu_{\Delta}$ \tablenotemark{d}\\
% 6 & Distance to Nearest Neighbor (kd-tree algorithm) \tablenotemark{e} \\
% 7 & Distance to Nearest Neighbor (projected cylindrical search) \tablenotemark{f}\\
% \hline
% \end{tabular}
% \tablenotetext{a}{Confidence class. 4 = certain, 3 = probable, 2 = possible, 1 = hint, and 0 = none. Same scheme as \citet{atkinson2013}.}
% \tablenotetext{b}{Type of feature seen. 1 = narrow and 2 = broad. 0 if detection confidence = 0.}
% \tablenotetext{c}{Star formation rate. See Section \ref{photom}.}
% \tablenotetext{d}{Morphological metric $\mu_{\Delta}$. See Section \ref{morphology}.}
% \tablenotetext{e}{Distance to nearest neighbor calculated using the kd-tree algorithm and suppressing peculiar velocities within groups. See Section \ref{envmetrics}}
% \tablenotetext{f}{Distance to nearest neighbor calculated as the projected distance to the nearest neighbor in a cylindrical volume within cz = 500 km/s of the main object. See Section \ref{envmetrics}}
% \end{table}
 
\begin{table*}[tp]
\caption{RESOLVE Tidal Feature Catalog} \label{tab:TFcat}
\begin{tabular}{l c p{0.8\textwidth}}
\hline
\hline
Column & Label & Description \\
\hline
1 & Name & RESOLVE ID \\
2 & Confidence & Detection confidence class. 4 = certain, 3 = probable, 2 = possible, 1 = hint, and 0 = none. Same scheme as \citet{atkinson2013}.\\
3 & Type &  Type of feature seen. 1 = narrow and 2 = broad. 0 if detection confidence = 0.\\
4 & SFR & Star formation rate ($M_{\sun}/yr$). See Section \ref{photom}.\\
5 & $\mu_{\Delta}$ & Morphological metric $\mu_{\Delta}$ ($M_{\sun}/kpc^{2}$). See Section \ref{morphology}.\\
6 & NNdist\_kd &  Distance to nearest neighbor (Mpc) calculated using the kd-tree algorithm and suppressing peculiar velocities within groups. See Section \ref{envmetrics}\\
7 & NNdist\_proj & Distance to nearest neighbor (Mpc) calculated as the projected distance to the nearest neighbor in a cylindrical volume within cz = 500 km/s of the main object. See Section \ref{envmetrics}.\\
\hline
\end{tabular}
\end{table*}
 
 \indent We have examined a multitude of galaxy properties in the RESOLVE catalog to compare their distributions for TF and NTF galaxies. A table containing our classifications and other information required to reproduce the analysis below (in combination with RESOLVE Data Releases 1 and 2; see online database\footnote{\url{http://resolve.astro.unc.edu/pages/database.php}}) is available in machine readable format in the online version of this paper. A summary of information included in the catalog is given in Table \ref{tab:detect}.
 
 \indent The distributions of gas-to-stellar mass ratios (G/S) for TF and NTF galaxies show the most statistically significant difference between the two samples (Figure~\ref{fig:gasfractionkde}). A K-S test comparing the G/S distributions in Figure~\ref{fig:gasfractionkde} yields p$_{KS}$ $\sim 10^{-9}$, meaning the gas fraction distributions of TF and NTF galaxies have a negligible probability of coming from the same parent distribution at $\sim 6.0\sigma$  significance.  We can see from the figure that TF galaxies in our sample tend to have higher gas fractions. We see the same result if including ``possible" detections of tidal features (those in class 2) in the TF sample, at a higher significance of $\sim 7.0\sigma$. \citet{lotz2010} found in simulations that galaxy mergers with high gas fractions exhibit disturbed morphologies for longer periods of time than their gas poor counterparts. Assuming the high surface brightness asymmetries studies by Lotz et al.\ are accompanied by low surface brightness features, it is not necessarily surprising that we tend to find more tidal features in gas-rich galaxies (but see Section \ref{observability}). Since our strong upper limits cluster near M$_{gas} < 0.05-0.1$M$_{*}$ \ as a matter of observational strategy, both galaxies with and without tidal features exhibit a false bimodality in gas fraction, with the upper limits causing a second peak slightly below the M$_{gas} \sim 0.1$M$_{*}$, or log(G/S) = $-$1, threshold.
 
\indent Other galaxy properties that produced slightly weaker results were u-r color, SSFR, and the morphology metric $\mu_{\Delta}$ introduced in \citet{kannappan2013} to distinguish quasi-bulgeless ($\mu_{\Delta}$ $<$ 8.6), bulged disk (8.6 $<$ $\mu_{\Delta}$ $<$ 9.5), and spheroid-dominated ($\mu_{\Delta}$ $>$ 9.5) galaxies. This list is somewhat unsurprising due to the tight correlation of colors and star formation histories with G/S ratios, and to a lesser degree with galaxy structure \citep{kannappan2004,franx2008,kannappan2013,lilly2016,saintonge2016}. Additional parameters (nearest-neighbor distance, stellar mass, group halo mass, axial ratio b/a, g-r color gradient, group virial radius, axial ratio of the inner disk region of the galaxy, HI asymmetry, group redshift, number of group members, and central status) inspected for correlations with tidal features did not yield statistically significant results, although we will see that some of these prove important when 
investigated by other means than direct global correlation. 

\subsection{Random Forest Analysis}\label{RF}
\indent Because of the possibility that gas-rich and gas-poor galaxies with tidal features might reflect different origins, we decided to look for parameters that might be important despite not showing a direct correspondence with tidal features, for example due to differing trends from distinct formation mechanisms in the gas-rich and gas-poor subpopulations. We applied an automated method to determine which parameters are most important for predicting tidal features. The Random Forest algorithm \citep{breiman2001}, implemented in this work with the public Python library scikit-learn \citep{pedregosa2011}, generates classifications based on a random ensemble of decision trees based on all provided input parameters. From this process, the algorithm also determines the score of the importance of each input feature in the resulting classifications. We trained the algorithm using our by-eye identification of tidal features discussed in Section \ref{detection}.  We judged the performance of our classifier with the F1 metric, which is the harmonic mean of precision (the number of correct positive results divided by the number of all positive results) and recall (the number of correct positive results divided by the number of positive results that should have been returned). An F1 score reaches its best value at 1 and worst at 0.  The best hyperparameters for our classifier, such as the number of generated decision trees, were optimized to maximize this F1 metric with ten-fold cross validation \citep{geisser1975}.

\begin{figure}[htb!]
\centering
\includegraphics[width=\linewidth]{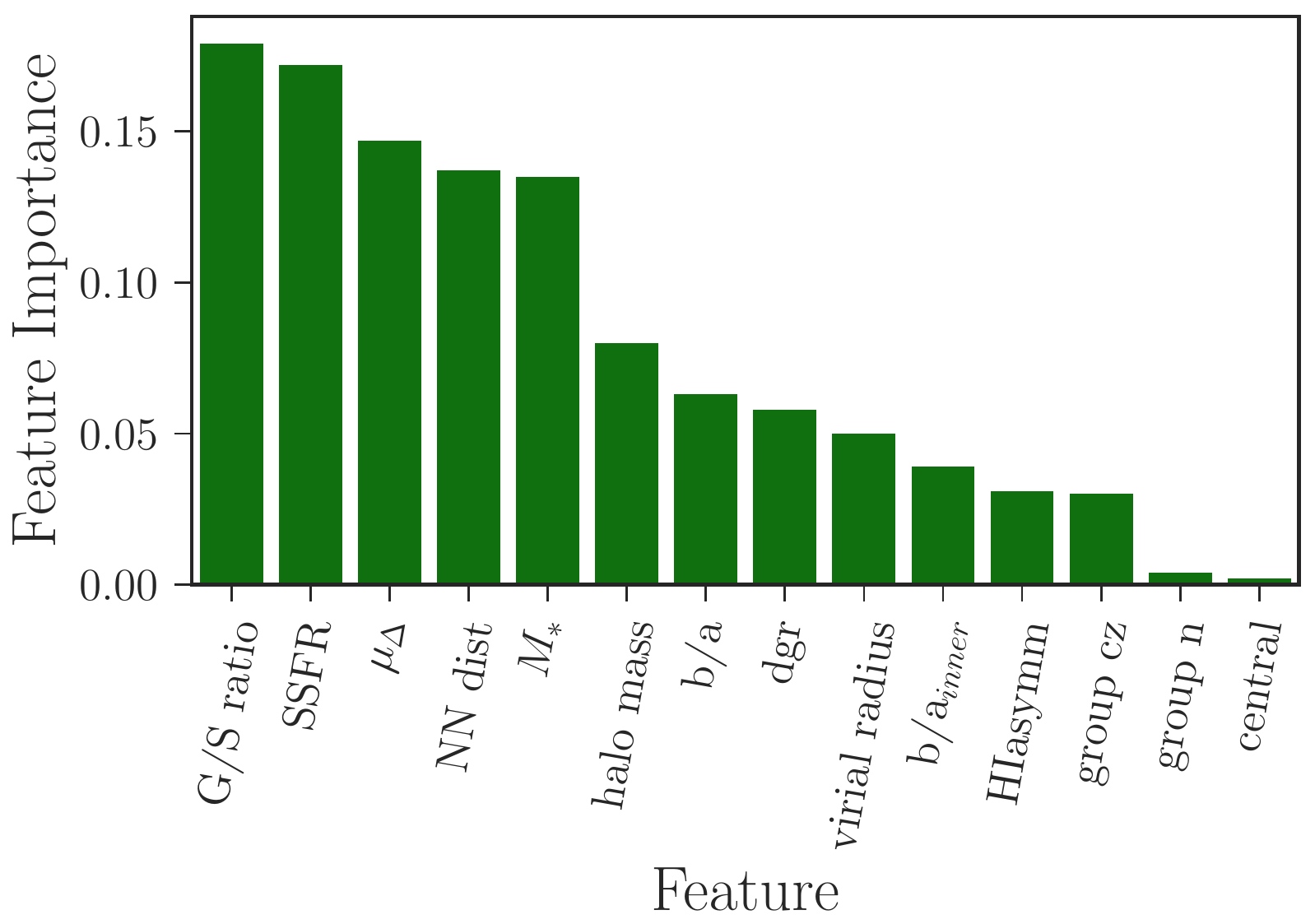}
  		\caption{Relative importance of each feature input into our random forest classifier in predicting whether a galaxy has a tidal feature. The top 5 most important parameters are: G/S ratio, SSFR, $\mu_{\Delta}$, nearest-neighbor distance, and stellar mass. The others correspond to, in decreasing order, group halo mass, axial ratio b/a, (g$-$r) color gradient, group virial radius, axial ratio of the inner disk region of the galaxy, HI asymmetry, group redshift, number of group members, and central status (1 or 0).}
        \label{fig:RFgraph}
\end{figure}

\indent The relative contribution of each parameter to our Random Forest classification is shown in Figure~\ref{fig:RFgraph}. Our best-performing Random Classifier only achieved an F1 score of 0.5; however, from the feature selection results of Figure~\ref{fig:RFgraph}, we were able to identify additional parameters to be further investigated. The top five most important parameters in descending order are as follows: the atomic gas-to-stellar mass ratio, the SSFR, the morphology metric $\mu_{\Delta}$, the distance to the galaxy's nearest neighbor in RESOLVE, and the stellar mass of the galaxy. Below we analyze these parameters in relation to our TF and NTF samples. Although group environment metrics (i.e. halo mass and number of group members) are deemed relatively unimportant by the Random Forest results, we analyze them as well to put our results in interpretive context.

 \subsection{Galaxy Properties and Tidal Features for Gas-Poor and Gas-Rich Galaxies Separately}\label{GPGRsep}
 
\indent The strong correlation of increased G/S with tidal features indicates that separating our samples into gas-poor and gas-rich galaxies may yield differing results in relation to the parameters identified in Section \ref{RF}. We make this divide at G/S= 0.1 in order to include all strong upper limits in the gas-poor population, as marked by the dashed line in Figure ~\ref{fig:gasfractionkde}. In our resulting gas-rich sample, 142 out of 747 galaxies ($\sim$19$^{\pm 2} \%$) have tidal features, $\sim$79$ \%$ of which are classified as ``narrow.'' In contrast, 38 out of 292 gas-poor galaxies ($\sim$13$^{\pm 3} \%$) have tidal features, $\sim$74$ \%$ of which are classified as ``narrow.''  Below we analyze the roles of galaxy properties in relation to our gas-rich and gas-poor TF and NTF samples.
 
 \subsubsection{Star Formation}\label{SF}
 \begin{figure*}[htb!]
\centering
\includegraphics[width=.9\linewidth]{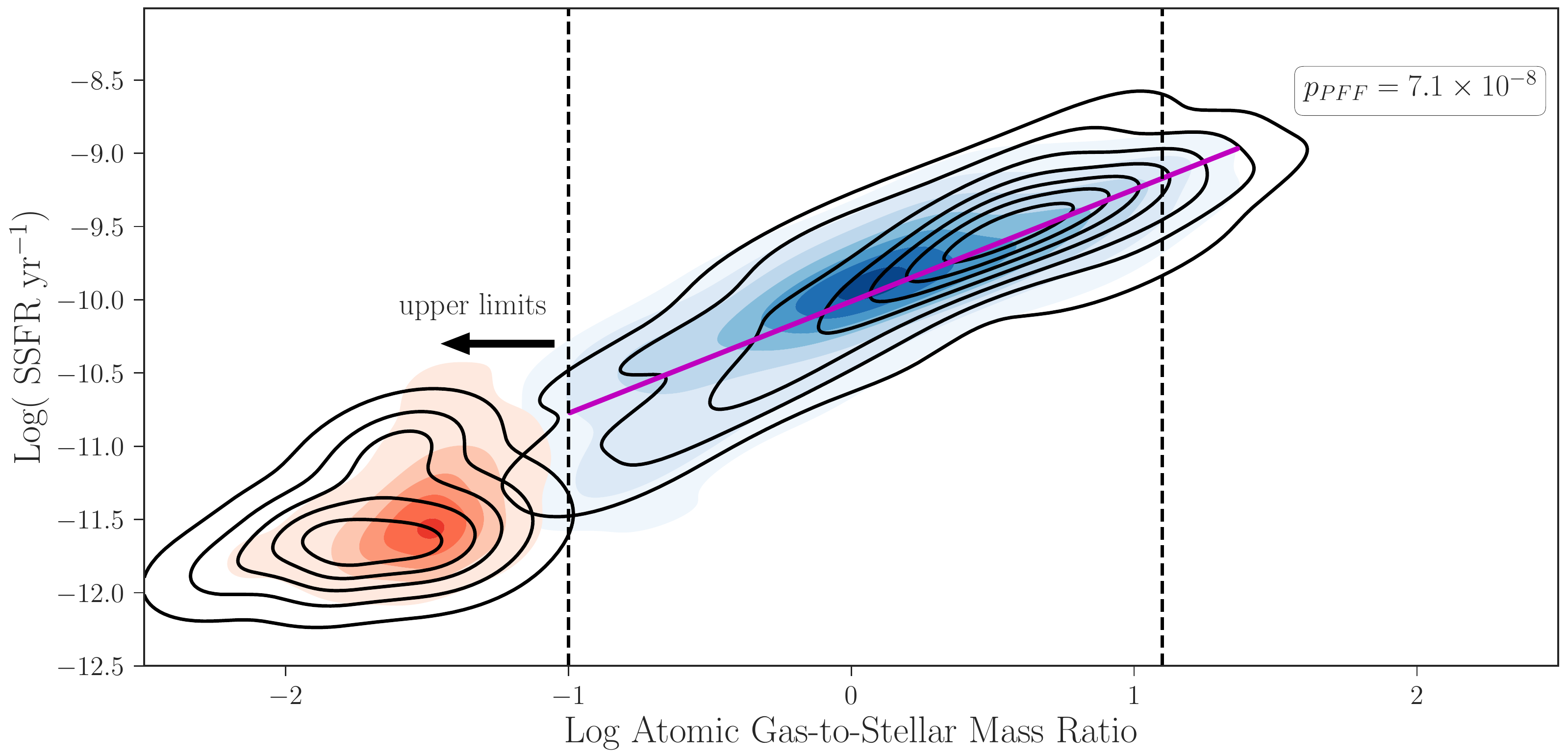}
  		\caption{Distributions of specific star formation rate versus gas-to-stellar mass ratio for TF (black contours) and NTF (blue/red contours) galaxies. The gas-rich galaxies with 21cm detections (the sample of gas-rich galaxies not including those with G/S measurements from the photometric gas fractions technique) are shown in blue, while the gas-poor galaxies are shown in red. For each subsample, the lowest contour contains at least 95\% of the data. The line of best fit for the log(SSFR) v. log(G/S) for the gas-rich galaxies is shown in purple.  The dashed lines mark the region over which the best fit line is calculated, corresponding to regions where the vertical scatter the line appears symmetrical. Gas-rich TF galaxies are more concentrated to higher G/S and higher SSFR, with the difference from the NTF distribution at a p$_{PFF}$ $\sim 10^{-8}$ significance.}
\label{fig:ssfrgr}
\end{figure*}

\begin{figure}[h!]
\centering
\includegraphics[width=0.95\linewidth]{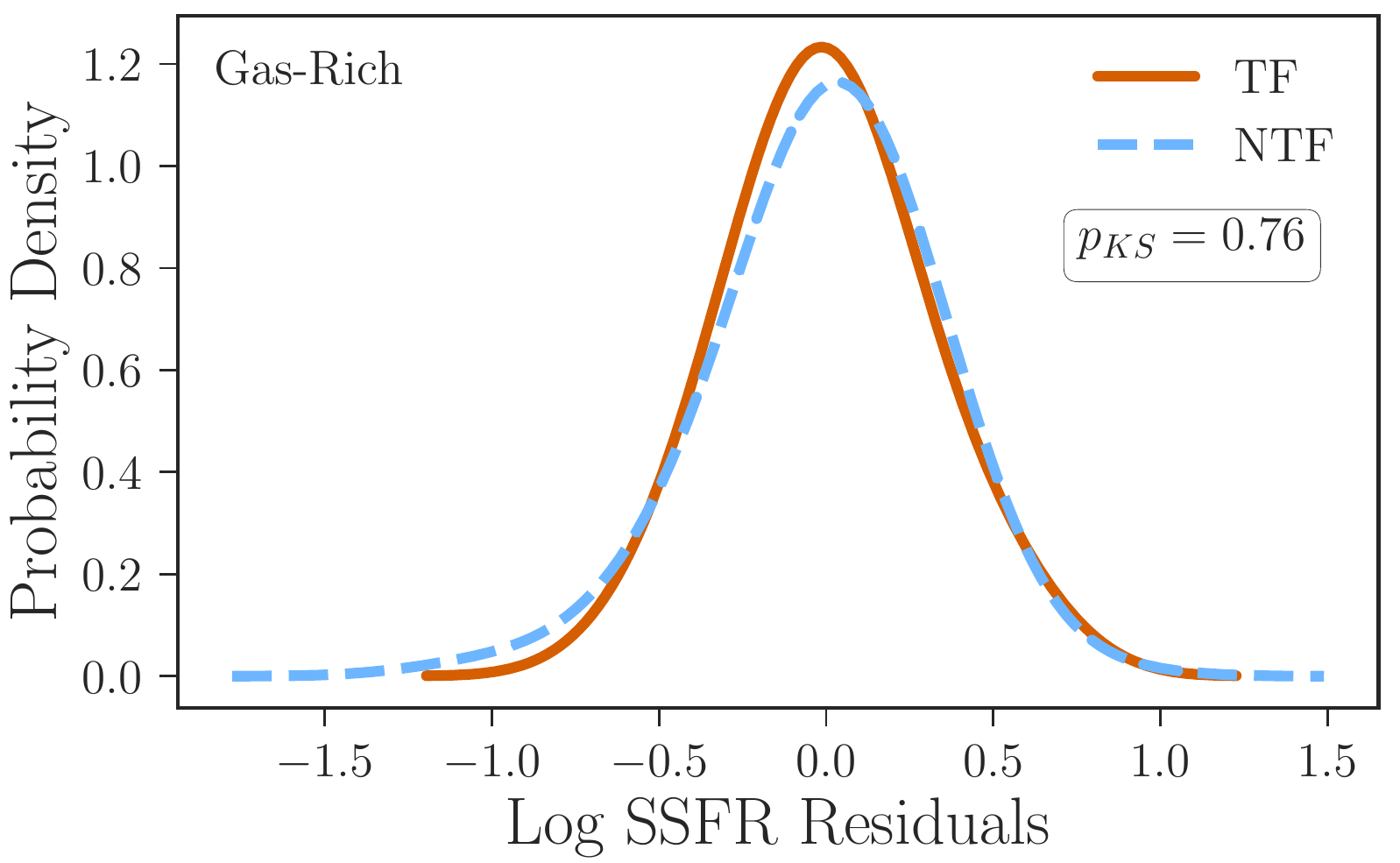}
  		\caption{Distributions of the residuals from the best fit line shown in Figure~\ref{fig:ssfrgr} for gas-rich galaxies with 21cm detections with (orange) and without (blue) tidal features. There is not a statistically significant difference between the two distributions as determined by the K-S test, implying the difference between the distributions found in Figure~\ref{fig:ssfrgr} is driven by the difference in G/S. The probability densities shown are kernel density estimations in logarithmic space created with a Gaussian kernel and a cross-validated optimal bandwidth of 0.23 dex, and thus have units of 1/ the x-axis unit.}
        \label{fig:SSFRresids}
\end{figure}

\begin{figure}[h!]
\centering
\includegraphics[width=0.95\linewidth]{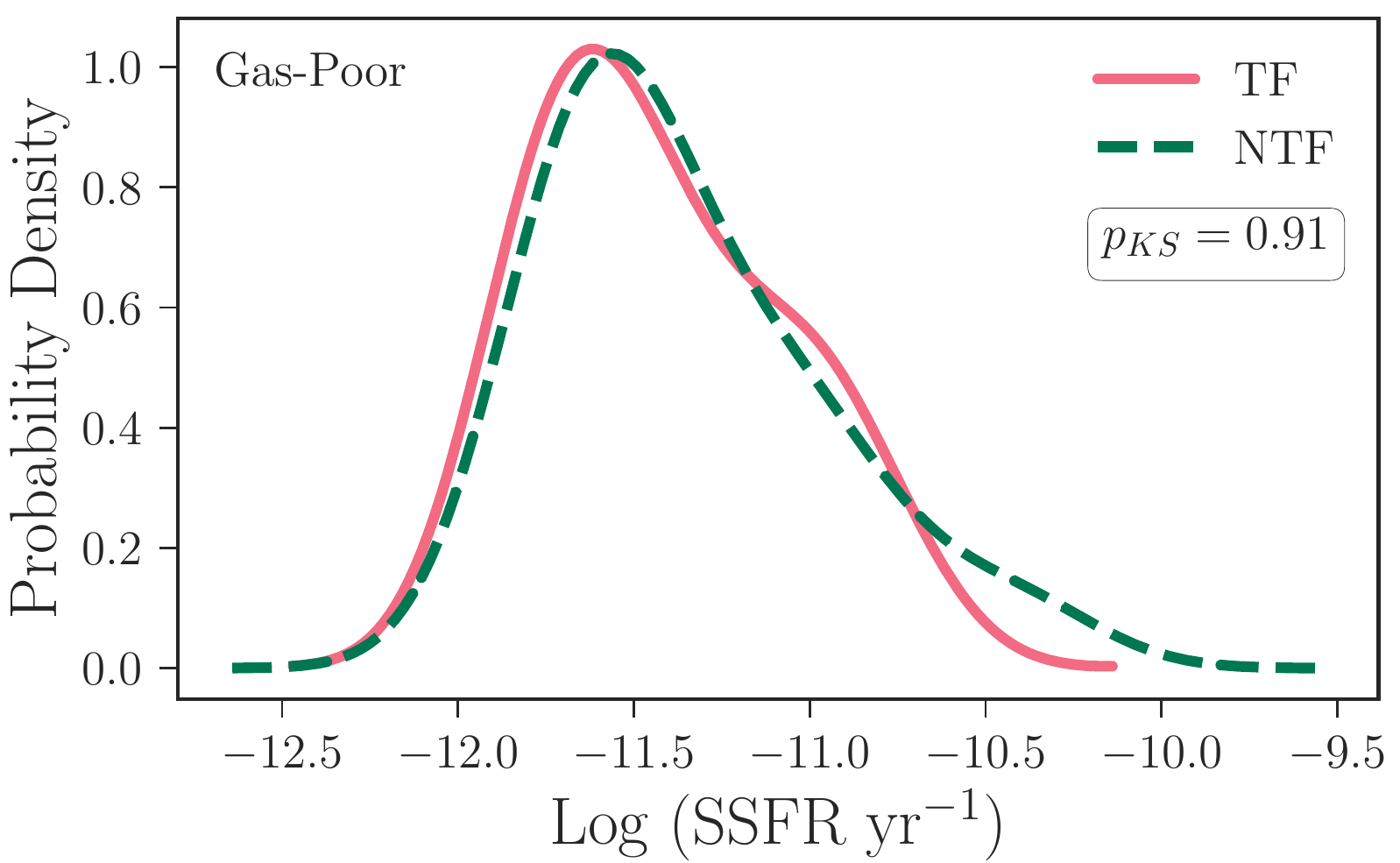}
  		\caption{Distributions of the specific star formation rates of gas-poor galaxies.  TF galaxies are shown in pink while NTF galaxies are shown in green. There is not a statistically significant difference between the two distributions as determined by the K-S test. The probability densities shown are kernel density estimations in logarithmic space created with a Gaussian kernel and a cross-validated optimal bandwidth of 0.25 dex, and thus have units of 1/ the x-axis unit.}
        \label{fig:SSFRlim}
\end{figure}

\indent To investigate the possible distinct origins of tidal features around gas-rich and gas-poor galaxies, we consider the distributions of the specific star formation rates (SSFRs). A comparison of the SSFR vs. G/S distributions for galaxies with and without tidal features is shown in Figure~\ref{fig:ssfrgr}. We separate the gas-rich galaxies with HI masses (blue) from the gas-poor galaxies (red). As the SSFR is closely related to the galaxy's color, which is used to calibrate the photometric gas fractions technique we use for confused galaxies and galaxies with weak upper limits, we do not include these galaxies in this plot (i.e. we only plot the gas-rich galaxies with 21cm detections). We have applied the \citet{fasano1987} variation of the Peacock test \citep{peacock1983}, an extension of the K-S test to two dimensions, to determine the statistical significance of the difference between the distributions of the TF and NTF galaxies for the gas-rich sample. Within this sample, galaxies with tidal features appear to be have both larger G/S (as seen in Section \ref{RF}) as well as higher SSFR at $\sim5.5\sigma$ significance. This trend remains the same when including ``possible" detections of tidal features (those in class 2) in the TF sample, with a slightly higher significance of $\sim6.1\sigma$.

\indent However, we want to test whether this difference is driven mostly by the correlation between G/S and SSFR, or if there is an additional tendency of gas-rich TF galaxies to have higher SSFRs independent of G/S. We fit a linear relationship to the total SSFR v. G/S sample (shown in red) for gas-rich galaxies with 21cm detections in the region in which the vertical scatter from the line appeared symmetrical (marked by the dashed lines). The SSFR residuals from this line are shown in Figure~\ref{fig:SSFRresids};  we do not find a statistically significant difference between the distributions of residuals for TF and NTF galaxies, which also holds if class 2 confidence TF detections are counted as TF rather than NTF galaxies.  Thus, we conclude the tendency of gas-rich TF galaxies to have higher SSFRs seen in Figure~\ref{fig:ssfrgr} is indeed driven primarily by G/S. 

\indent We have similarly compared the SSFR distributions for TF and NTF galaxies in the gas-poor sample as shown in Figure~\ref{fig:SSFRlim}.  We do not find a statistically significant difference between the two distributions, which also holds if class 2 confidence TF detections are counted as TF rather than NTF galaxies.

\subsubsection{Morphology}\label{morphology}
\begin{figure*}[htb!]
\centering
\includegraphics[width=.9\linewidth]{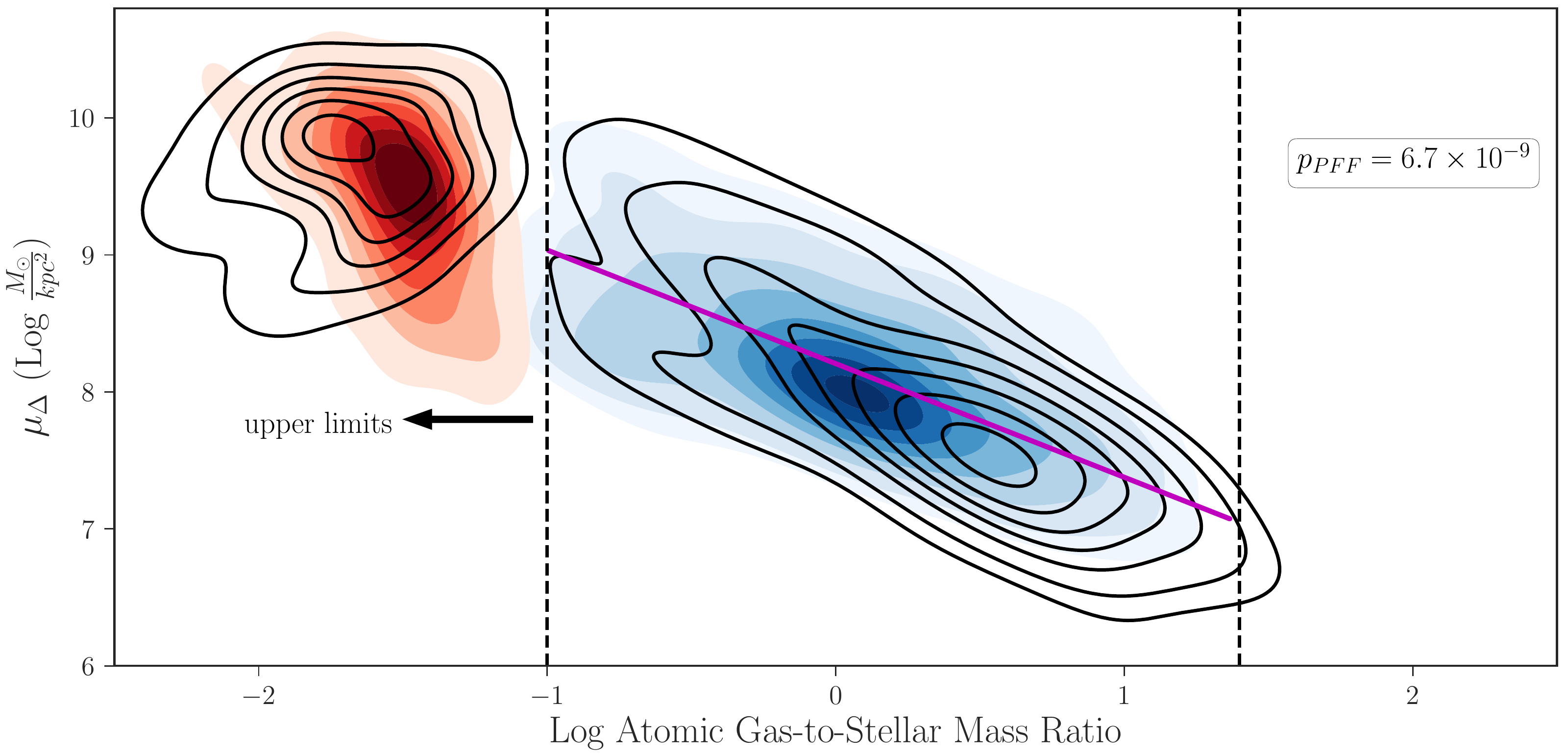}
  		\caption{ Distributions of $\mu_{\Delta}$ versus gas-to-stellar mass ratio for TF (black contours) and NTF (blue/red contours) galaxies. The gas-rich galaxies (including those estimated using the photometric gas fractions technique; see Section \ref{HI}) are shown in blue, while the gas-poor galaxies are shown in red.  For each subsample, the lowest contour contains at least 95\% of the data. The linear fit of $\mu_{\Delta}$ v. log(G/S) for gas-poor galaxies is shown in purple.  The dashed lines mark the region over which the best fit line is calculated, corresponding to regions where the vertical scatter from the line appears symmetrical. Gas-rich TF galaxies are more concentrated to higher G/S and lower $\mu_{\Delta}$, appearing more disky, with the difference from the distribution for NTF galaxies at a p$_{PFF}$ $\sim 10^{-9}$ significance.}
\label{fig:mudgf}
\end{figure*}

\begin{figure}[h!]
\centering
\includegraphics[width=\linewidth]{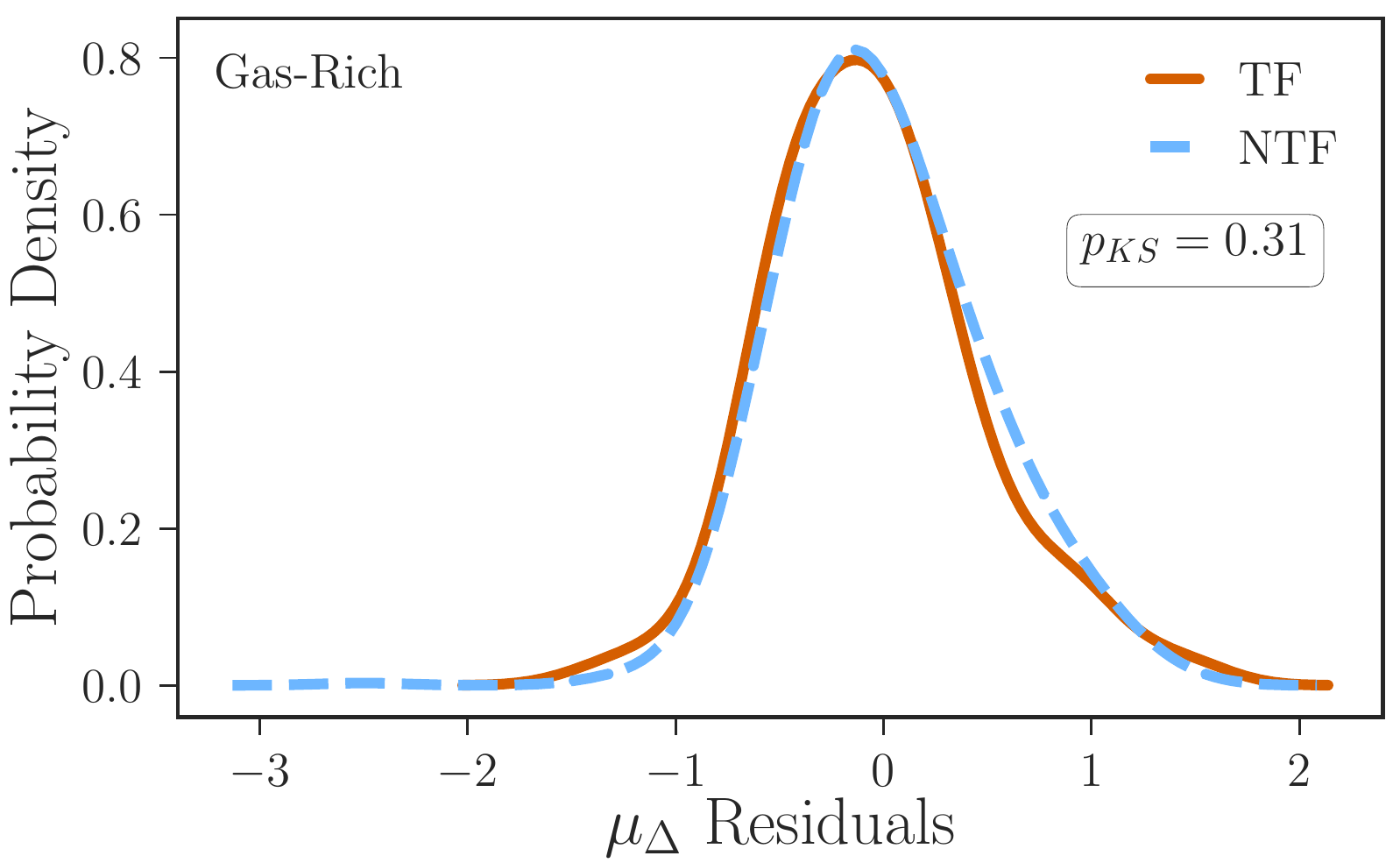}
  		\caption{Distributions of the residuals from the best fit line shown in Figure~\ref{fig:mudgf} for gas-rich galaxies with (orange) and without (blue) tidal features. There is not a statistically significant difference between the two distributions as determined by the K-S test, implying the difference between the distributions found in Figure~\ref{fig:mudgf} is driven by the difference in G/S. The probability densities shown are kernel density estimations created with a Gaussian kernel and a cross-validated optimal bandwidth of 0.21 dex, and thus have units of 1/ the x-axis unit.}
        \label{fig:mudresids}
\end{figure}

\begin{figure}[h!]
\centering
\includegraphics[width=\linewidth]{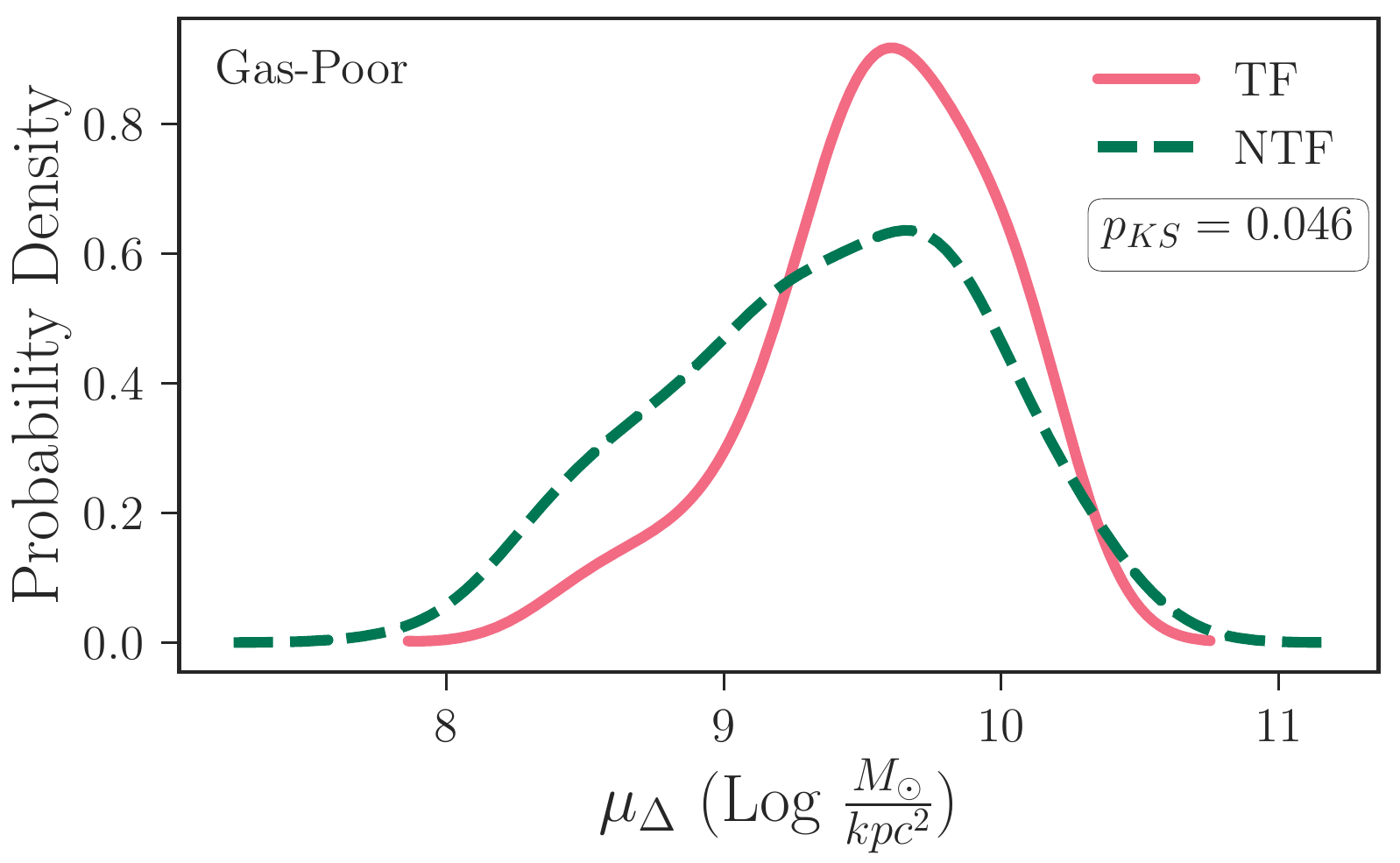}
  		\caption{Distributions of the $\mu_{\Delta}$ values of gas-poor galaxies.  TF galaxies are shown in pink while NTF galaxies are shown in green. There is a weak, 2$\sigma$ significant difference between the two distributions as determined by the K-S test. The probability densities shown are kernel density estimations created with a Gaussian kernel and a cross-validated optimal bandwidth of 0.21 dex, and thus have units of 1/ the x-axis unit.}
        \label{fig:mudlim}
\end{figure}

\indent To examine the relationship of morphology to our TF and NTF samples, we use the structure metric $\mu_{\Delta}$ introduced in \citet{kannappan2013} to distinguish quasi-bulgeless ($\mu_{\Delta}$ $<$ 8.6), bulged disk (8.6 $<$ $\mu_{\Delta}$ $<$ 9.5), and spheroid-dominated ($\mu_{\Delta}$ $>$ 9.5) galaxies. Kannappan et al. define $\mu_{\Delta}$ as
\begin{equation}
\mu_{\Delta}=\mu_{90}+1.7\Delta \mu
\end{equation}
combing an overall surface mass density
\begin{equation}
\mu_{90}=\text{log} \frac{0.9 M_{*}}{\pi r^{2}_{90,r}}
\end{equation}
with a surface mass density contrast
\begin{equation}
\Delta \mu = \text{log} \frac{0.5 M_{*}}{\pi r^{2}_{50,r}}-\text{log} \frac{0.4 M_{*}}{\pi r^{2}_{90,r}-\pi r^{2}_{50,r}}
\end{equation}
where all radii are converted to physical kpc units.

\indent A comparison of the $\mu_{\Delta}$ versus G/S distributions for TF and NTF galaxies is shown in Figure~\ref{fig:mudgf}. Similarly to Section ~\ref{SF}, we separate the gas-rich galaxies (blue) from the gas-poor galaxies (red), although here we include confused galaxies and galaxies with weak upper limits for which G/S has been calculated with the photometric gas fractions technique.  We calculate the significance of the difference between the distributions for gas-rich TF and NTF galaxies to be $\sim 5.8\sigma$, with TF galaxies tending to be more disky than their NTF counterparts. We see a similar trend when including ``possible" detections of tidal features (those in class 2) in the TF sample, but with a slightly higher significance of $\sim6.2\sigma$. The distributions of $\mu_{\Delta}$ residuals from the linear fit to the $\mu_{\Delta}$ v. G/S data (shown in red in Figure~\ref{fig:mudgf}) are plotted in Figure~\ref{fig:mudresids}. As for SSFR, we do not find a statistically significant difference between the distributions of $\mu_{\Delta}$ residuals for gas-rich TF and NTF galaxies, regardless of whether we include class 2 confidence detections in the TF sample.  Thus, we conclude the tendency of gas-rich TF galaxies to have diskier morphologies seen in Figure~\ref{fig:mudgf} is driven primarily by G/S. 

\indent We also compared the $\mu_{\Delta}$ distributions for gas-poor TF and NTF galaxies as shown in Figure~\ref{fig:mudlim}.  The TF galaxies in this sample do seem to be slightly concentrated at higher $\mu_{\Delta}$ (more bulged morphologies) than their NTF counterparts, but the significance of the difference is only $\sim2.0\sigma$.  We see similar trends if we include class 2 confidence TF detections as TF galaxies rather than NTF galaxies with a higher significance of $\sim2.7\sigma$.

\subsubsection{Stellar Mass}\label{mstar}

\begin{figure}
\centering
\includegraphics[width=\linewidth]{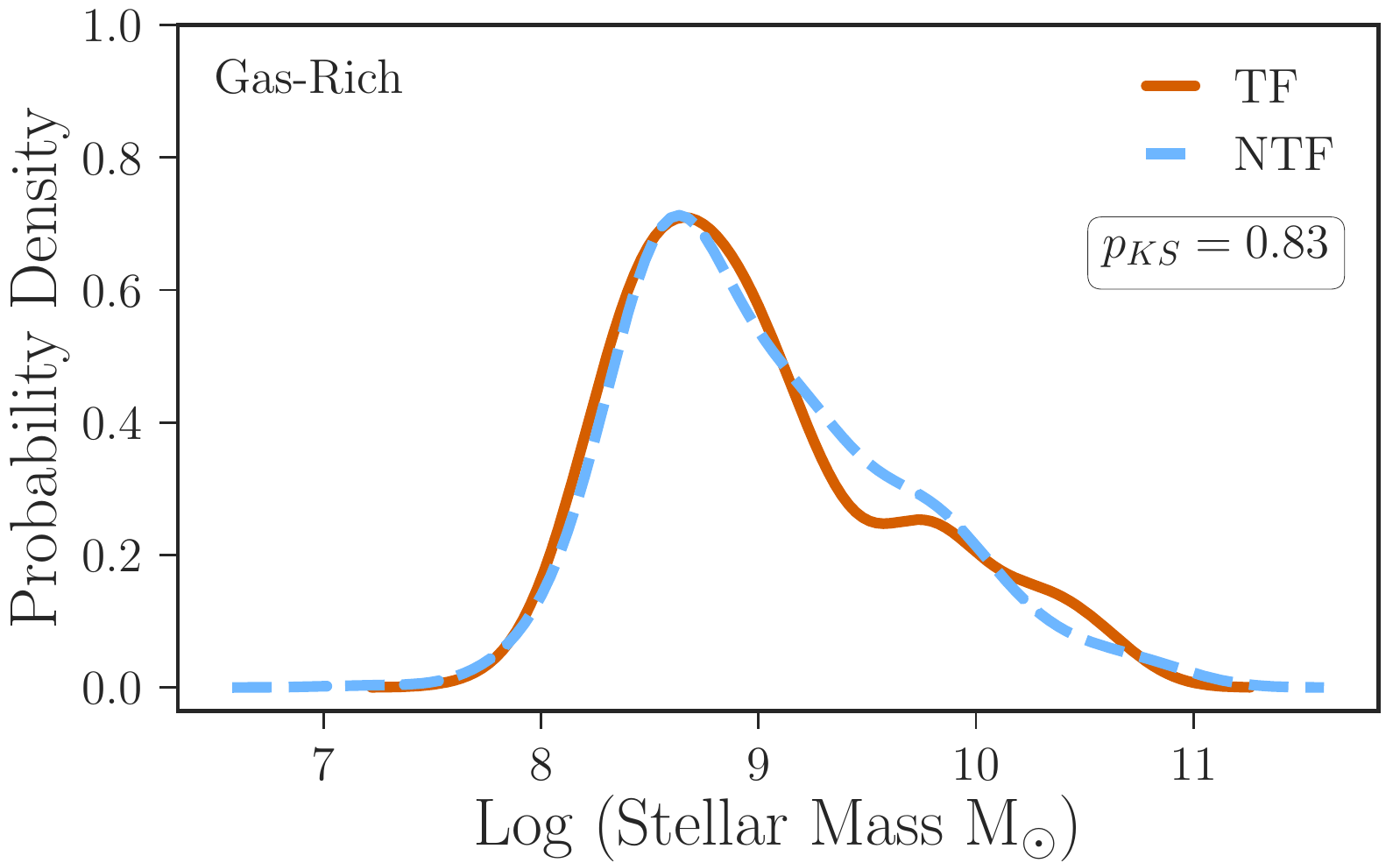}
  		\caption{Distributions of stellar mass for gas-rich TF (orange) and NTF (blue) galaxies. The K-S test does not find a statistically significant difference between the two distributions. The probability densities shown are kernel density estimations in logarithmic space created with a Gaussian kernel and a cross-validated optimal bandwidth of 0.25 dex, and thus have units of 1/ the x-axis unit.}
        \label{fig:grstellar}
\end{figure}

\begin{figure}
\centering
\includegraphics[width=\linewidth]{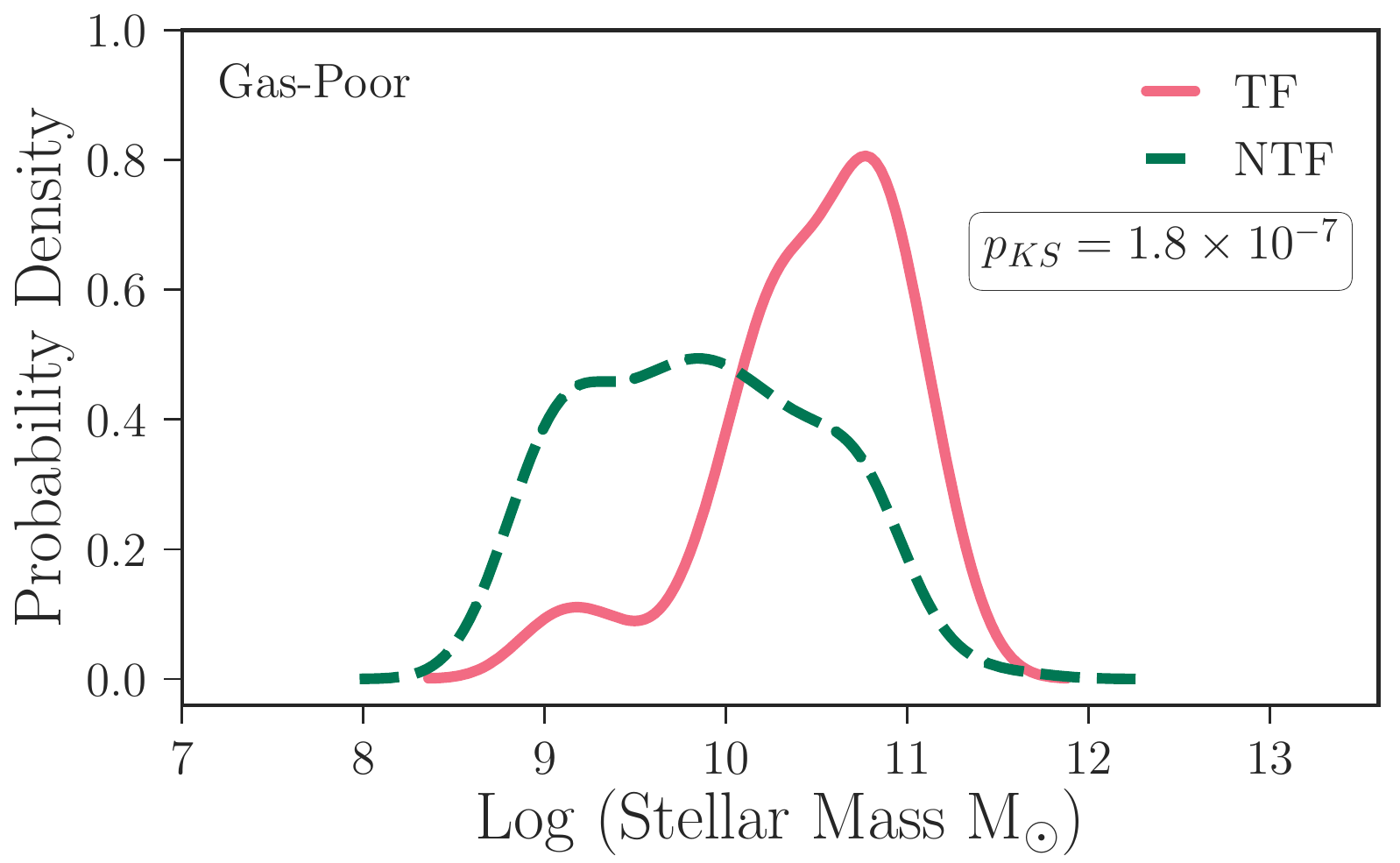}
  		\caption{Distributions of stellar mass for gas-poor TF (pink) and NTF (teal) galaxies. The gas-poor galaxies with tidal features are more massive than their counterparts without tidal features at $\sim 5.2\sigma$ significance. The probability densities shown are kernel density estimations in logarithmic space created with a Gaussian kernel and a cross-validated optimal bandwidth of 0.25 dex, and thus have units of 1/ the x-axis unit.}
        \label{fig:gpstellar}
\end{figure}

\indent The stellar mass distributions for our gas-rich and gas-poor galaxies are shown in Figures~\ref{fig:grstellar} and ~\ref{fig:gpstellar}, respectively. We do not find a statistically significant difference between the stellar mass distributions of gas-rich TF and NTF galaxies, regardless of whether we include ``possible" detections (class 2) in the TF sample. In contrast, gas-poor galaxies show a much more prominent relationship between tidal features and stellar mass. Gas-poor galaxies with tidal features are more massive than their NTF counterparts at $\sim5.2\sigma$  significance. The same trend is seen (at a slightly lower significance of $\sim4.9\sigma$) when counting class 2 confidence TF detections as TF rather than NTF galaxies.

\subsection{Environment and Tidal Features for Gas-Poor and Gas-Rich Galaxies Separately}\label{environment}
 To analyze the environments of our galaxies, we consider both nearest neighbor distance and group mass and richness to assess environmental influences on local and halo scales.
 
\indent As a statistical detail, in this section the Mann-Whitney-U (MWU) test has been applied to each pair of distributions in addition to the K-S test. Both the MWU and K-S tests return p-values that test the null hypothesis that the two samples have the same distribution. One difference is that the MWU test is mainly sensitive to discrepancies between the two medians while the K-S test is more sensitive to general differences between the two distributions (shape, spread, median, etc.).  Another difference is that the K-S test is not suited for samples with repeated values. Since we expect pairs of duplicate nearest-neighbor distances, as well as multiple groups of galaxies with the same group parameters, the MWU test may provide a more accurate p-value.

 \subsubsection{Nearest Neighbor Distance}\label{nnd}
\begin{figure*}
\centering
\gridline{\fig{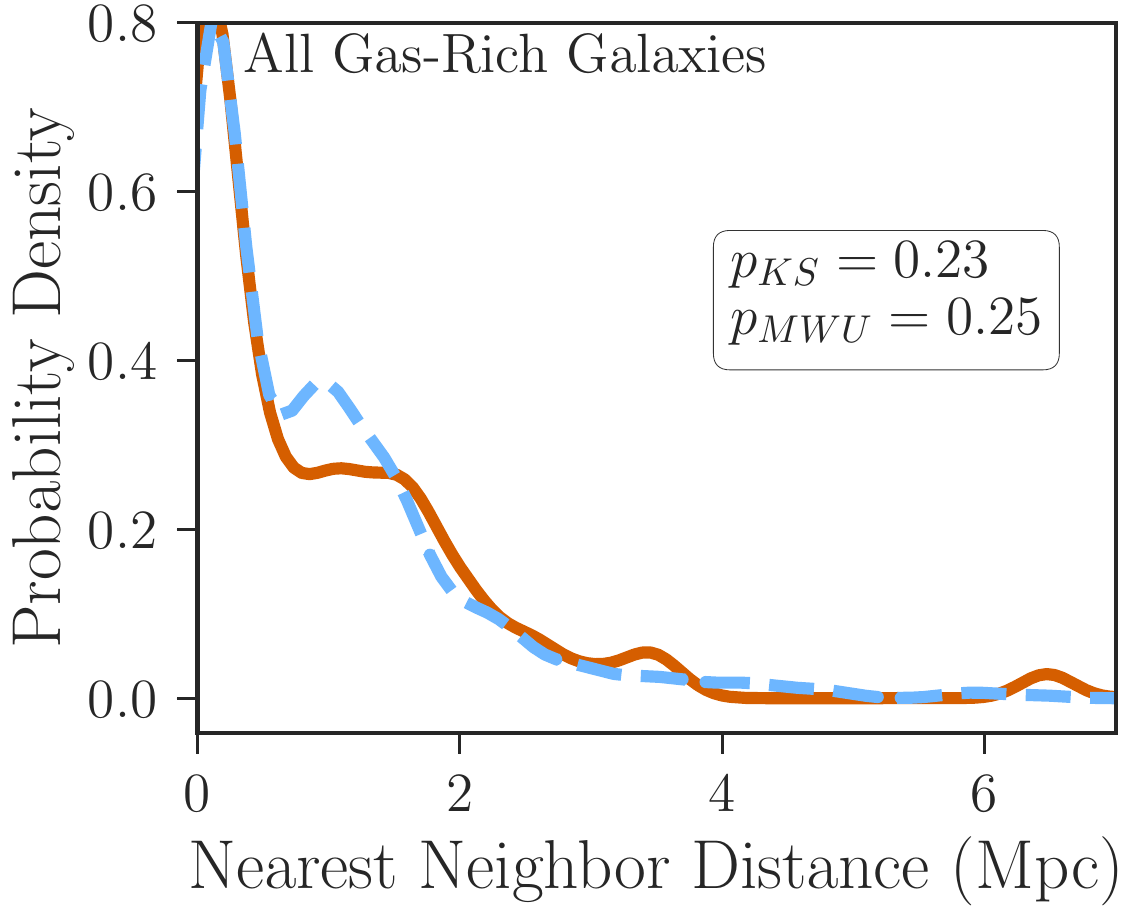}{0.32\textwidth}{(a)}
          \fig{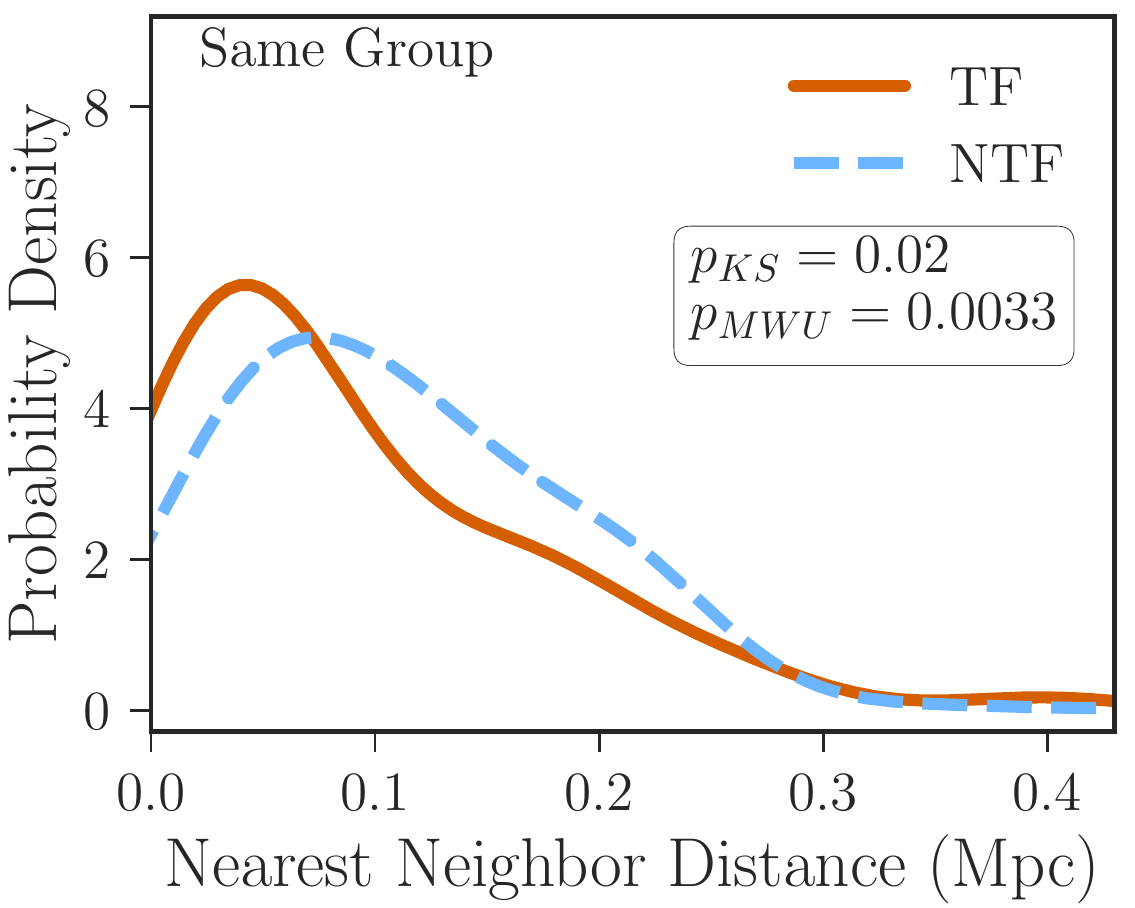}{0.32\textwidth}{(b)}
          \fig{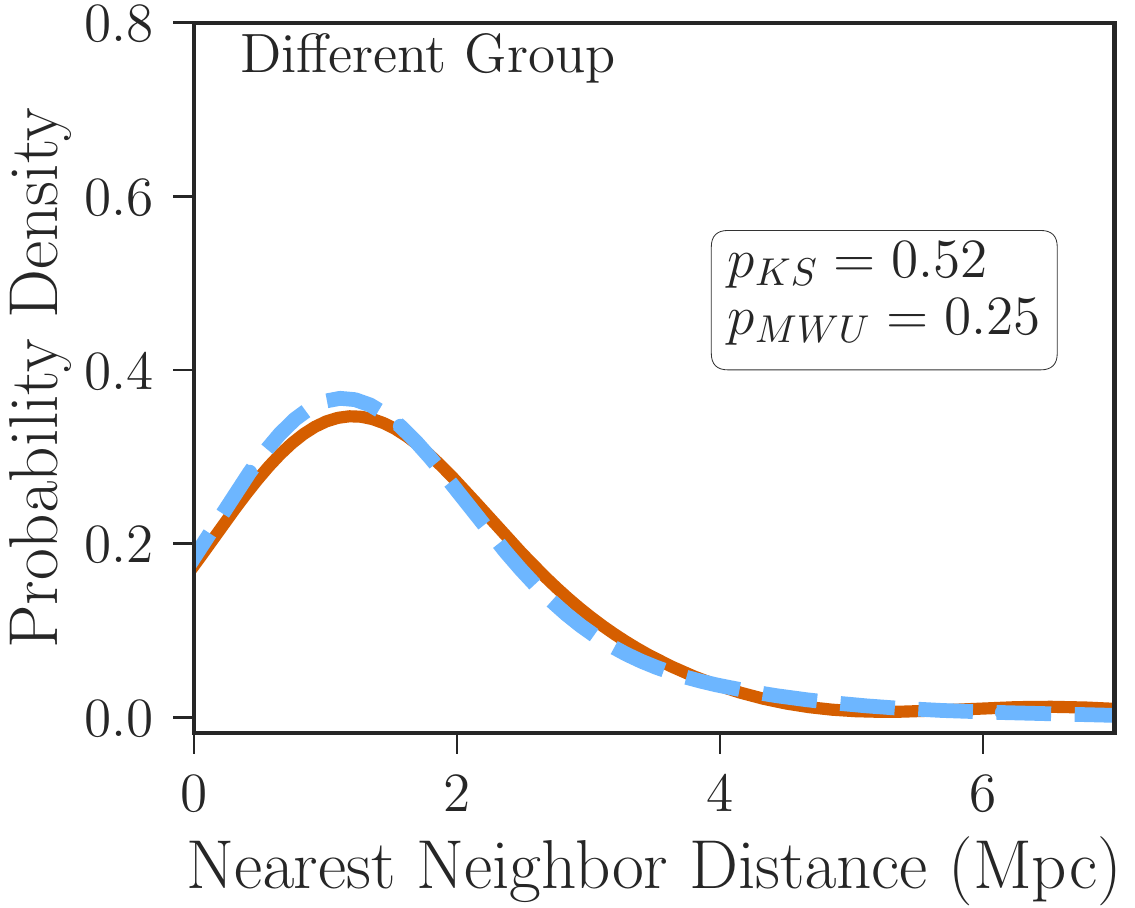}{0.32\textwidth}{(c)}}
\caption{Distributions of nearest neighbor distances for gas-rich TF (orange) and NTF (blue) galaxies. Panel (a) shows the two distributions, whose difference is not statistically-significant.  These populations are further divided into galaxies whose nearest neighbors are within the same groups (b) and galaxies with neighbors outside of their groups (c). The gas-rich TF galaxies with neighbors within their group are closer to their nearest neighbors than their NTF counterparts at $\sim2.9\sigma$ confidence. Gas-rich TF and NTF galaxies with nearest neighbors outside of their group do not show any difference in nearest neighbor distance. The probability densities shown in panels (a), (b), and (c) are kernel density estimations created with Gaussian kernels and cross-validated optimal bandwidths of 0.20, 0.04, and 0.80 Mpc, respectively, and thus each have units of 1/ the x-axis unit.}
\label{fig:nngr}
\end{figure*}

\begin{figure*}
\centering
\gridline{\fig{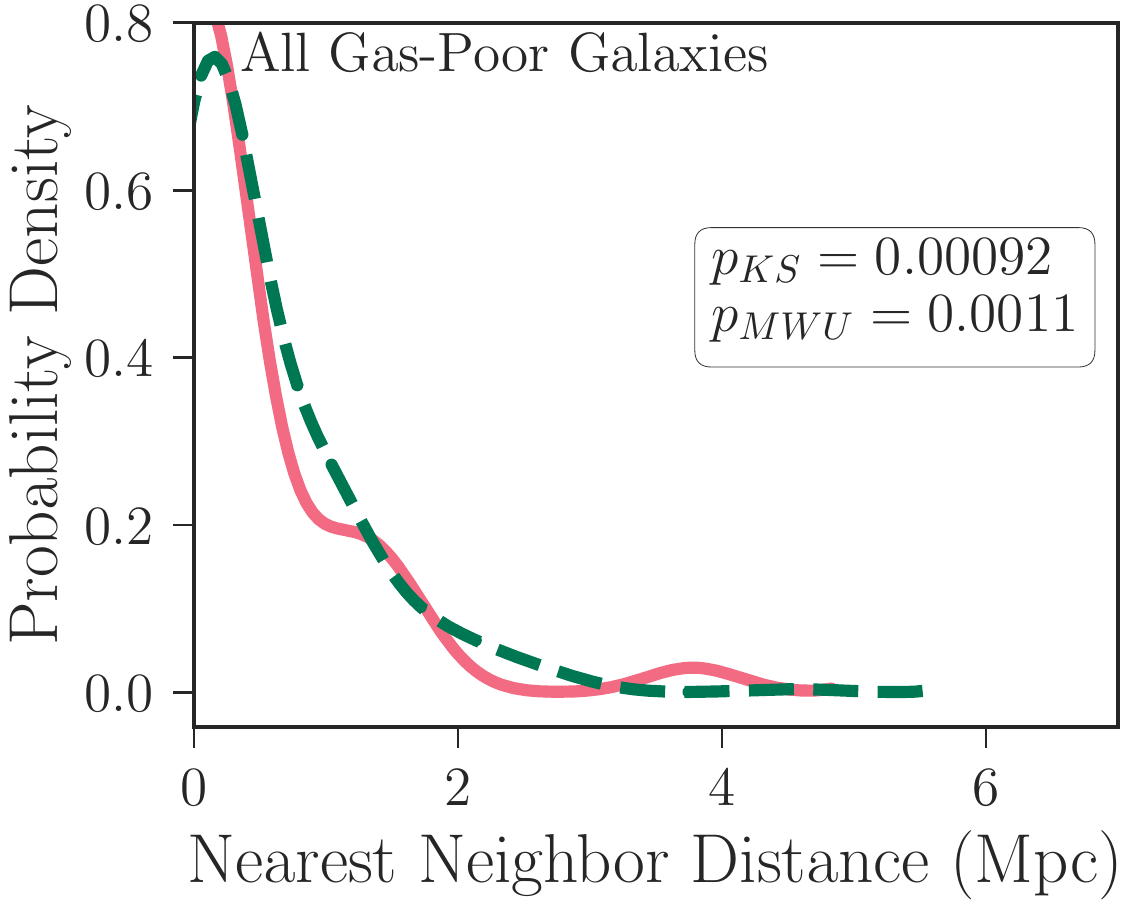}{0.32\textwidth}{(a)}
          \fig{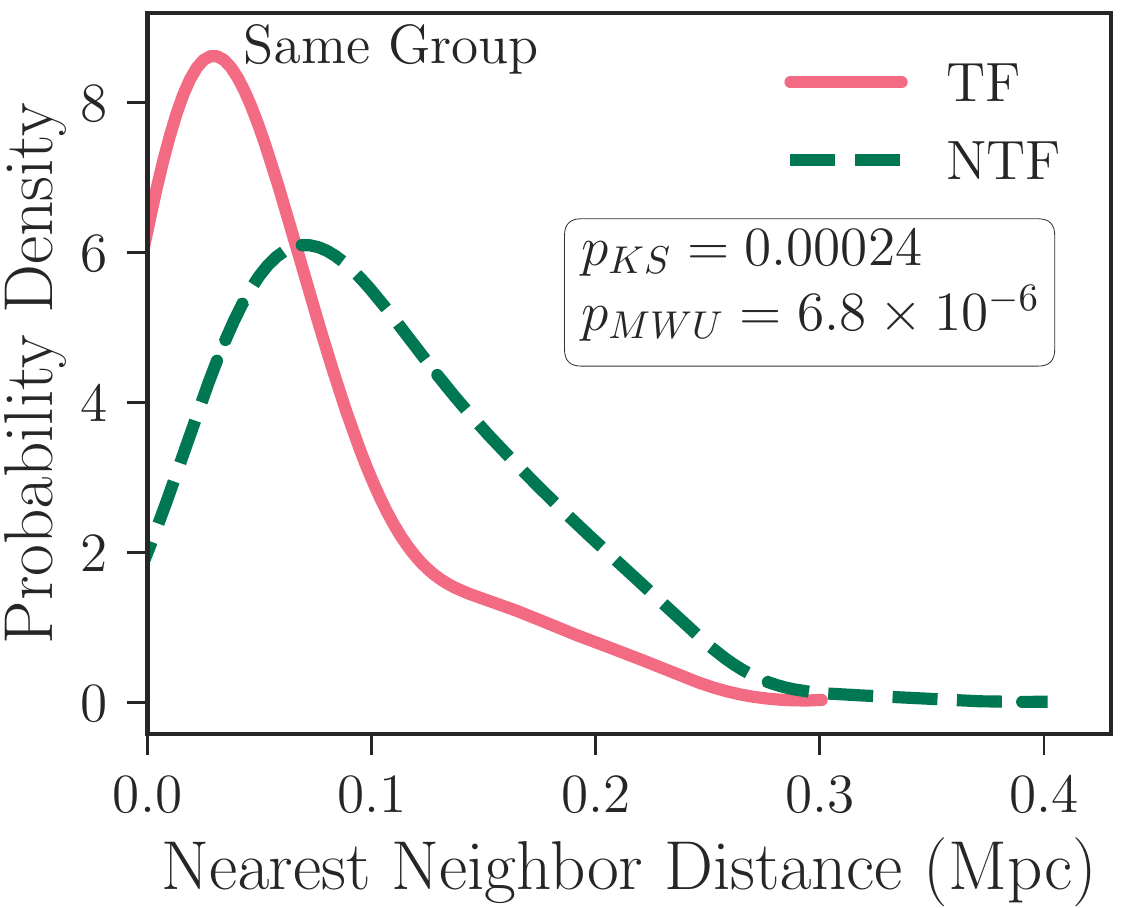}{0.32\textwidth}{(b)}
          \fig{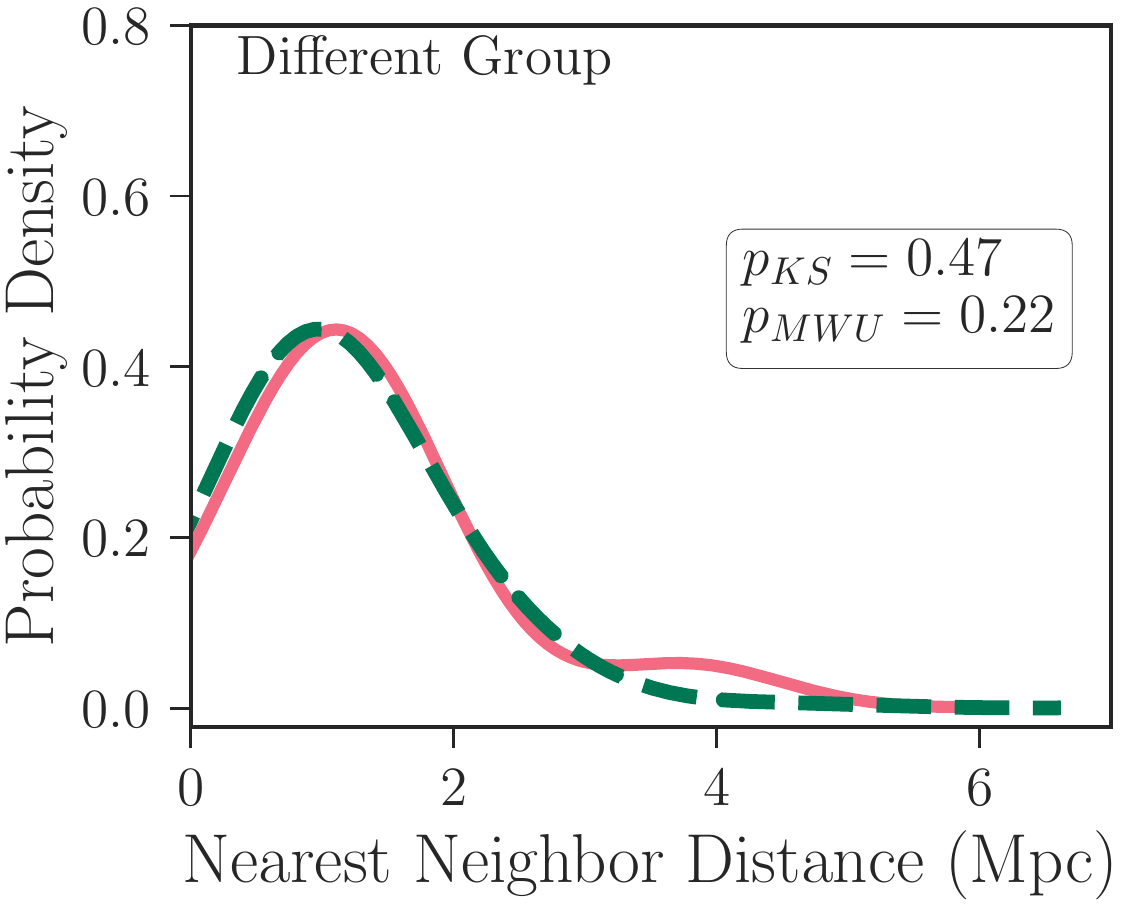}{0.32\textwidth}{(c)}}
\caption{Distributions of nearest neighbor distances for gas-poor TF (pink) and NTF (green) galaxies. Panel (a) shows the two distributions; TF gas-poor galaxies appear to be closer to their nearest neighbor at $\sim3.3\sigma$  confidence.  These populations are further divided into galaxies whose nearest neighbors are within the same groups (b) and galaxies with neighbors outside of their groups (c). The gas-poor TF galaxies with neighbors within their group are closer to their nearest neighbors than the gas-poor NTF galaxies at $\sim4.5\sigma$ confidence. Gas-poor galaxies with nearest neighbors outside of their group do not show a similarly significant result. The probability densities shown in panels (a), (b), and (c) are kernel density estimations created with Gaussian kernels and cross-validated optimal bandwidths of 0.35, 0.03, and 0.70 Mpc, respectively, and thus each have units of 1/ the x-axis unit.}
\label{fig:nngp}
\end{figure*}

\indent For each galaxy in RESOLVE, we have calculated the projected distance to its nearest neighbor using the kd-tree algorithm described in section \ref{envmetrics} and suppressing peculiar velocities within groups (i.e. assuming a single group cz for galaxies in the same group). We compare the distributions of projected distances for TF and NTF galaxies within our gas-rich and gas-poor subsamples in Figures~\ref{fig:nngr}a and~\ref{fig:nngp}a, respectively.  In panels b and c, we further divide each subsample based on whether the nearest neighbor is a member of the same group (it may not be if the galaxy is isolated, i.e. in an N=1 group). 

\indent For gas-rich galaxies with neighbors in the same group, those with/without tidal features have median separation of 0.06/0.1 Mpc; the difference between the two gas-rich distributions is significant at $\sim2.9\sigma$. For gas-poor galaxies with neighbors in the same group, those with/without tidal features have median separation of 0.03/0.09 Mpc; the difference between the two gas-poor distributions is significant at $\sim4.5\sigma$ (a notably stronger result than for gas-rich galaxies).  However, regardless of gas content, galaxies with nearest neighbors outside of their own group do not show a correlation between the presence of tidal features and nearest neighbor distance. We obtain results similar to these kd-tree results if we instead calculate nearest neighbor distances as projected distances to the nearest neighbor in a cylindrical volume within cz = 500 km/s of the main object. 

\indent Thus, regardless of gas content, we find that galaxies with neighbors in the same group are closer to their nearest neighbors if they have tidal features than if they do not, with a stronger result observed for gas-poor galaxies than gas-rich ones. This result persists if we include ``possible" detections of tidal features (those in class 2) in the TF sample, though at slightly higher significances of $\sim3.2\sigma$ and $\sim4.7\sigma$ for gas-rich and gas-poor galaxies, respectively. We infer that a substantial fraction of detected tidal features are the result of ongoing interactions, or early-stage mergers, especially in gas-poor galaxies.

 \subsubsection{Group Mass and Richness}\label{groupmass}

\begin{figure}
\centering
\includegraphics[width=\linewidth]{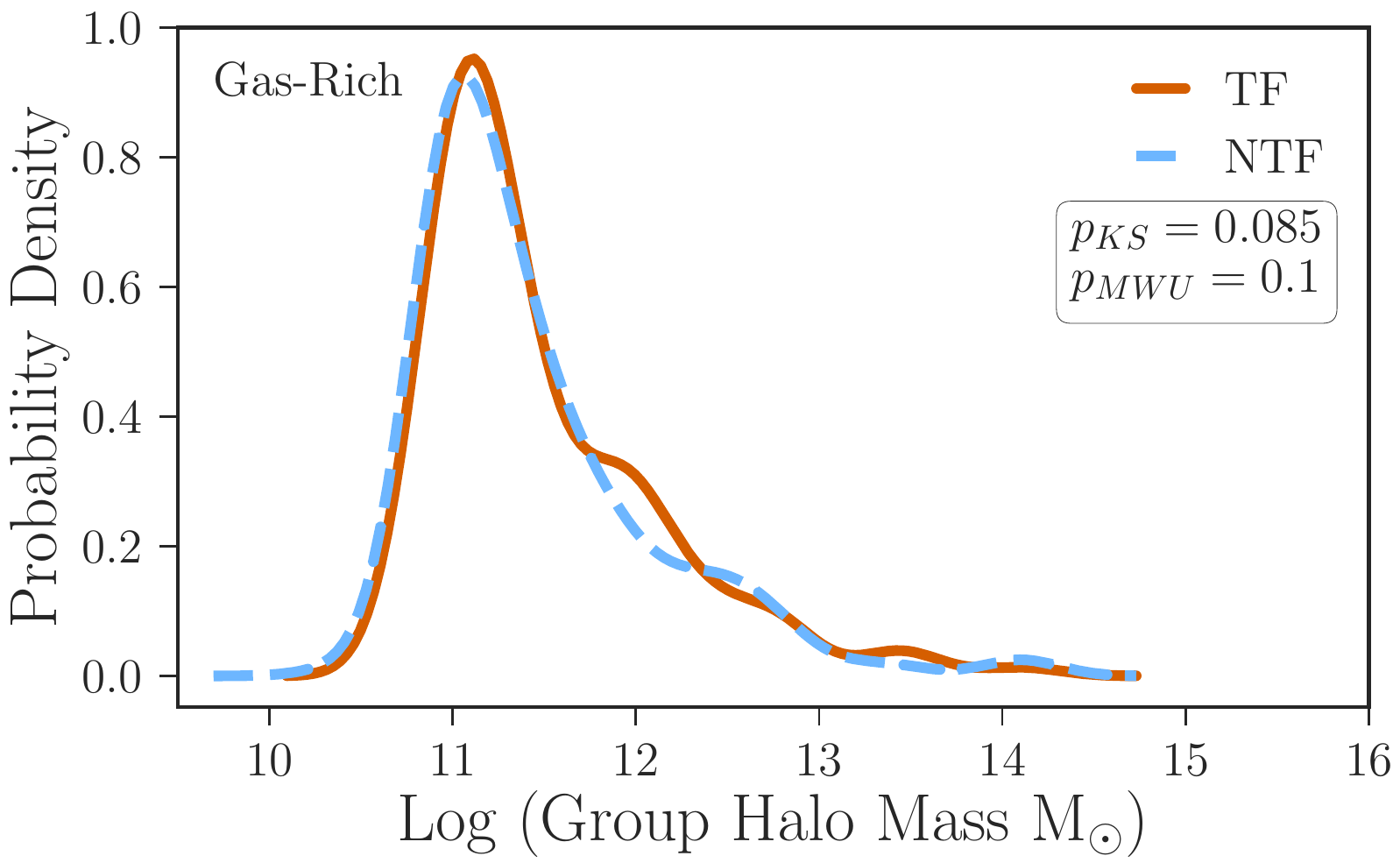}
  		\caption{Distributions of the group halo masses of gas-rich TF (orange) and NTF (blue) galaxies. We do not find a statistically-significant difference between the two distributions. The probability densities shown are kernel density estimations in logarithmic space created with a Gaussian kernel and a cross-validated optimal bandwidth of 0.21 dex, and thus have units of 1/ the x-axis unit.}
        \label{fig:halomgr}
\end{figure}

\begin{figure}
\centering
\includegraphics[width=\linewidth]{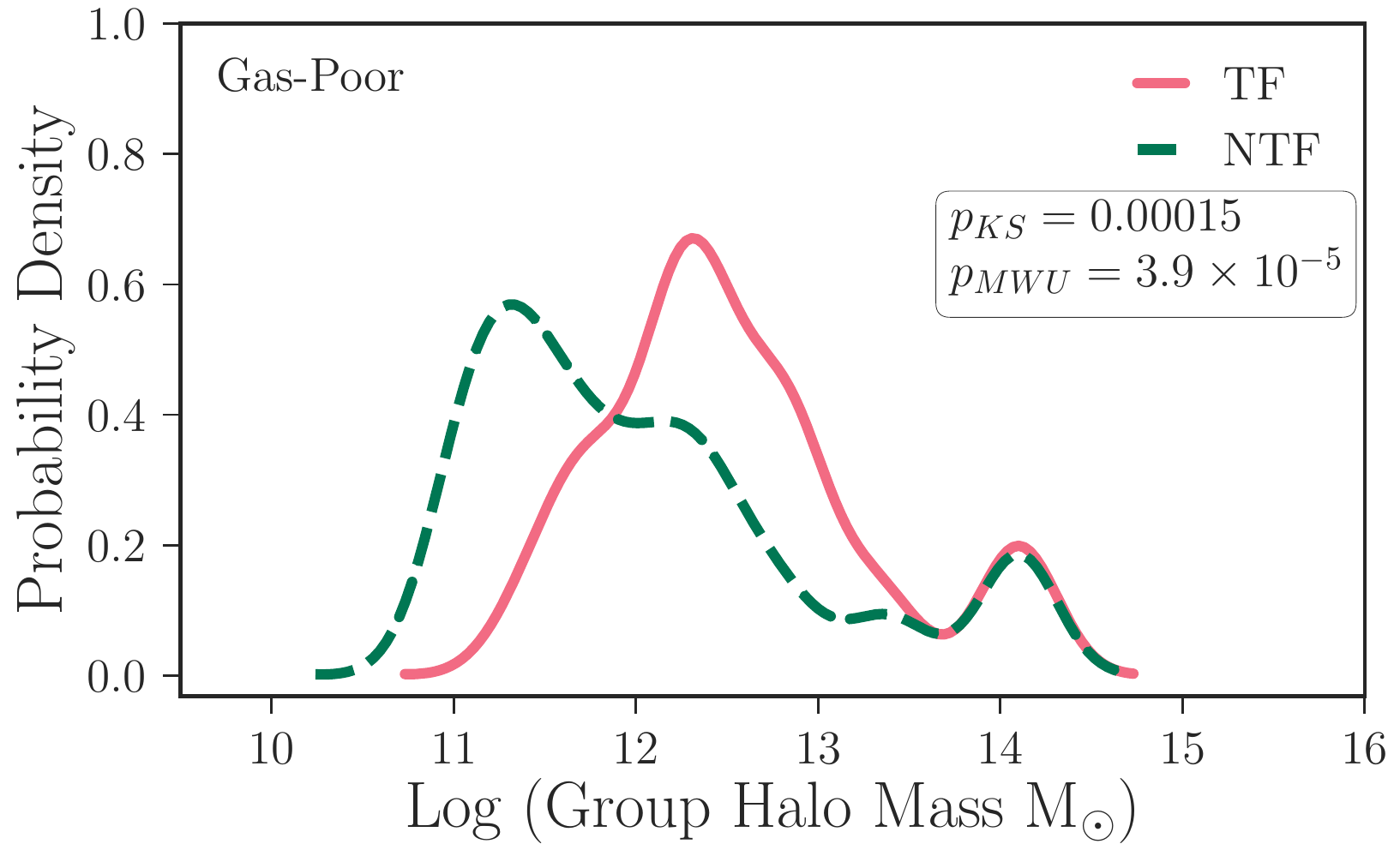}
  		\caption{Distributions of the group halo masses of gas-poor TF (pink) and NTF (green) tidal features. The gas-poor TF galaxies tend to reside in larger group halos with a $\sim4.1\sigma$ significance. The probability densities shown are kernel density estimations in logarithmic space created with a Gaussian kernel and a cross-validated optimal bandwidth of 0.21 dex, and thus have units of 1/ the x-axis unit.}
        \label{fig:halomgp}
\end{figure}

\indent Although group halo mass and number of group members appeared to be relatively insignificant indicators of tidal features according to the Random Forest results in section \ref{RF}, the relationship between tidal features and nearest-neighbor distance shown in Figure~\ref{fig:nngp} implies an environmental dependence, especially for our gas-poor sample.  Group halo mass distributions for our gas-rich and gas-poor subsamples are shown in Figures ~\ref{fig:halomgr} and~\ref{fig:halomgp}, respectively, with each subdivided into TF and NTF galaxies . Though we do not find a statistically significant difference between the halo masses of gas-rich TF and NTF galaxies, the halo masses of gas-\textit{poor} TF galaxies are more massive than those of their NTF counterparts at $\sim4.1\sigma$ significance. The same trend is seen (at a slightly lower significance of $\sim3.7\sigma$) when including less confident detections of tidal features (those in class 2) in the TF sample.

\begin{figure}
\centering
		\includegraphics[width=\linewidth]{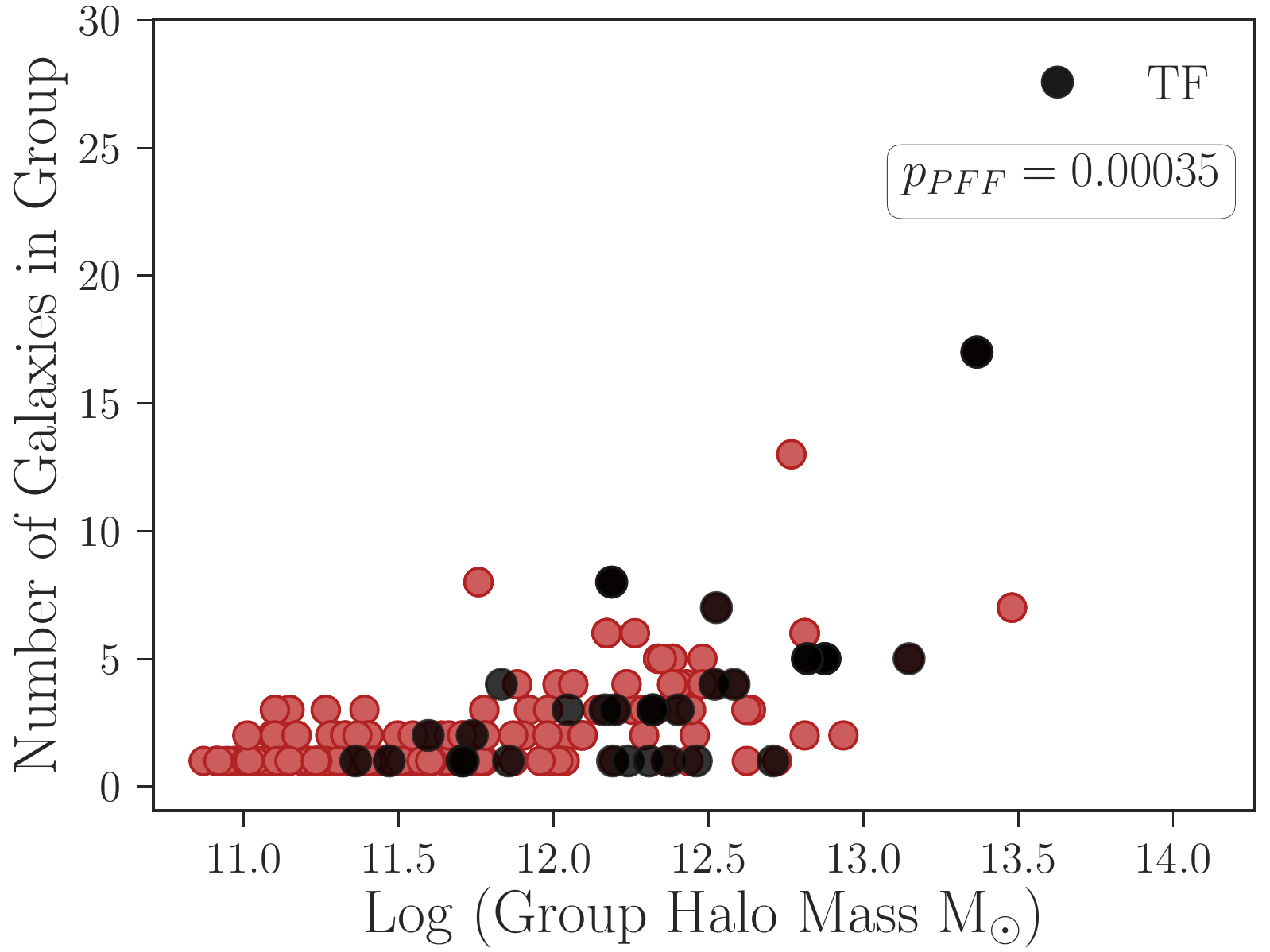}
  		\caption{ Distributions of the number of group members versus group halo mass for our gas-poor NTF (red) and TF (black) galaxies. Gas-poor TF galaxies show a different distribution of the number of group members versus group halo mass than seen for gas-poor NTF galaxies at $\sim3.6\sigma$ significance.}
        \label{fig:halongp}
\end{figure}

Moreover, we find a difference in the relation of group number versus group halo mass for gas-poor TF and NTF galaxies (Figure~\ref{fig:halongp}). A PFF test (Section \ref{SF}) shows $\sim3.6\sigma$ significance for the differences between the two distributions. We see the same trend (at a lower significance of $\sim3.3\sigma$) when including less confident detections of tidal features (those in class 2) in the gas-poor TF sample. This result may be consistent with gas-poor TF galaxies residing in groups with fewer members at fixed group halo mass than gas-poor NTF galaxies, which could point to an enhanced fraction of merger remnants among gas-poor galaxies with tidal features.\footnote{A more direct comparison of the distributions of the number of group members for gas-poor TF and NTF galaxies in fixed halo mass bins is infeasible due to the small size of the gas-poor TF sample (39 galaxies).} Merger remnants would remain in the same halo but have fewer companions than before the merger.  These completed mergers would complement the galaxy interactions and ongoing mergers evidenced by the closer nearest-neighbor distances for TF galaxies in Figure~\ref{fig:nngp}.  Overall, gas-poor galaxies with tidal features do appear to live in different environments than those without.  

\indent A similar analysis of gas-rich galaxies does not yield any significant differences for TF and NTF environments.  We lack any compelling evidence for a difference in group halo masses or number of members between gas-rich galaxies that do and do not have tidal features. 

\section{Discussion}\label{discussion}
\subsection{Comparison to Literature}\label{literature}
\indent As previously mentioned, past surveys of faint substructure do not agree on the frequency of these features; see Table 1 in \citetalias{atkinson2013} for an overview of results from published surveys. It is difficult to compare these surveys with each other and with the present work due to differences in surface brightness limits and sample selection.  For example, \citet{martinez2010} find many tidal features around a small sample ($\sim 8$) of nearby galaxies but their sample was not intended to be statistical and deliberately included galaxies already known to have tidal features.

\indent Both \citet{tal2009} and \citet{vandokkum2005} find frequencies of tidal features factors of 2$-$3 times higher than we find in this paper, but they consider only early-type galaxies.  If we limit our sample to galaxies with $\mu_{\Delta}$ > 9.5 (bulged galaxies, as defined in Section \ref{morphology}), we find that $18.0\substack{+7 \\ -6}$\% have tidal features, which is still much lower than the rates reported in these studies. \citet{vandokkum2005} uses images with limiting surface brightnesses of 29.5 mag arcsec $^{-2}$ in the $\mu$ band and may therefore be able to see many faint features that are not recoverable in the DECaLS images.  Furthermore, the luminosity cut on their sample disfavors low mass early type galaxies, which may be less likely to host tidal features as suggested by Figure \ref{fig:gpstellar}.  Similarly, although \citet{tal2009} use slightly shallower images than in the present work (V$_{vega}$ < 27 mag arcsec$^{-2}$), like van Dokkum et al.\ they consider only luminous ellipticals so their focus is on relatively massive systems, which we have found to be more likely to host tidal features (Figure \ref{fig:gpstellar}.) 

\indent \citet{duc2015} look for tidal features around early-type galaxies with K-band magnitudes brighter than $-$21.5 (i.e. stellar masses above $\sim 10^{9.8} M_{\sun}$) in 35\% of the ATLAS$^{\text{3D}}$ volume limited sample. They find that 66\% of their sample galaxies have streams, tails, or shells. Their high tidal feature fraction could additionally be due to differences in image depth, as their images reach a limiting surface brightness of 28.5 mag arcsec$^{-2}$ in the g band, as well as a bias towards higher mass galaxies from the magnitude cut. Additionally, \citet{duc2015} inspect residual images obtained by subtracting model galaxies as well as the original thumbnail images, which may reveal more substructure than the method presented here.

\indent In contrast with the previous three studies, \citet{miskolczi2011} examine face-on disk galaxies (> 2' in diameter) and find strong evidence for stream-like features around $\sim$6\% of their sample and less distinct features around a total of $\sim$19\% of their sample. They detect individual features down to 28 mag arcsec$^{-2}$, slightly deeper than the limit of our own images. Their rate of tidal features is similar to our own, but the authors do not include broader features such as shells in their detections.  If we restrict our sample to disk galaxies ($\mu_{\Delta}$ < 9.5) and only include ``narrow'' features, we find that $13^{\pm2}$\% show tidal features. Their slightly higher detection rate could be due to a difference in detection strength; in fact, if we include galaxies with a confidence level of 2, our fraction of disk galaxies with narrow tidal features rises to $17\substack{+3 \\ -2}$\%, in agreement with the fraction from \citet{miskolczi2011}.

\indent We can also compare our results with theoretical predictions for spiral galaxies from \citet{byrd1992}.  They find that a galaxy in their studied sample has a 80 $-$ 90\% chance of having tidal arms from close passages of small companions, taking into account companion decay time and arm lifetime. However, Byrd and Howard do not take feature visibility into account, so even though we only find $13^{\pm2}$\% of the disk galaxies in our sample have ``narrow'' tidal features, this rate can be taken as a lower limit still possibly consistent with the much higher occurrence rate calculated in 	\citet{byrd1992}.  See Section \ref{observability} for a more thorough discussion on the effect of feature observability on our results. 

\indent \citet{atkinson2013} (\citetalias{atkinson2013}) possibly provides the best comparison to the present work. As previously mentioned, our confidence classes are based on the same scheme used in \citetalias{atkinson2013}.  As shown in Table \ref{tab:detect}, we find many more features in the ``probable" confidence class (3) than in the ``certain" class (4), in contrast to Atkinson et al.\ who label many more features as class 4 than 3; see Figure 8 of \citetalias{atkinson2013}. They were able to detect tidal features around 12$\%$ of their 1781 galaxy sample with the strictest confidence (4); the fraction increased to 17.6$^{\pm1}\%$ when including probable but slightly less convincing features (including class 3). Atkinson et al.\ measure a mean limiting surface brightness of 27.7 mag arcsec$^{−2}$ for their g'-band images, but possibly reach a depth similar to that of our DECaLS images with their stacked g', r', and i' images. Their more inclusive $17.6\%$ rate agrees with our own fraction; however, our results differ when looking at subsamples of galaxies.  We find that gas-rich galaxies are more likely to be tidally disturbed (Section \ref{GPGRsep}), while they find that red galaxies, presumably gas-poor, are two times more likely to host tidal features than their blue counterparts (22.4$^{\pm1.5}$\% vs. 11.6$^{\pm1.7}$\%). 

\indent To investigate the source of this discrepancy, we first tested for any systematic differences in classifications. We inspected 120 galaxies from the sample studied in \citetalias{atkinson2013} in stacked g', r', and i' images from the CHFTLS-Wide fields (30 galaxies from each field). Our classifications agreed in confidence exactly with those of Atkinson and his colleagues for 91 galaxies, with most of the disagreements about the remaining 29 galaxies occurring between ``probable" and ``certain" features (classes 3 and 4).  Converting into a binary TF and NTF classification as used in this paper (where TF includes galaxies with features in the 3 and 4 confidence classes), we compared our classifications with those in \citetalias{atkinson2013} and found a Cohen's kappa score of 0.899, indicating a ``substantial strength of agreement" \citep{landis1977}. 
    
\indent In addition, we inspected 107 galaxies from the \citetalias{atkinson2013} sample in the r-band DECaLS DR3 coadds used to inspect RESOLVE galaxies in this work. Our classifications agreed in confidence for 85 of these galaxies. Again converting these confidences into TF/NTF flags, we find a Cohen's kappa score of 0.811 when comparing the two classifications. In particular, we flag 3 galaxies as TF that were not detected with much confidence by Atkinson and colleagues, all of which were blue. We may be slightly more likely to detect tidal features around blue galaxies than Atkinson, although the effect is slight. 
    
\begin{figure}
\centering
\includegraphics[width=\linewidth]{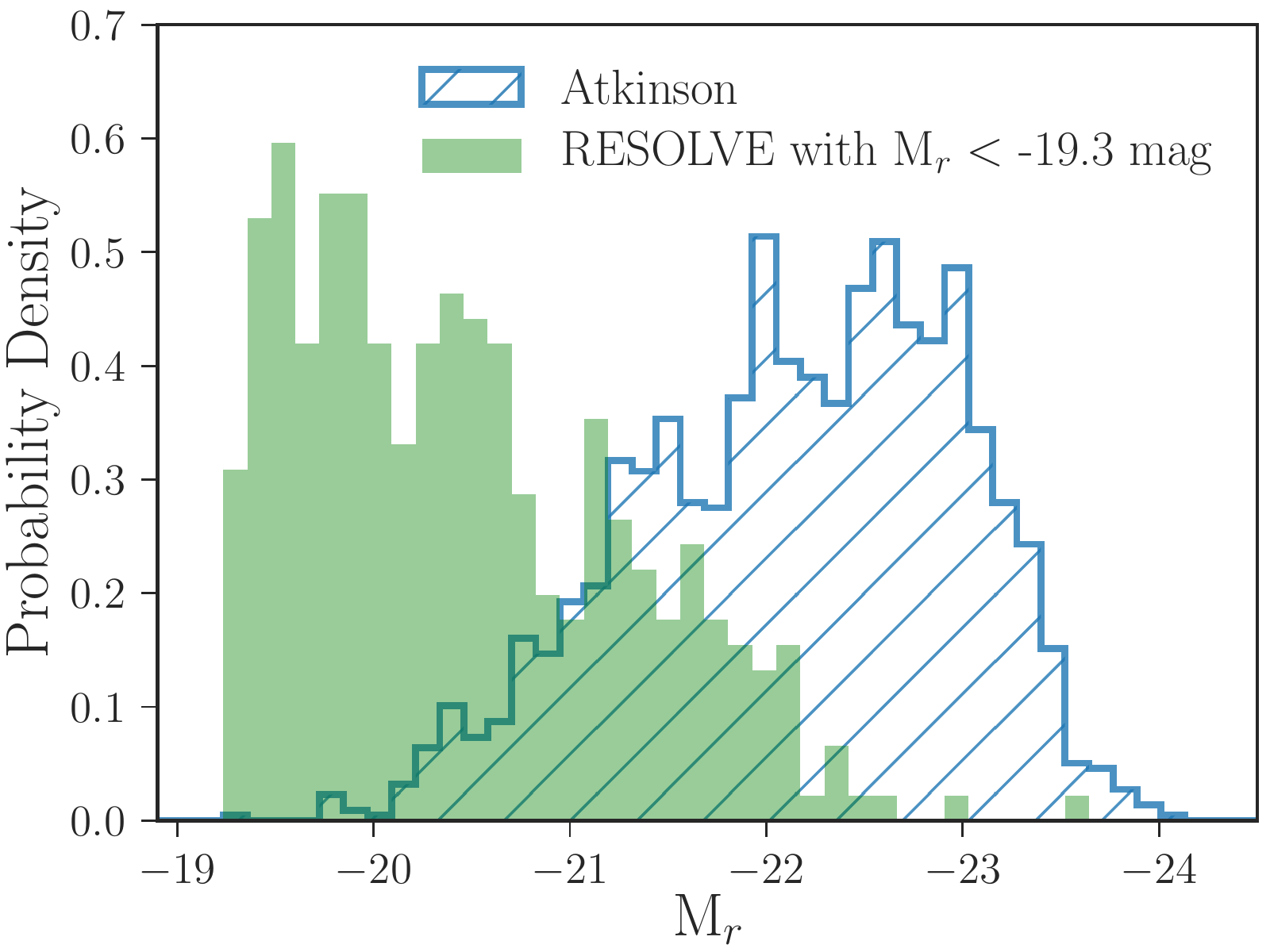}
  		\caption{Distributions of the r-band absolute magnitudes of galaxies studied by \citet{atkinson2013} (blue cross-hatch) and in the RESOLVE survey (green), with the latter sample limited to M$_{r} < - 19.3$ mag. Since the \citetalias{atkinson2013} sample is magnitude-limited, they are heavily biased towards bright systems in comparison to the sample studied in this work.}
        \label{fig:absrdist}
\end{figure}  
    
\indent As with the other surveys discussed, differences in sample selection must be taken into account. Atkinson and colleagues study 1781 galaxies in set apparent magnitude (15.5 mag < r <  17 mag) and redshift (0.04 < z < 0.2) ranges; their sample thus prioritizes bright red systems like the other magnitude-limited samples discussed previously.  In contrast, RESOLVE is a volume-limited survey and reaches r-band completeness limits of M$_{r} < -17$ for RESOLVE-B and M$_{r} < -17.33$ for RESOLVE-A; as a result, this work is naturally weighted toward smaller and more gas-rich galaxies. If we limit our sample in absolute magnitude near the faint-galaxy limit of \citetalias{atkinson2013}, i.e. apply M$_{r} < - 19.3$ mag, our sample decreases to only 370 galaxies. However, due to RESOLVE's volume-limited nature, this subsample still represents a generally fainter luminosity distribution than the luminosity-biased \citetalias{atkinson2013} sample, as shown in Figure \ref{fig:absrdist}. For this subset of brighter RESOLVE galaxies, we find that $22\substack{+5 \\ -4}$\% of these images show tidal features, which agrees with the Atkinson rate within uncertainties. Splitting this sample by gas content yields measured TF fractions of $20\substack{+7 \\ -6}$\% and $23\substack{+6 \\ -6}$\% for gas-poor and gas-rich galaxies, respectively. We converted the SDSS magnitudes of these galaxies into MegaCam filters using the equations from \citet{gwyn2008} and then applied the same color cut given in Eqn. 1 of \citetalias{atkinson2013}. 98\% of our gas-poor galaxies were labeled ``red" as expected, but only 61\% of our gas-rich galaxies were labeled ``blue" under Atkinson's system. Using these color labels, we find that $23\substack{+6 \\ -5}$\% and $19\substack{+8 \\ -6}$\% of our red and blue galaxies, respectively, have tidal features when looking at RESOLVE galaxies with M$_{r} < - 19.3$ mag. Thus, although our fraction of blue galaxies with tidal features is slightly higher, our results still agree with those in \citetalias{atkinson2013} within our uncertainties. 

\subsection{Effect of Observability}\label{observability}

\indent Although we discussed the effects of surface brightness limitations in Section \ref{frequency}, other detection biases affect our ability to identify tidal features. Principally, the time-scale over which a feature is detectable depends on the internal properties of the progenitor \citep{darg2010}. Simulations of equal-mass gas-rich mergers show time-scales depend strongly on geometric parameters like the pericentric distance and relative orientation of the galaxies \citep{lotz2008}. In addition, gas fraction correlates strongly with tidal feature longevity; \citet{lotz2010} found that asymmetry was detectable for $\leq$ 300 Myr for f$_{gas} \sim 20\%$ or log(G/S)$=-0.6$ and to $\geq$ 1 Gyr for f$_{gas} \sim 50\%$ or log(G/S)$=0.0$.  Thus, gas-poor galaxies we classified as lacking tidal features may have had recent mergers or interactions $>$300 Myr ago but appear relaxed.  Conversely, the higher rate of tidal features in our gas-rich sample could be just due to this observational effect.

\begin{figure*}[htb!]
\centering
\includegraphics[width=.9\linewidth]{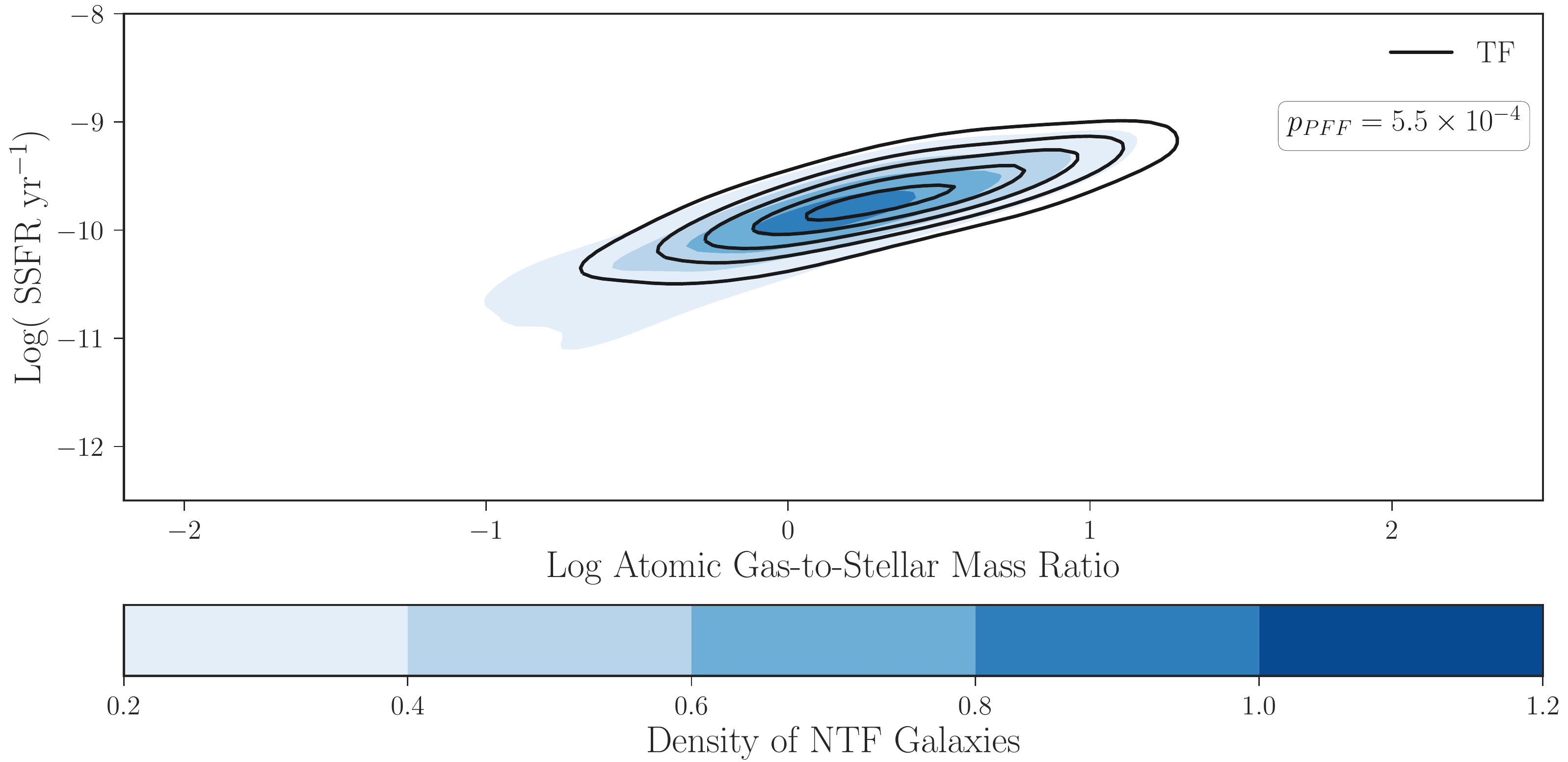}
  		\caption{ Median SSFR v. G/S distribution for 5000 mock TF and TF samples based on merger rates and visibility timescales as a function of gas fraction.  Though we see a slight increase in concentration towards higher G/S and higher SSFR, similar to our true TF and NTF samples, the median significance of the difference between these distributions is much smaller than that seen in Figure~\ref{fig:ssfrgr}. }
\label{fig:ssfrsimkde}
\end{figure*}

\begin{figure}[htb!]
\centering
\includegraphics[width=\linewidth]{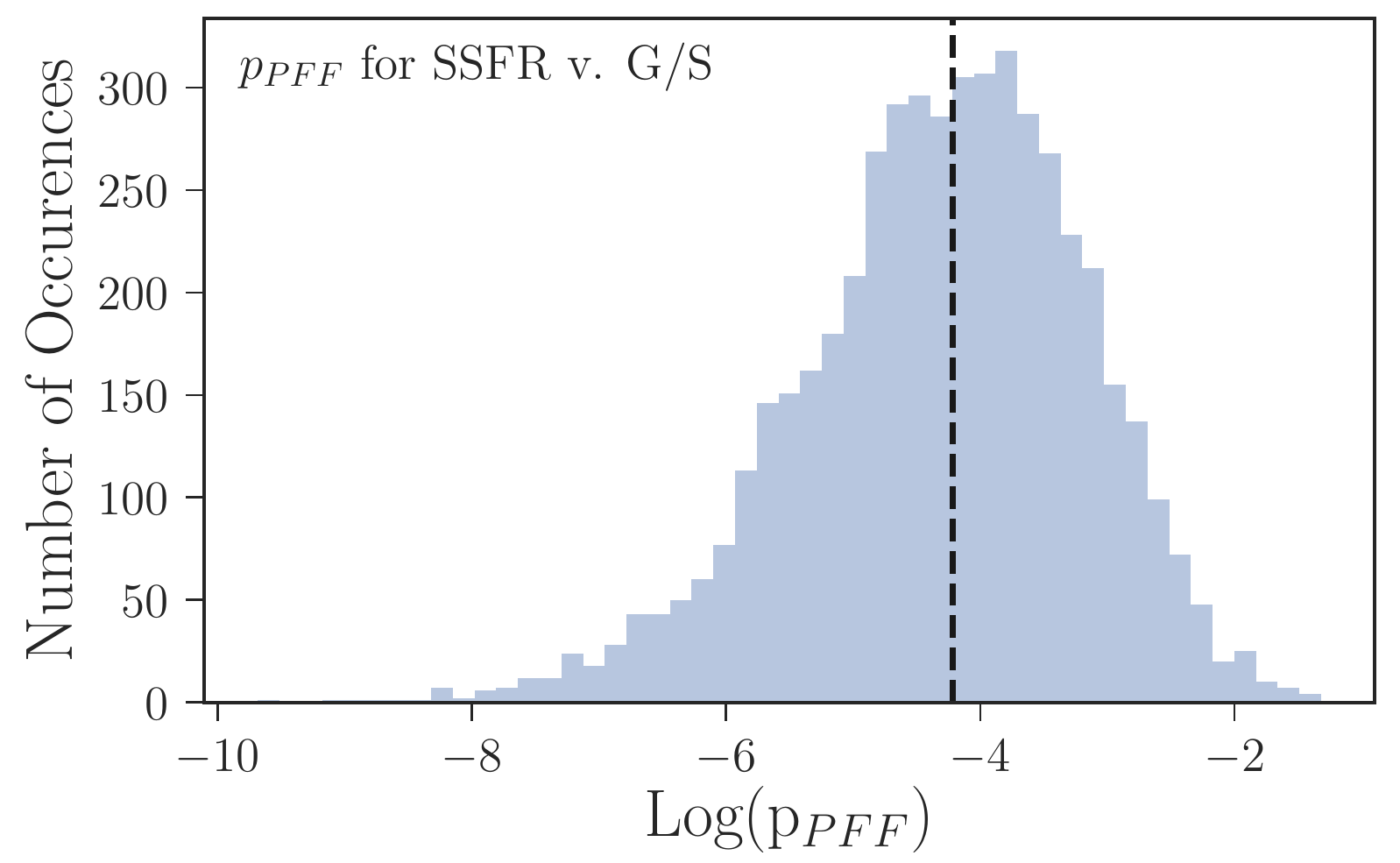}
  		\caption{Distribution of the p$_{PFF}$ comparing TF and NTF SSFR v. G/S distributions for 5000 mock TF and NTF samples based on merger rates and visibility timescales as a function of gas fraction. The median $p_{PFF}$ value is $\sim 10^{-4}$, marked by a vertical dashed line.}
        \label{fig:ssfrppff}
\end{figure}

\begin{figure*}[htb!]
\centering
\includegraphics[width=.95\linewidth]{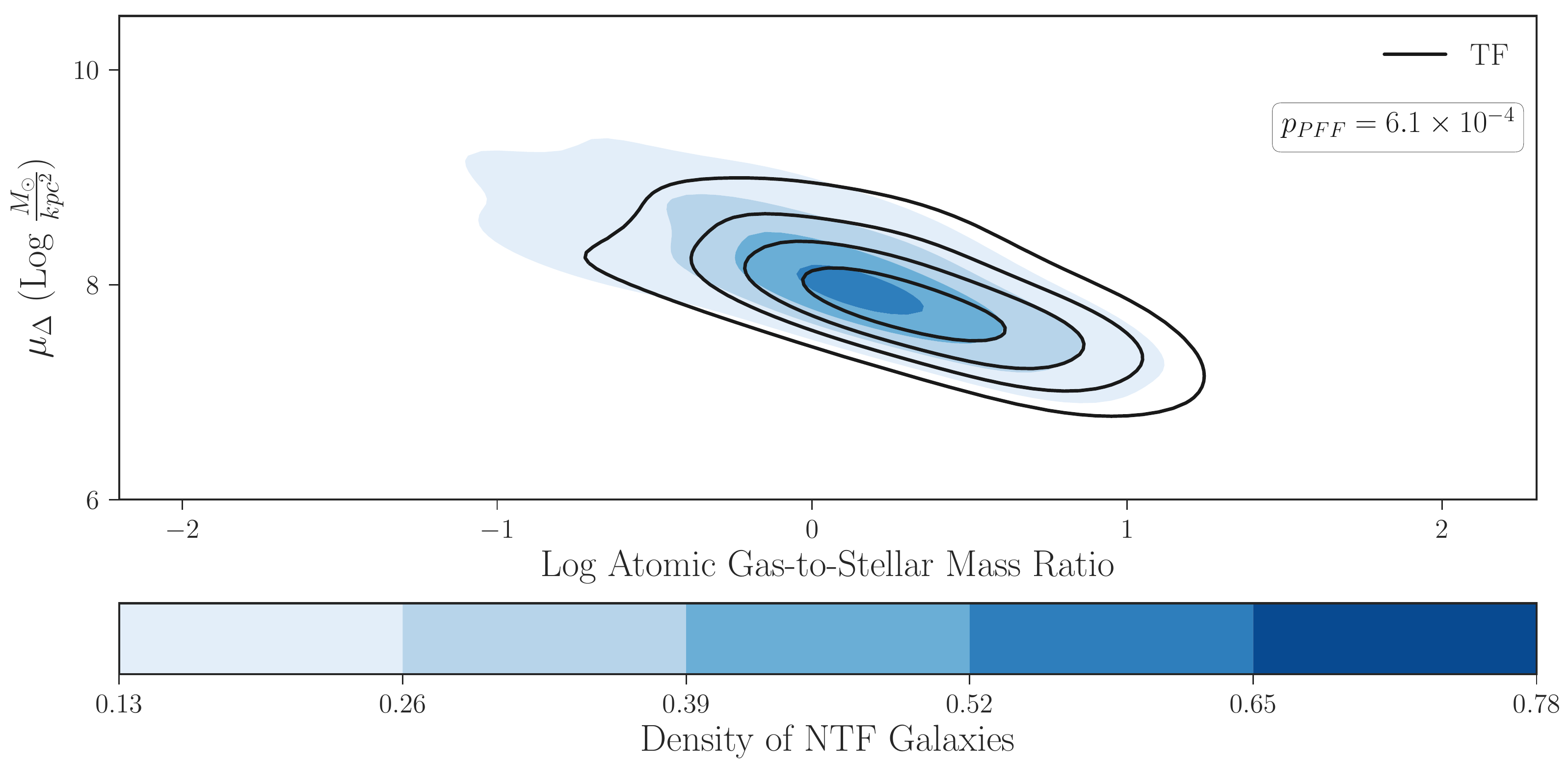}
  		\caption{Median $\mu_{\Delta}$ v. G/S distribution for 5000 mock TF and TF samples based on merger rates and visibility timescales as a function of gas fraction.  Though we see a slight increase in concentration towards higher G/S and lower $\mu_{\Delta}$, similar to our true TF and NTF samples, the median significance of the difference between these distributions is much smaller than that seen in Figure~\ref{fig:mudgf}. }
\label{fig:mudsimkde}
\end{figure*}

\begin{figure}[h!]
\centering
\includegraphics[width=\linewidth]{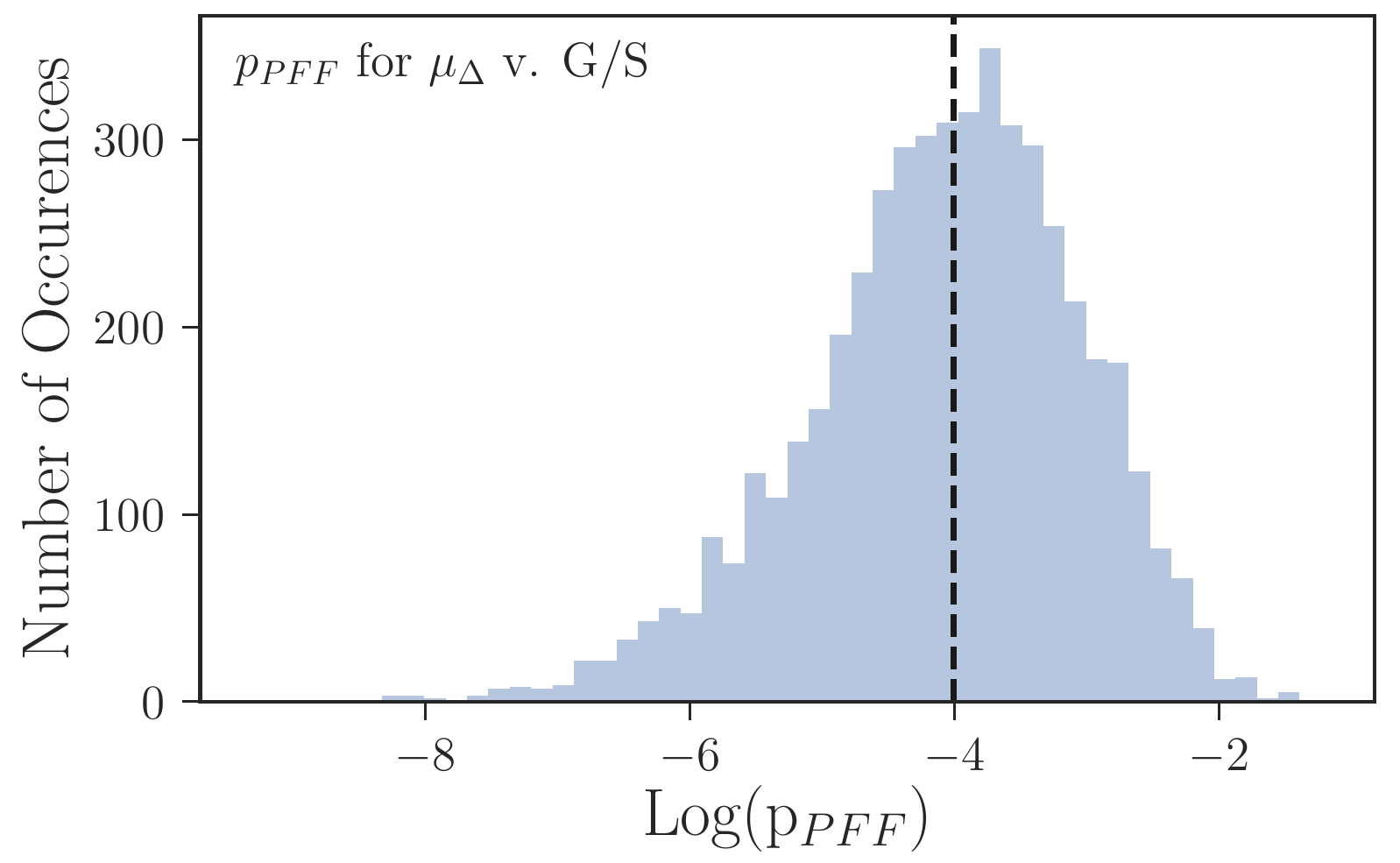}
  		\caption{Distribution of the p$_{PFF}$ comparing TF and NTF $\mu_{\Delta}$ v. G/S distributions for 5000 mock TF and NTF samples based on merger rates and visibility timescales as a function of gas fraction. The median $p_{PFF}$ value is $\sim 10^{-4}$, marked by a vertical dashed line.}
        \label{fig:mudppff}
\end{figure}

\indent To test whether our results that gas-rich TF galaxies have higher G/S ratios, increased SSFR, and more disky morphologies are mainly just a result of this relationship between G/S and TF observability, we have created mock TF and NTF samples from our inspected RESOLVE galaxies based on an estimate of the merger rate and visibility timescales as a function of gas fraction. We assume the minor merger rate at z=0 for mass ratios of 1:30 and higher is 0.2 mergers per halo per Gyr \citep{fakhouri2008}.  In addition, we use estimates of merger visibility timescales as a function of gas fraction from \citet{lotz2010}. Each stellar mass bin, ranging from log($M_{*}$)$=8.5-11.5$ in steps of 0.5 dex, is assigned a merger visibility timescale using the median gas fraction of our gas-rich galaxies in that bin. We limit our sample of galaxies in the same ways as in Sections \ref{SF} and \ref{morphology} for comparison. Thus, in Figure~\ref{fig:ssfrsimkde}, only gas-rich galaxies with 21cm detections are used, while the whole gas-rich sample is shown in Figure~\ref{fig:mudsimkde}. Using this timescale, we calculate the expected number of visible mergers in each stellar mass bin, then randomly assign RESOLVE galaxies in that bin to our TF sample accordingly. We repeat this process 5000 times, comparing the SSFR vs. G/S and $\mu_{\Delta}$ vs. G/S distributions for each iteration and recording the p-value that the TF and NTF distributions come from the same parent distribution as in Sections \ref{SF} and \ref{morphology}. Histograms of these $p_{PFF}$-values for SSFR v. G/S and $\mu_{\Delta}$ v. G/S are shown in Figures ~\ref{fig:ssfrppff} and ~\ref{fig:mudppff}, respectively. The median 2D KDEs of these distributions are shown in Figures ~\ref{fig:ssfrsimkde} and ~\ref{fig:mudsimkde}, respectively. From these figures, one can see that the dependence of visibility on gas fraction probed by these mock samples can contribute to the trends we see for the real TF and NTF samples in Figures ~\ref{fig:ssfrgr} and ~\ref{fig:mudgf}. However, the median $p_{PFF}$ when comparing TF and NTF galaxies for these mock SSFR v. G/S and $\mu_{\Delta}$ v. G/S distributions is $\sim 10^{-4}$ for each parameter, much larger than the $p_{pff}$ $\sim 10^{-8}$ and $\sim 10^{-9}$ seen for our TF and NTF classifications. Thus, we conclude that there is higher G/S, along with higher SSFR and diskier morphologies, for our gas-rich TF galaxies when compared to the NTF sample, independent of the visibility bias towards more tidal features with higher gas fractions.

\subsection{Interpretation}
\indent Due to the difference in observability discussed above, studying gas-rich and gas-poor galaxies separately helps restrict our analyses to galaxies with similar feature time-scales which could correspond to different formation scenarios. In Section \ref{results}, we studied the relationship between the presence of tidal features and other progenitor properties such as morphology, specific star formation rate, stellar mass, and environment. Here, we will argue that different types of tidal features may dominate in the two populations. The tidal features around our gas-poor galaxies are possibly connected to the cannibalism of small satellites as well as mergers at various stages.  In contrast, while our gas-rich TF galaxies may also be in various stages of merging, we will argue that accretion of gas or a gas-rich satellite could also explain the features seen in this population. 
 
\indent To recap, only our gas-rich galaxies showed strong differences between TF and NTF galaxies when looking at the morphology metric $\mu_{\Delta}$ as well as SSFR (Figures~\ref{fig:ssfrgr} and~\ref{fig:mudgf}) and these trends are derivative of the even stronger difference in G/S (Figure~\ref{fig:gasfractionkde}), such that TF galaxies have higher G/S, diskier morphology, and higher SSFR. In contrast, only our gas-poor galaxies show significantly different stellar masses and group halo masses between the TF and NTF galaxies (Figures~\ref{fig:gpstellar} and~\ref{fig:halomgp}), with TF galaxies having higher masses and a possible tendency to reside in groups with fewer members at a fixed halo mass (Figure~\ref{fig:halongp}). Both gas-rich and gas-poor TF galaxies with nearest neighbors in the same group were closer to their neighbors than their NTF counterparts (Figures~\ref{fig:nngr} and~\ref{fig:nngp}), although this result was stronger for gas-poor galaxies.
 
\subsubsection{Interpretation of Tidal Features around Gas-Poor Galaxies}\label{IntGP}
\indent Both cannibalism of small satellites and ongoing interactions and mergers are possible origin scenarios for tidal features around our gas-poor galaxies. Since gas-poor galaxies are typically supported by dispersion, they would less often form the tidal tails associated with interactions between rotating disk galaxies; instead, tidal features in gas-poor galaxies are more often likely to be tidal streams from disrupted companions or broader, diffuse features indicative of dry mergers. We find both ``narrow'' and ``broad'' features around our gas-poor galaxies, consistent with multiple formation mechanisms for these tidal features. Since gas-poor TF galaxies appear to be $\sim3\times$ closer to their nearest neighbors than their NTF counterparts, these tidal features must often result from ongoing interactions with these neighbors. 

\indent For massive primary galaxies, these neighbors could be satellites in the process of being stripped that are massive enough to fall above the RESOLVE survey limit. We find that gas-poor TF galaxies have $\sim 5$x higher median stellar masses than gas-poor NTF galaxies. This difference could indicate a large number of tidal streams from disrupted satellites: if we can detect these streams down to a certain floor in pre-disruption companion mass, we should see more of these features for higher primary masses because the merger rate increases with larger host halo mass and smaller satellite mass ratio.  As a toy calculation to demonstrate this point, consider a comparison of the mean merger rates per halo for primaries at the median stellar masses for our gas-poor TF and NTF subsamples. We assume we can only see debris from a disrupted companion with stellar mass 10$^{8} M_{\sun}$ and above in both cases. We estimate the mean merger rate per halo for a primary of a certain stellar mass using the merger rates calculated for the Millennium simulations \citep{fakhouri2010}. We relate halo masses to stellar masses using the central galaxy stellar mass-halo mass relation given by \citet{eckert2016}, under the assumption that our hypothetical primary galaxies are centrals. Integrating over the range of companion masses $\geq 10^{8} M_{\sun}$ for a primary of stellar mass 3.8 x $10^{10} M_{\sun}$ (the median stellar mass of our gas-poor TF galaxies), we obtain a mean merger rate of 1.0 $\frac{\text{mergers}}{\text{halo dz}}$. In contrast, integrating over the range of companion masses $\geq 10^{8} M_{\sun}$ for a primary with a stellar mass of 7.1 x $10^{9} M_{\sun}$ (the median stellar mass of our gas-poor NTF galaxies), we obtain a mean merger rate of 0.5 $\frac{\text{mergers}}{\text{halo dz}}$.  The implication that tidal streams should be more prevalent around higher stellar mass galaxies may also explain the higher group halo masses of gas-poor TF galaxies (Figure~\ref{fig:halomgp}). The halo mass difference is weaker than the stellar mass difference (Figure~\ref{fig:gpstellar}), suggesting the halo mass result likely follows from the stellar mass-halo mass relation, which has substantial scatter \citep{eckert2016}. 
 
\indent Alternatively, gas-poor TF galaxies of any mass may be closer to their nearest neighbors if they are in an ongoing dry merger or flyby interaction. These dry mergers and interactions would be more likely than resonant disk interactions to form ``broad'' features such as shells, which account for $\sim 26\substack{+17 \\ -13}$\% of our gas-poor TF sample. As a logical extension, it makes sense that some gas-poor TF galaxies would also reside in groups with fewer members at a fixed halo mass, as hinted by Figure~\ref{fig:halongp}, implying not only ongoing interactions but also merger remnants in this population. When two galaxies merge, the group halo mass remains the same while the number of members decreases. The result shown in Figure \ref{fig:halongp} is consistent with the weak (2.0$\sigma$) result that poor TF galaxies are slightly more bulged than their NTF counterparts, potentially reflecting spheroid growth during mergers.  

\indent We conclude that gas-poor TF galaxies may reflect a mix of satellite and flyby stripping and early-to-late stages of merging, i.e. both interacting and completely merged. 

\subsubsection{Interpretation of Tidal Features around Gas-Rich Galaxies}
\indent A larger role for gas accretion through cannibalization of gas-rich satellites or fresh gas infall from the intergalactic medium  may explain the very different trends we see for tidal features around gas-rich galaxies. Since our gas-rich population is disk-dominated (Figure \ref{fig:mudgf}), these galaxies can show both tidal tails torn out from their disks and tidal streams from disrupted satellites. The majority of tidal features in our gas-rich population are narrow features such as tidal tails/arms or streams ($\sim 79\substack{+6 \\ -7}$\%). Gas-rich galaxies with and without tidal features have similar stellar and halo mass distributions, implying an underlying common galaxy population that is experiencing sporadic events that produce tidal features. If these events were primarily satellite stripping events, creating tidal streams, the similarity of the stellar mass distributions for our gas-rich TF and NTF galaxies would be surprising given the higher expected frequency of stripping events for higher stellar mass primaries as discussed in Section \ref{IntGP}.  
 
\indent In fact, we performed the same toy calculation as in Section \ref{IntGP} to estimate the mean merger rate per halo for primaries in our gas-rich population, and it is only 0.18 $\frac{\text{mergers}}{\text{halo dz}}$ at the population median mass of 7.8 x $10^{8} M_{\sun}$. This low merger rate implies that most satellite mergers for galaxies in our gas-rich population are below the assumed $10^{8} M_{\sun}$ floor for observable stripped companions.  Though we do not know the actual floor in pre-disruption companion mass above which we can detect tidal streams, it is possible that most satellite stripping events involving our gas-rich sample involve satellites below this observability floor and do not contribute to our TF classification.  On the other hand, tidal \textit{tails/arms} torn out from the disk of the primary (rather than from the disrupted companion) due to minor mergers and interactions may still be visible even when the satellite is absorbed or torn apart to the point where it is unable to be detected above our survey floor. These features form when the primary has a rotating disk so is subject to the spin-orbit coupling resonance during the interaction \citep[e.g.][]{byrd1992}. For example, NGC 5996 is a spiral galaxy with clear tidal arms while little distortion is seen in its small companion NGC 5994 \citep{sengupta2012}. 

\indent Furthermore, though our gas-rich galaxies are less massive overall than the gas-poor sample (Figures~\ref{fig:grstellar} and~\ref{fig:gpstellar}), we find 1.5$\times$ more tidal features in the gas-rich sample (19\% vs. 13\%); this high fraction is not easily explained by the halo-halo merger rate in \citet{fakhouri2010}. To estimate the contribution from major mergers, we combine the major merger rate at z=0 integrated over mass ratios of 1:10 and higher (0.05 mergers per halo per Gyr for a primary stellar mass of 7.8 x $10^{8} M_{\sun}$ from \citealt{fakhouri2010}) with estimates of merger visibility timescales as a function of gas fraction as used in Section \ref{observability} \citep{lotz2010} to approximate the expected fraction of high asymmetry galaxies induced by major mergers in our gas-rich sample as $\sim$10\%.  Since in fact we find that 19\% of our gas-rich galaxies have tidal features, major mergers can not be the sole origin of our gas-rich TF galaxies. 

\indent Gas accretion from infalling gas-rich satellites may be functionally indistinguishable from cosmic gas infall, and both could explain the higher gas fractions and SSFRs of gas-rich TF galaxies while maintaining similar stellar masses group environments to their NTF counterparts. Furthermore, gas infall can supply the angular momentum needed for the rapid growth of disks \citep{stewart2011}, explaining the diskier morphologies of TF galaxies in the gas-rich population. As shown in Figure~\ref{fig:halomgr}, the majority of our gas-rich galaxies both with and without tidal features reside in haloes with masses below the halo mass scale at which cold-mode accretion is expected to dominate \citep{keres2009}.  This scale also corresponds to the ``gas-richness threshold scale" \citep{kannappan2013} below which gas-dominated galaxies are the norm.
 
\indent Observable interactions with companions above the RESOLVE survey limit also play a role for our gas-rich TF subsample. Gas-rich TF galaxies with neighbors within the same group have $\sim1.7$x closer nearest-neighbor distances than their NTF counterparts. This result is not as strong as for the gas-poor population, but it is still marginally significant ($\sim 2.9\sigma$).
 
\indent We conclude that tidal features around gas-rich galaxies likely arise from mergers and interactions as well as gas accretion, where the latter includes gas-rich satellite accretion below the RESOLVE survey limit.

\section{Conclusions}
In this work, we have performed a census of tidal features around galaxies in the RESOLVE survey using images primarily from DECaLS. Of the 1048 RESOLVE galaxies visually inspected for tidal features, $17^{\pm 2} \%$ of the galaxies show faint substructure in the DECaLS images. However, due to limitations of survey depth, this percentage should be seen as a lower limit. 

\indent We have used RESOLVE supporting data to study the relationship between tidal features and gas content, star formation, morphology, stellar mass, and galaxy environment. Our key results are as follows.

\begin{enumerate}
  \item Galaxies with tidal features tend to have higher atomic gas-to-stellar-mass ratios (G/S) than those without, for a ``gas-rich" sample defined by G/S $>$0.1. The ``gas-poor" sample with G/S $<$ 0.1 mostly has only upper limits on G/S.  The frequency of tidal features in our gas-rich sample, $\sim$19\%, is higher than in our gas-poor sample, $\sim$ 13\%.
  \item We observe elevated short-term specific star formation rates, diskier morphologies, and marginally closer distances to nearest neighbors (when the neighbor is in the same group as the galaxy) for gas-rich galaxies with tidal features as opposed to gas-rich galaxies without tidal features. However, the first two of these results are primarily driven by the correlation of SSFR and morphology with G/S.  We do not find any statistically significant differences between the stellar or halo mass distributions of galaxies with and without tidal features in our gas-rich subsample.
  \item In contrast, gas-poor galaxies with tidal features do not show significantly different SSFRs compared to gas-poor galaxies without tidal features, and seem to show slightly more bulged morphologies (at low statistical significance). Moreover, at high significance, gas-poor galaxies with tidal features have higher stellar and halo masses than those without. Even more so than for gas-rich galaxies, gas-poor galaxies with tidal features are closer to their nearest-neighbor (if the neighbor is in the same group).  We also observe that gas-poor galaxies with tidal features possibly reside in groups with fewer members at a fixed halo mass, suggestive of a post-merger remnant subpopulation.
\end{enumerate}

These results lend support to different origin scenarios for tidal features around gas-rich and gas-poor galaxies. Gas-rich galaxies with tidal features show higher gas fractions, diskier morphologies, and higher SSFRs than those without, pointing towards interactions and mergers but also other causes, i.e. accretion of gas or gas-rich satellites. Gas-poor galaxies with tidal features show stronger environmental differences when compared to those without, pointing to dry interactions, mergers, and satellite stripping as the main origin scenarios for tidal features in this subsample.  Future work on the observability of tidal tails and streams in cosmological hydrodynamic simulations, and observational work to differentiate tails from streams, would provide a foundation to better understand these results. 

\acknowledgments
We thank our referee, Roberto Abraham, for his insightful and constructive comments. We are grateful for helpful comments from Adrienne Erickcek. We would like to thank Alexie Leauthaud and the CS82 team for data access that helped launch this work, even though the data was not ultimately used. This work was supported a North Carolina Space Grant Undergraduate Research Scholarship. RESOLVE was funded by NSF CAREER award 0955368. This research uses services or data provided by the Science Data Archive at NOAO. NOAO is operated by the Association of Universities for Research in Astronomy (AURA), Inc. under a cooperative agreement with the National Science Foundation. 

\indent Funding for SDSS-III has been provided by the Alfred P.
Sloan Foundation, the Participating Institutions, the National Science
Foundation and the US Department of Energy Office of Science.
The SDSS-III web site is http://www.sdss3.org/. SDSS-III is
managed by the Astrophysical Research Consortium for the Participating
Institutions of the SDSS-III Collaboration including the University
of Arizona, the Brazilian Participation Group, Brookhaven
National Laboratory, University of Cambridge, Carnegie Mellon
University, University of Florida, the French Participation Group,
the German Participation Group, Harvard University, the Instituto
de Astrof'ısica de Canarias, the Michigan State/Notre Dame/JINA
Participation Group, Johns Hopkins University, Lawrence Berkeley
National Laboratory, Max Planck Institute for Astrophysics, Max
Planck Institute for Extraterrestrial Physics, New Mexico State
University, New York University, Ohio State University, Pennsylvania
State University, University of Portsmouth, Princeton University,
the Spanish Participation Group, University of Tokyo, University
of Utah, Vanderbilt University, University of Virginia, University
of Washington and Yale University.

\indent We acknowledge the use of DECaLS data; full acknowledgments can be found here: \url{http://legacysurvey.org/acknowledgment/}.

%\facility{facility ID}

\bibliographystyle{aasjournal}
\bibliography{bibliography.bib}

\begin{thebibliography}{}
\expandafter\ifx\csname natexlab\endcsname\relax\def\natexlab#1{#1}\fi
\providecommand{\url}[1]{\href{#1}{#1}}

\bibitem[{{Abazajian} {et~al.}(2009){Abazajian}, {Adelman-McCarthy},
  {Ag{\"u}eros}, {Allam}, {Allende Prieto}, {An}, {Anderson}, {Anderson},
  {Annis}, {Bahcall}, \& et~al.}]{abazajian2009}
{Abazajian}, K.~N., {Adelman-McCarthy}, J.~K., {Ag{\"u}eros}, M.~A., {et~al.}
  2009, \apjs, 182, 543

\bibitem[{{Adams} {et~al.}(2012){Adams}, {Zaritsky}, {Sand}, {Graham},
  {Bildfell}, {Hoekstra}, \& {Pritchet}}]{adams2012}
{Adams}, S.~M., {Zaritsky}, D., {Sand}, D.~J., {et~al.} 2012, \aj, 144, 128

\bibitem[{{Aihara} {et~al.}(2011){Aihara}, {Allende Prieto}, {An}, {Anderson},
  {Aubourg}, {Balbinot}, {Beers}, {Berlind}, {Bickerton}, {Bizyaev}, {Blanton},
  {Bochanski}, {Bolton}, {Bovy}, {Brandt}, {Brinkmann}, {Brown}, {Brownstein},
  {Busca}, {Campbell}, {Carr}, {Chen}, {Chiappini}, {Comparat}, {Connolly},
  {Cortes}, {Croft}, {Cuesta}, {da Costa}, {Davenport}, {Dawson}, {Dhital},
  {Ealet}, {Ebelke}, {Edmondson}, {Eisenstein}, {Escoffier}, {Esposito},
  {Evans}, {Fan}, {Femen{\'{\i}}a Castell{\'a}}, {Font-Ribera}, {Frinchaboy},
  {Ge}, {Gillespie}, {Gilmore}, {Gonz{\'a}lez Hern{\'a}ndez}, {Gott}, {Gould},
  {Grebel}, {Gunn}, {Hamilton}, {Harding}, {Harris}, {Hawley}, {Hearty}, {Ho},
  {Hogg}, {Holtzman}, {Honscheid}, {Inada}, {Ivans}, {Jiang}, {Johnson},
  {Jordan}, {Jordan}, {Kazin}, {Kirkby}, {Klaene}, {Knapp}, {Kneib},
  {Kochanek}, {Koesterke}, {Kollmeier}, {Kron}, {Lampeitl}, {Lang}, {Le Goff},
  {Lee}, {Lin}, {Long}, {Loomis}, {Lucatello}, {Lundgren}, {Lupton}, {Ma},
  {MacDonald}, {Mahadevan}, {Maia}, {Makler}, {Malanushenko}, {Malanushenko},
  {Mandelbaum}, {Maraston}, {Margala}, {Masters}, {McBride}, {McGehee},
  {McGreer}, {M{\'e}nard}, {Miralda-Escud{\'e}}, {Morrison}, {Mullally},
  {Muna}, {Munn}, {Murayama}, {Myers}, {Naugle}, {Neto}, {Nguyen}, {Nichol},
  {O'Connell}, {Ogando}, {Olmstead}, {Oravetz}, {Padmanabhan},
  {Palanque-Delabrouille}, {Pan}, {Pandey}, {P{\^a}ris}, {Percival},
  {Petitjean}, {Pfaffenberger}, {Pforr}, {Phleps}, {Pichon}, {Pieri}, {Prada},
  {Price-Whelan}, {Raddick}, {Ramos}, {Reyl{\'e}}, {Rich}, {Richards}, {Rix},
  {Robin}, {Rocha-Pinto}, {Rockosi}, {Roe}, {Rollinde}, {Ross}, {Ross},
  {Rossetto}, {S{\'a}nchez}, {Sayres}, {Schlegel}, {Schlesinger}, {Schmidt},
  {Schneider}, {Sheldon}, {Shu}, {Simmerer}, {Simmons}, {Sivarani}, {Snedden},
  {Sobeck}, {Steinmetz}, {Strauss}, {Szalay}, {Tanaka}, {Thakar}, {Thomas},
  {Tinker}, {Tofflemire}, {Tojeiro}, {Tremonti}, {Vandenberg}, {Vargas
  Maga{\~n}a}, {Verde}, {Vogt}, {Wake}, {Wang}, {Weaver}, {Weinberg}, {White},
  {White}, {Yanny}, {Yasuda}, {Yeche}, \& {Zehavi}}]{aihara2011}
{Aihara}, H., {Allende Prieto}, C., {An}, D., {et~al.} 2011, \apjs, 193, 29

\bibitem[{{Atkinson} {et~al.}(2013){Atkinson}, {Abraham}, \&
  {Ferguson}}]{atkinson2013}
{Atkinson}, A.~M., {Abraham}, R.~G., \& {Ferguson}, A.~M.~N. 2013, \apj, 765,
  28

\bibitem[{{Berlind} {et~al.}(2006){Berlind}, {Frieman}, {Weinberg}, {Blanton},
  {Warren}, {Abazajian}, {Scranton}, {Hogg}, {Scoccimarro}, {Bahcall},
  {Brinkmann}, {Gott}, {Kleinman}, {Krzesinski}, {Lee}, {Miller}, {Nitta},
  {Schneider}, {Tucker}, {Zehavi}, \& {SDSS Collaboration}}]{berlind2006}
{Berlind}, A.~A., {Frieman}, J., {Weinberg}, D.~H., {et~al.} 2006, \apjs, 167,
  1

\bibitem[{{Bertin} \& {Arnouts}(1996)}]{bertin1996}
{Bertin}, E., \& {Arnouts}, S. 1996, \aaps, 117, 393

\bibitem[{{Blum} {et~al.}(2016){Blum}, {Burleigh}, {Dey}, {Schlegel},
  {Meisner}, {Levi}, {Myers}, {Lang}, {Moustakas}, {Patej}, {Valdes}, {Kneib},
  {Huanyuan}, {Nord}, {Olsen}, {Delubac}, {Saha}, {James}, {Walker}, \& {DECaLS
  Team}}]{blum2016}
{Blum}, R.~D., {Burleigh}, K., {Dey}, A., {et~al.} 2016, in American
  Astronomical Society Meeting Abstracts, Vol. 228, American Astronomical
  Society Meeting Abstracts \#228, 317.01

\bibitem[{{Bournaud} {et~al.}(2005){Bournaud}, {Combes}, {Jog}, \&
  {Puerari}}]{bournaud2005}
{Bournaud}, F., {Combes}, F., {Jog}, C.~J., \& {Puerari}, I. 2005, \aap, 438,
  507

\bibitem[{Breiman(2001)}]{breiman2001}
Breiman, L. 2001, Machine Learning, 45, 5.
\newblock \url{https://doi.org/10.1023/A:1010933404324}

\bibitem[{{Bridge} {et~al.}(2010){Bridge}, {Carlberg}, \&
  {Sullivan}}]{bridge2010}
{Bridge}, C.~R., {Carlberg}, R.~G., \& {Sullivan}, M. 2010, \apj, 709, 1067

\bibitem[{{Bullock} \& {Johnston}(2005)}]{bullock2005}
{Bullock}, J.~S., \& {Johnston}, K.~V. 2005, \apj, 635, 931

\bibitem[{{Byrd} \& {Howard}(1992)}]{byrd1992}
{Byrd}, G.~G., \& {Howard}, S. 1992, \aj, 103, 1089

\bibitem[{{Colless} {et~al.}(2001){Colless}, {Dalton}, {Maddox}, {Sutherland},
  {Norberg}, {Cole}, {Bland-Hawthorn}, {Bridges}, {Cannon}, {Collins}, {Couch},
  {Cross}, {Deeley}, {De Propris}, {Driver}, {Efstathiou}, {Ellis}, {Frenk},
  {Glazebrook}, {Jackson}, {Lahav}, {Lewis}, {Lumsden}, {Madgwick}, {Peacock},
  {Peterson}, {Price}, {Seaborne}, \& {Taylor}}]{colless2001}
{Colless}, M., {Dalton}, G., {Maddox}, S., {et~al.} 2001, \mnras, 328, 1039

\bibitem[{{Conselice} {et~al.}(2008){Conselice}, {Rajgor}, \&
  {Myers}}]{conselice2008}
{Conselice}, C.~J., {Rajgor}, S., \& {Myers}, R. 2008, \mnras, 386, 909

\bibitem[{{Cooper} {et~al.}(2010){Cooper}, {Cole}, {Frenk}, {White}, {Helly},
  {Benson}, {De Lucia}, {Helmi}, {Jenkins}, {Navarro}, {Springel}, \&
  {Wang}}]{cooper2010}
{Cooper}, A.~P., {Cole}, S., {Frenk}, C.~S., {et~al.} 2010, \mnras, 406, 744

\bibitem[{{Darg} {et~al.}(2010){Darg}, {Kaviraj}, {Lintott}, {Schawinski},
  {Sarzi}, {Bamford}, {Silk}, {Andreescu}, {Murray}, {Nichol}, {Raddick},
  {Slosar}, {Szalay}, {Thomas}, \& {Vandenberg}}]{darg2010}
{Darg}, D.~W., {Kaviraj}, S., {Lintott}, C.~J., {et~al.} 2010, \mnras, 401,
  1552

\bibitem[{{D'Onghia} {et~al.}(2010){D'Onghia}, {Vogelsberger},
  {Faucher-Giguere}, \& {Hernquist}}]{donghia2010}
{D'Onghia}, E., {Vogelsberger}, M., {Faucher-Giguere}, C.-A., \& {Hernquist},
  L. 2010, \apj, 725, 353

\bibitem[{{Driver} {et~al.}(2011){Driver}, {Hill}, {Kelvin}, {Robotham},
  {Liske}, {Norberg}, {Baldry}, {Bamford}, {Hopkins}, {Loveday}, {Peacock},
  {Andrae}, {Bland-Hawthorn}, {Brough}, {Brown}, {Cameron}, {Ching}, {Colless},
  {Conselice}, {Croom}, {Cross}, {de Propris}, {Dye}, {Drinkwater}, {Ellis},
  {Graham}, {Grootes}, {Gunawardhana}, {Jones}, {van Kampen}, {Maraston},
  {Nichol}, {Parkinson}, {Phillipps}, {Pimbblet}, {Popescu}, {Prescott},
  {Roseboom}, {Sadler}, {Sansom}, {Sharp}, {Smith}, {Taylor}, {Thomas},
  {Tuffs}, {Wijesinghe}, {Dunne}, {Frenk}, {Jarvis}, {Madore}, {Meyer},
  {Seibert}, {Staveley-Smith}, {Sutherland}, \& {Warren}}]{driver2011}
{Driver}, S.~P., {Hill}, D.~T., {Kelvin}, L.~S., {et~al.} 2011, \mnras, 413,
  971

\bibitem[{{Duarte} \& {Mamon}(2014)}]{duarte2014}
{Duarte}, M., \& {Mamon}, G.~A. 2014, \mnras, 440, 1763

\bibitem[{{Dubinski} {et~al.}(1996){Dubinski}, {Mihos}, \&
  {Hernquist}}]{dubinski1996}
{Dubinski}, J., {Mihos}, J.~C., \& {Hernquist}, L. 1996, \apj, 462, 576

\bibitem[{{Duc} \& {Renaud}(2013)}]{duc2013}
{Duc}, P.-A., \& {Renaud}, F. 2013, in Lecture Notes in Physics, Berlin
  Springer Verlag, Vol. 861, Lecture Notes in Physics, Berlin Springer Verlag,
  ed. J.~{Souchay}, S.~{Mathis}, \& T.~{Tokieda}, 327

\bibitem[{{Duc} {et~al.}(2015){Duc}, {Cuillandre}, {Karabal}, {Cappellari},
  {Alatalo}, {Blitz}, {Bournaud}, {Bureau}, {Crocker}, {Davies}, {Davis}, {de
  Zeeuw}, {Emsellem}, {Khochfar}, {Krajnovi{\'c}}, {Kuntschner}, {McDermid},
  {Michel-Dansac}, {Morganti}, {Naab}, {Oosterloo}, {Paudel}, {Sarzi}, {Scott},
  {Serra}, {Weijmans}, \& {Young}}]{duc2015}
{Duc}, P.-A., {Cuillandre}, J.-C., {Karabal}, E., {et~al.} 2015, \mnras, 446,
  120

\bibitem[{{Ebrova}(2013)}]{ebrova2013}
{Ebrova}, I. 2013, ArXiv e-prints, arXiv:1312.1643

\bibitem[{{Eckert} {et~al.}(2016){Eckert}, {Kannappan}, {Stark}, {Moffett},
  {Berlind}, \& {Norris}}]{eckert2016}
{Eckert}, K.~D., {Kannappan}, S.~J., {Stark}, D.~V., {et~al.} 2016, \apj, 824,
  124

\bibitem[{{Eckert} {et~al.}(2015){Eckert}, {Kannappan}, {Stark}, {Moffett},
  {Norris}, {Snyder}, \& {Hoversten}}]{eckert2015}
---. 2015, \apj, 810, 166

\bibitem[{{Ellison} {et~al.}(2015){Ellison}, {Fertig}, {Rosenberg}, {Nair},
  {Simard}, {Torrey}, \& {Patton}}]{ellison2015}
{Ellison}, S.~L., {Fertig}, D., {Rosenberg}, J.~L., {et~al.} 2015, \mnras, 448,
  221

\bibitem[{{Fakhouri} \& {Ma}(2008)}]{fakhouri2008}
{Fakhouri}, O., \& {Ma}, C.-P. 2008, \mnras, 386, 577

\bibitem[{{Fakhouri} {et~al.}(2010){Fakhouri}, {Ma}, \&
  {Boylan-Kolchin}}]{fakhouri2010}
{Fakhouri}, O., {Ma}, C.-P., \& {Boylan-Kolchin}, M. 2010, \mnras, 406, 2267

\bibitem[{{Falco} {et~al.}(1999){Falco}, {Kurtz}, {Geller}, {Huchra}, {Peters},
  {Berlind}, {Mink}, {Tokarz}, \& {Elwell}}]{falco1999}
{Falco}, E.~E., {Kurtz}, M.~J., {Geller}, M.~J., {et~al.} 1999, \pasp, 111, 438

\bibitem[{{Fasano} \& {Franceschini}(1987)}]{fasano1987}
{Fasano}, G., \& {Franceschini}, A. 1987, \mnras, 225, 155

\bibitem[{{Feldmann} {et~al.}(2008){Feldmann}, {Mayer}, \&
  {Carollo}}]{feldmann2008}
{Feldmann}, R., {Mayer}, L., \& {Carollo}, C.~M. 2008, \apj, 684, 1062

\bibitem[{{Fliri} \& {Trujillo}(2016)}]{fliri2016}
{Fliri}, J., \& {Trujillo}, I. 2016, \mnras, 456, 1359

\bibitem[{{Franx} {et~al.}(2008){Franx}, {van Dokkum}, {F{\"o}rster Schreiber},
  {Wuyts}, {Labb{\'e}}, \& {Toft}}]{franx2008}
{Franx}, M., {van Dokkum}, P.~G., {F{\"o}rster Schreiber}, N.~M., {et~al.}
  2008, \apj, 688, 770

\bibitem[{Geisser(1975)}]{geisser1975}
Geisser, S. 1975, Journal of the American Statistical Association, 70, 320

\bibitem[{{Gwyn}(2008)}]{gwyn2008}
{Gwyn}, S.~D.~J. 2008, \pasp, 120, 212

\bibitem[{{Hambly} {et~al.}(2008){Hambly}, {Collins}, {Cross}, {Mann}, {Read},
  {Sutorius}, {Bond}, {Bryant}, {Emerson}, {Lawrence}, {Rimoldini}, {Stewart},
  {Williams}, {Adamson}, {Hirst}, {Dye}, \& {Warren}}]{hambly2008}
{Hambly}, N.~C., {Collins}, R.~S., {Cross}, N.~J.~G., {et~al.} 2008, \mnras,
  384, 637

\bibitem[{{Haynes} {et~al.}(2011){Haynes}, {Giovanelli}, {Martin}, {Hess},
  {Saintonge}, {Adams}, {Hallenbeck}, {Hoffman}, {Huang}, {Kent}, {Koopmann},
  {Papastergis}, {Stierwalt}, {Balonek}, {Craig}, {Higdon}, {Kornreich},
  {Miller}, {O'Donoghue}, {Olowin}, {Rosenberg}, {Spekkens}, {Troischt}, \&
  {Wilcots}}]{haynes2011}
{Haynes}, M.~P., {Giovanelli}, R., {Martin}, A.~M., {et~al.} 2011, \aj, 142,
  170

\bibitem[{{Hopkins} {et~al.}(2009){Hopkins}, {Cox}, {Younger}, \&
  {Hernquist}}]{hopkins2009}
{Hopkins}, P.~F., {Cox}, T.~J., {Younger}, J.~D., \& {Hernquist}, L. 2009,
  \apj, 691, 1168

\bibitem[{{Ji} {et~al.}(2014){Ji}, {Peirani}, \& {Yi}}]{ji2014}
{Ji}, I., {Peirani}, S., \& {Yi}, S.~K. 2014, \aap, 566, A97

\bibitem[{{Jog} \& {Combes}(2009)}]{jog2009}
{Jog}, C.~J., \& {Combes}, F. 2009, \physrep, 471, 75

\bibitem[{{Johnston} {et~al.}(2008){Johnston}, {Bullock}, {Sharma}, {Font},
  {Robertson}, \& {Leitner}}]{johnston2008}
{Johnston}, K.~V., {Bullock}, J.~S., {Sharma}, S., {et~al.} 2008, \apj, 689,
  936

\bibitem[{{Jones} {et~al.}(2009){Jones}, {Read}, {Saunders}, {Colless},
  {Jarrett}, {Parker}, {Fairall}, {Mauch}, {Sadler}, {Watson}, {Burton},
  {Campbell}, {Cass}, {Croom}, {Dawe}, {Fiegert}, {Frankcombe}, {Hartley},
  {Huchra}, {James}, {Kirby}, {Lahav}, {Lucey}, {Mamon}, {Moore}, {Peterson},
  {Prior}, {Proust}, {Russell}, {Safouris}, {Wakamatsu}, {Westra}, \&
  {Williams}}]{jones2009}
{Jones}, D.~H., {Read}, M.~A., {Saunders}, W., {et~al.} 2009, \mnras, 399, 683

\bibitem[{{Joye} \& {Mandel}(2003)}]{joye2003}
{Joye}, W.~A., \& {Mandel}, E. 2003, in Astronomical Society of the Pacific
  Conference Series, Vol. 295, Astronomical Data Analysis Software and Systems
  XII, ed. H.~E. {Payne}, R.~I. {Jedrzejewski}, \& R.~N. {Hook}, 489

\bibitem[{{Kannappan}(2004)}]{kannappan2004}
{Kannappan}, S.~J. 2004, \apjl, 611, L89

\bibitem[{{Kannappan} \& {Gawiser}(2007)}]{kannappan2007}
{Kannappan}, S.~J., \& {Gawiser}, E. 2007, \apjl, 657, L5

\bibitem[{{Kannappan} \& {Wei}(2008)}]{kannappan2008}
{Kannappan}, S.~J., \& {Wei}, L.~H. 2008, in American Institute of Physics
  Conference Series, Vol. 1035, The Evolution of Galaxies Through the Neutral
  Hydrogen Window, ed. R.~{Minchin} \& E.~{Momjian}, 163--168

\bibitem[{{Kannappan} {et~al.}(2013){Kannappan}, {Stark}, {Eckert}, {Moffett},
  {Wei}, {Pisano}, {Baker}, {Vogel}, {Fabricant}, {Laine}, {Norris}, {Jogee},
  {Lepore}, {Hough}, \& {Weinberg-Wolf}}]{kannappan2013}
{Kannappan}, S.~J., {Stark}, D.~V., {Eckert}, K.~D., {et~al.} 2013, \apj, 777,
  42

\bibitem[{{Kere{\v s}} {et~al.}(2009){Kere{\v s}}, {Katz}, {Fardal},
  {Dav{\'e}}, \& {Weinberg}}]{keres2009}
{Kere{\v s}}, D., {Katz}, N., {Fardal}, M., {Dav{\'e}}, R., \& {Weinberg},
  D.~H. 2009, \mnras, 395, 160

\bibitem[{{Kim} {et~al.}(2012){Kim}, {Sheth}, {Hinz}, {Lee}, {Zaritsky},
  {Gadotti}, {Knapen}, {Schinnerer}, {Ho}, {Laurikainen}, {Salo},
  {Athanassoula}, {Bosma}, {de Swardt}, {Mu{\~n}oz-Mateos}, {Madore},
  {Comer{\'o}n}, {Regan}, {Men{\'e}ndez-Delmestre}, {Gil de Paz}, {Seibert},
  {Laine}, {Erroz-Ferrer}, \& {Mizusawa}}]{kim2012}
{Kim}, T., {Sheth}, K., {Hinz}, J.~L., {et~al.} 2012, \apj, 753, 43

\bibitem[{{Lagos} {et~al.}(2017){Lagos}, {Stevens}, {Bower}, {Davis},
  {Contreras}, {Padilla}, {Obreschkow}, {Croton}, {Trayford}, {Welker}, \&
  {Theuns}}]{lagos2017}
{Lagos}, C.~d.~P., {Stevens}, A.~R.~H., {Bower}, R.~G., {et~al.} 2017, ArXiv
  e-prints, arXiv:1701.04407

\bibitem[{Landis \& Koch(1977)}]{landis1977}
Landis, J.~R., \& Koch, G.~G. 1977, Biometrics, 33, 159.
\newblock \url{http://www.jstor.org/stable/2529310}

\bibitem[{{Lilly} \& {Carollo}(2016)}]{lilly2016}
{Lilly}, S.~J., \& {Carollo}, C.~M. 2016, \apj, 833, 1

\bibitem[{{Lisenfeld} {et~al.}(2011){Lisenfeld}, {Espada}, {Verdes-Montenegro},
  {Kuno}, {Leon}, {Sabater}, {Sato}, {Sulentic}, {Verley}, \&
  {Yun}}]{lisenfeld2011}
{Lisenfeld}, U., {Espada}, D., {Verdes-Montenegro}, L., {et~al.} 2011, \aap,
  534, A102

\bibitem[{{Lotz} {et~al.}(2008){Lotz}, {Jonsson}, {Cox}, \&
  {Primack}}]{lotz2008}
{Lotz}, J.~M., {Jonsson}, P., {Cox}, T.~J., \& {Primack}, J.~R. 2008, \mnras,
  391, 1137

\bibitem[{{Lotz} {et~al.}(2010){Lotz}, {Jonsson}, {Cox}, \&
  {Primack}}]{lotz2010}
---. 2010, \mnras, 404, 590

\bibitem[{{Malin} \& {Carter}(1983)}]{malin1983}
{Malin}, D.~F., \& {Carter}, D. 1983, \apj, 274, 534

\bibitem[{{Maneewongvatana} \& {Mount}(1999)}]{man1999}
{Maneewongvatana}, S., \& {Mount}, D.~M. 1999, eprint arXiv:cs/9901013,
  cs/9901013

\bibitem[{{Mapelli} {et~al.}(2008){Mapelli}, {Moore}, \&
  {Bland-Hawthorn}}]{mapelli2008}
{Mapelli}, M., {Moore}, B., \& {Bland-Hawthorn}, J. 2008, \mnras, 388, 697

\bibitem[{{Mart{\'{\i}}nez-Delgado} {et~al.}(2015){Mart{\'{\i}}nez-Delgado},
  {D'Onghia}, {Chonis}, {Beaton}, {Teuwen}, {GaBany}, {Grebel}, \&
  {Morales}}]{martinez2015}
{Mart{\'{\i}}nez-Delgado}, D., {D'Onghia}, E., {Chonis}, T.~S., {et~al.} 2015,
  \aj, 150, 116

\bibitem[{{Mart{\'{\i}}nez-Delgado} {et~al.}(2008){Mart{\'{\i}}nez-Delgado},
  {Pe{\~n}arrubia}, {Gabany}, {Trujillo}, {Majewski}, \&
  {Pohlen}}]{martinez2008}
{Mart{\'{\i}}nez-Delgado}, D., {Pe{\~n}arrubia}, J., {Gabany}, R.~J., {et~al.}
  2008, \apj, 689, 184

\bibitem[{{Mart{\'{\i}}nez-Delgado} {et~al.}(2010){Mart{\'{\i}}nez-Delgado},
  {Gabany}, {Crawford}, {Zibetti}, {Majewski}, {Rix}, {Fliri},
  {Carballo-Bello}, {Bardalez-Gagliuffi}, {Pe{\~n}arrubia}, {Chonis}, {Madore},
  {Trujillo}, {Schirmer}, \& {McDavid}}]{martinez2010}
{Mart{\'{\i}}nez-Delgado}, D., {Gabany}, R.~J., {Crawford}, K., {et~al.} 2010,
  \aj, 140, 962

\bibitem[{{Miskolczi} {et~al.}(2011){Miskolczi}, {Bomans}, \&
  {Dettmar}}]{miskolczi2011}
{Miskolczi}, A., {Bomans}, D.~J., \& {Dettmar}, R.-J. 2011, \aap, 536, A66

\bibitem[{{Moffett} {et~al.}(2015){Moffett}, {Kannappan}, {Berlind}, {Eckert},
  {Stark}, {Hendel}, {Norris}, \& {Grogin}}]{moffett2015}
{Moffett}, A.~J., {Kannappan}, S.~J., {Berlind}, A.~A., {et~al.} 2015, \apj,
  812, 89

\bibitem[{{Morrissey} {et~al.}(2007){Morrissey}, {Conrow}, {Barlow}, {Small},
  {Seibert}, {Wyder}, {Budav{\'a}ri}, {Arnouts}, {Friedman}, {Forster},
  {Martin}, {Neff}, {Schiminovich}, {Bianchi}, {Donas}, {Heckman}, {Lee},
  {Madore}, {Milliard}, {Rich}, {Szalay}, {Welsh}, \& {Yi}}]{morrissey2007}
{Morrissey}, P., {Conrow}, T., {Barlow}, T.~A., {et~al.} 2007, \apjs, 173, 682

\bibitem[{{Naab} {et~al.}(2006){Naab}, {Jesseit}, \& {Burkert}}]{naab2006}
{Naab}, T., {Jesseit}, R., \& {Burkert}, A. 2006, \mnras, 372, 839

\bibitem[{{Nair} \& {Abraham}(2010)}]{nair2010}
{Nair}, P.~B., \& {Abraham}, R.~G. 2010, \apjs, 186, 427

\bibitem[{{Oh} {et~al.}(2008){Oh}, {Kim}, {Lee}, \& {Kim}}]{oh2008}
{Oh}, S.~H., {Kim}, W.-T., {Lee}, H.~M., \& {Kim}, J. 2008, \apj, 683, 94

\bibitem[{{Paturel} {et~al.}(2003){Paturel}, {Petit}, {Prugniel}, {Theureau},
  {Rousseau}, {Brouty}, {Dubois}, \& {Cambr{\'e}sy}}]{paturel2003}
{Paturel}, G., {Petit}, C., {Prugniel}, P., {et~al.} 2003, \aap, 412, 45

\bibitem[{{Peacock}(1983)}]{peacock1983}
{Peacock}, J.~A. 1983, \mnras, 202, 615

\bibitem[{Pedregosa {et~al.}(2011)Pedregosa, Varoquaux, Gramfort, Michel,
  Thirion, Grisel, Blondel, Prettenhofer, Weiss, Dubourg,
  {et~al.}}]{pedregosa2011}
Pedregosa, F., Varoquaux, G., Gramfort, A., {et~al.} 2011, Journal of Machine
  Learning Research, 12, 2825

\bibitem[{{Richer} {et~al.}(2003){Richer}, {Georgiev}, {Rosado}, {Bullejos},
  {Valdez-Guti{\'e}rrez}, \& {Dultzin-Hacyan}}]{richer2003}
{Richer}, M.~G., {Georgiev}, L., {Rosado}, M., {et~al.} 2003, \aap, 397, 99

\bibitem[{{Robitaille} \& {Bressert}(2012)}]{aplpy}
{Robitaille}, T., \& {Bressert}, E. 2012, {APLpy: Astronomical Plotting Library
  in Python}, Astrophysics Source Code Library, , , ascl:1208.017

\bibitem[{{Saintonge} {et~al.}(2016){Saintonge}, {Catinella}, {Cortese},
  {Genzel}, {Giovanelli}, {Haynes}, {Janowiecki}, {Kramer}, {Lutz},
  {Schiminovich}, {Tacconi}, {Wuyts}, \& {Accurso}}]{saintonge2016}
{Saintonge}, A., {Catinella}, B., {Cortese}, L., {et~al.} 2016, \mnras, 462,
  1749

\bibitem[{{Schlegel} {et~al.}(2015){Schlegel}, {Blum}, {Castander}, {Dey},
  {Finkbeiner}, {Foucaud}, {Honscheid}, {James}, {Lang}, {Levi}, {Moustakas},
  {Myers}, {Newman}, {Nord}, {Nugent}, {Patej}, {Reil}, {Rudnick}, {Rykoff},
  {Ford Schlafly}, {Stark}, {Valdes}, {Walker}, {Weaver}, \& {DECam Legacy
  Survey Collaboration}}]{schlegel2015}
{Schlegel}, D.~J., {Blum}, R.~D., {Castander}, F.~J., {et~al.} 2015, in
  American Astronomical Society Meeting Abstracts, Vol. 225, American
  Astronomical Society Meeting Abstracts, 336.07

\bibitem[{{Schweizer} \& {Seitzer}(1988)}]{schweizer1988}
{Schweizer}, F., \& {Seitzer}, P. 1988, \apj, 328, 88

\bibitem[{{Schweizer} \& {Seitzer}(1992)}]{schweizer1992}
---. 1992, \aj, 104, 1039

\bibitem[{{Sengupta} {et~al.}(2012){Sengupta}, {Saikia}, \&
  {Dwarakanath}}]{sengupta2012}
{Sengupta}, C., {Saikia}, D.~J., \& {Dwarakanath}, K.~S. 2012, \mnras, 420, 2

\bibitem[{{Sheen} {et~al.}(2012){Sheen}, {Yi}, {Ree}, \& {Lee}}]{sheen2012}
{Sheen}, Y.-K., {Yi}, S.~K., {Ree}, C.~H., \& {Lee}, J. 2012, \apjs, 202, 8

\bibitem[{{Skrutskie} {et~al.}(2006){Skrutskie}, {Cutri}, {Stiening},
  {Weinberg}, {Schneider}, {Carpenter}, {Beichman}, {Capps}, {Chester},
  {Elias}, {Huchra}, {Liebert}, {Lonsdale}, {Monet}, {Price}, {Seitzer},
  {Jarrett}, {Kirkpatrick}, {Gizis}, {Howard}, {Evans}, {Fowler}, {Fullmer},
  {Hurt}, {Light}, {Kopan}, {Marsh}, {McCallon}, {Tam}, {Van Dyk}, \&
  {Wheelock}}]{skrutskie2006}
{Skrutskie}, M.~F., {Cutri}, R.~M., {Stiening}, R., {et~al.} 2006, \aj, 131,
  1163

\bibitem[{{Stark} {et~al.}(2013){Stark}, {Kannappan}, {Wei}, {Baker}, {Leroy},
  {Eckert}, \& {Vogel}}]{stark2013}
{Stark}, D.~V., {Kannappan}, S.~J., {Wei}, L.~H., {et~al.} 2013, \apj, 769, 82

\bibitem[{{Stark} {et~al.}(2016){Stark}, {Kannappan}, {Eckert}, {Florez},
  {Hall}, {Watson}, {Hoversten}, {Burchett}, {Guynn}, {Baker}, {Moffett},
  {Berlind}, {Norris}, {Haynes}, {Giovanelli}, {Leroy}, {Pisano}, {Wei},
  {Gonzalez}, \& {Calderon}}]{stark2016}
{Stark}, D.~V., {Kannappan}, S.~J., {Eckert}, K.~D., {et~al.} 2016, \apj, 832,
  126

\bibitem[{{Stewart} {et~al.}(2011){Stewart}, {Kaufmann}, {Bullock}, {Barton},
  {Maller}, {Diemand}, \& {Wadsley}}]{stewart2011}
{Stewart}, K.~R., {Kaufmann}, T., {Bullock}, J.~S., {et~al.} 2011, \apj, 738,
  39

\bibitem[{{Tal} {et~al.}(2009){Tal}, {van Dokkum}, {Nelan}, \&
  {Bezanson}}]{tal2009}
{Tal}, T., {van Dokkum}, P.~G., {Nelan}, J., \& {Bezanson}, R. 2009, \aj, 138,
  1417

\bibitem[{{Toomre} \& {Toomre}(1972)}]{toomre1972}
{Toomre}, A., \& {Toomre}, J. 1972, \apj, 178, 623

\bibitem[{{van Dokkum}(2005)}]{vandokkum2005}
{van Dokkum}, P.~G. 2005, \aj, 130, 2647

\bibitem[{{Warren} {et~al.}(2006){Warren}, {Abazajian}, {Holz}, \&
  {Teodoro}}]{warren2006}
{Warren}, M.~S., {Abazajian}, K., {Holz}, D.~E., \& {Teodoro}, L. 2006, \apj,
  646, 881

\bibitem[{{Wilkins} {et~al.}(2012){Wilkins}, {Gonzalez-Perez}, {Lacey}, \&
  {Baugh}}]{wilkins2012}
{Wilkins}, S.~M., {Gonzalez-Perez}, V., {Lacey}, C.~G., \& {Baugh}, C.~M. 2012,
  \mnras, 427, 1490

\end{thebibliography}
% \bibliography{references}

\appendix
\vspace{0.8cm}
\begin{figure*}[h!]
\centering
\includegraphics[width=1\linewidth]{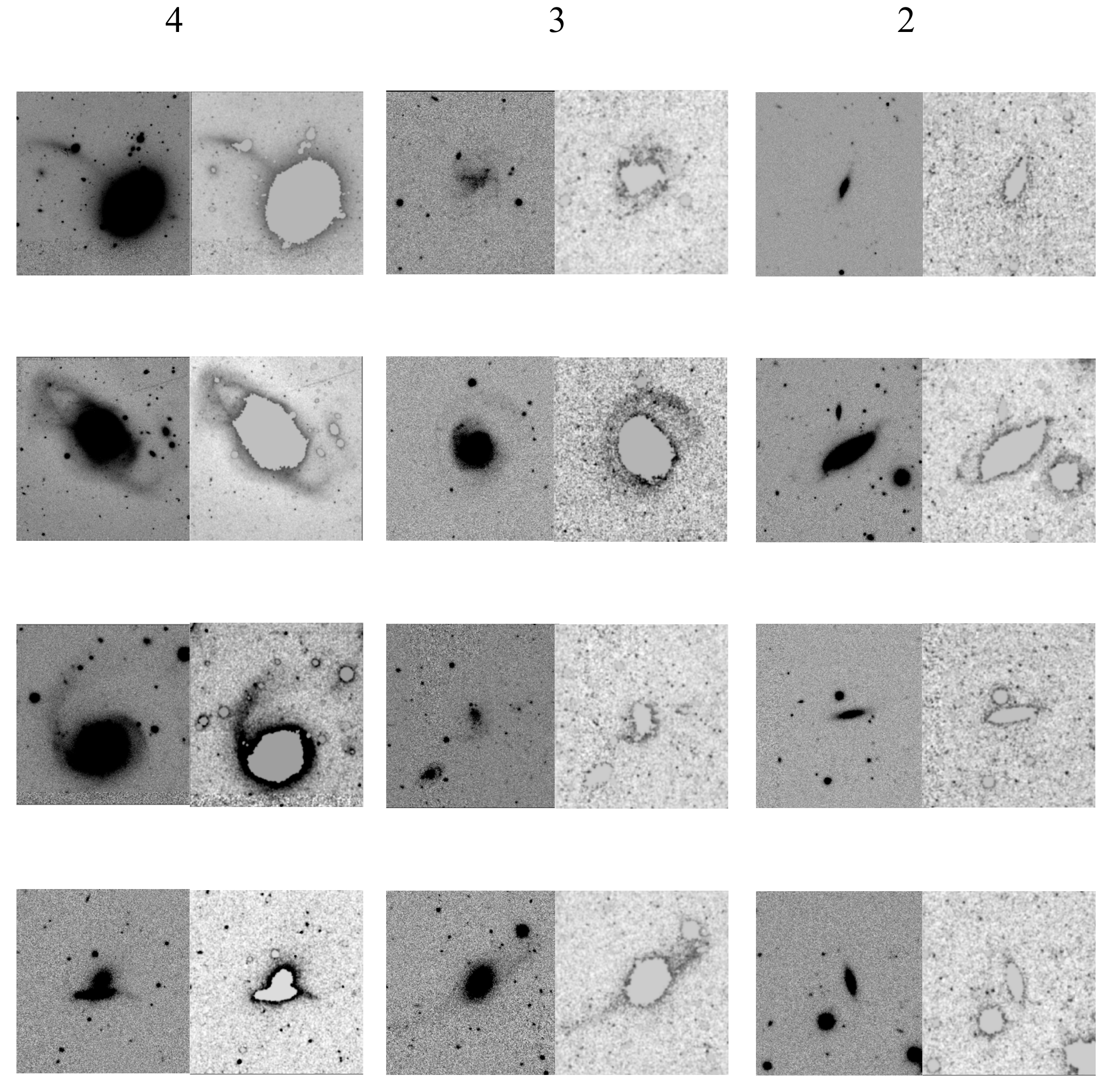}
  		\caption{Examples of ``narrow'' tidal features identified around RESOLVE galaxies. Four features with each confidence level from 4 (certain) to 2 (possible) are shown.}
\end{figure*}

\newpage
\begin{figure*}[h!]
\centering
\includegraphics[width=1\linewidth]{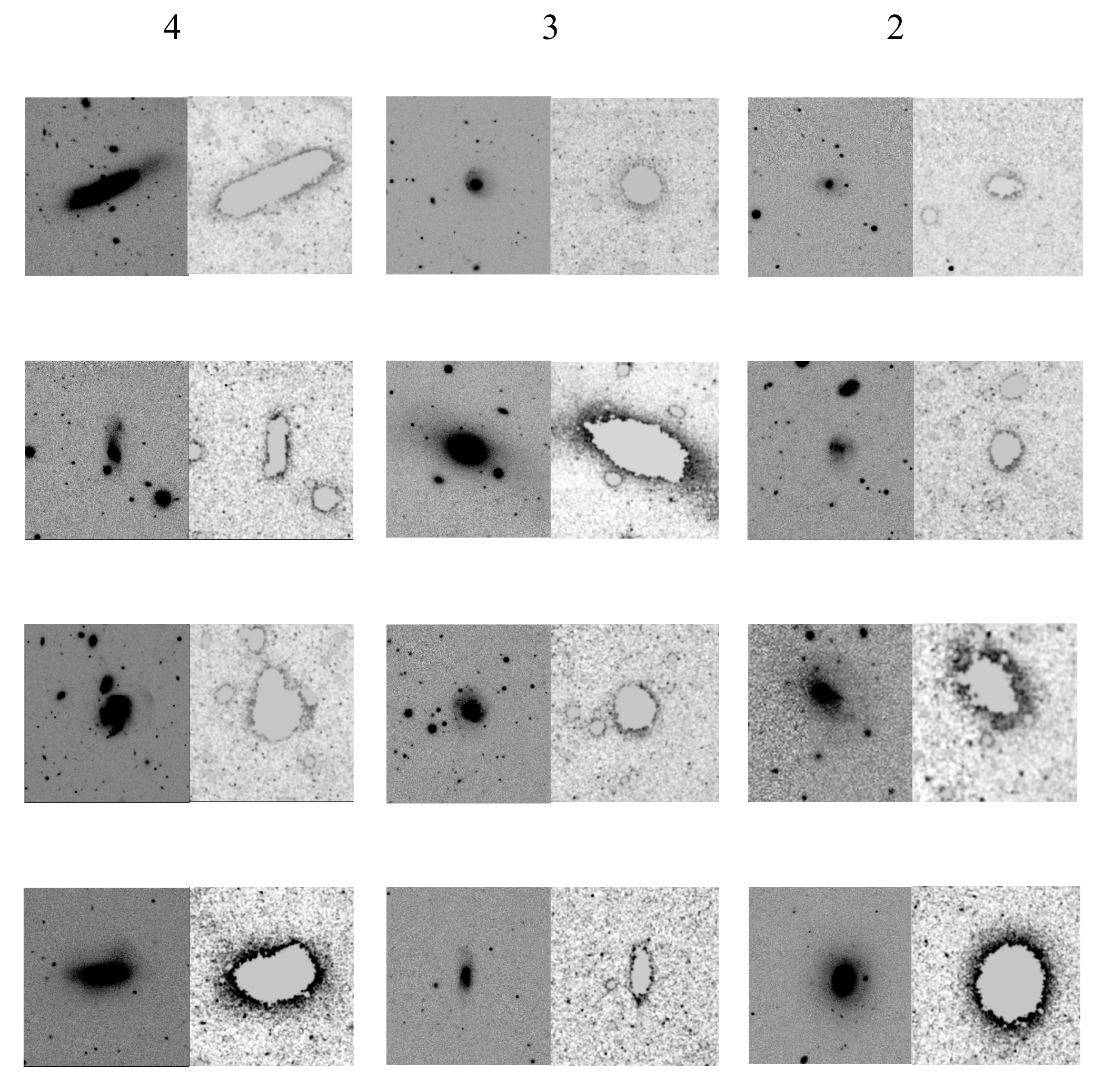}
  		\caption{Examples of ``broad'' tidal features identified around RESOLVE galaxies. Four features with each confidence level from 4 (certain) to 2 (possible) are shown.}
\end{figure*}
% \begin{figure*}[h!]
% \centering
% \includegraphics[width=.65\linewidth]{rf0138comp.pdf}
%   		\caption{Original DECaLS image as well as the masked image of rf0138. A stellar stream is clearly visible coming out of the left side of the galaxy.}
% \end{figure*}

% \begin{figure*}[h!]
% \centering
% \includegraphics[width=.65\linewidth]{rs0112comp.pdf}
%   		\caption{Original DECaLS image as well as the masked image of rs0112. This galaxy hosts a stellar stream which loops around the upper left of the image then crosses to the lower left.}
% \end{figure*}

% \begin{figure*}[h!]
% \centering
% \includegraphics[width=.65\linewidth]{rs0259comp.pdf}
%   		\caption{Original DECaLS image as well as the masked image of rs0259. A tidal tail can be seen extending out of the left side of the galaxy.}
% \end{figure*}

\end{document}